\documentclass[useAMS,fleqn,usenatbib]{mn2e}

\usepackage{amsmath}
\usepackage{amssymb}
\usepackage{graphicx}
\usepackage{xfrac}
\usepackage{txfonts}
\usepackage{natbib}
\usepackage{hyperref}
\hypersetup{
    pdftoolbar=true,        % show Acrobat’s toolbar?
    pdfmenubar=true,        % show Acrobat’s menu?
    pdffitwindow=false,     % window fit to page when opened
    pdfstartview={FitH},    % fits the width of the page to the window
    pdftitle={Order statistic},    % title
    pdfauthor={Jean-Claude Waizmann},     % author
    pdfsubject={Draft},   % subject of the document
    pdfcreator={Jean-Claude Waizmann},   % creator of the document
    pdfproducer={Jean-Claude Waizmann}, % producer of the document
    colorlinks=true,       % false: boxed links; true: colored links
    linkcolor=blue,          % color of internal links
    citecolor=blue,        % color of links to bibliography
    filecolor=blue,      % color of file links
    urlcolor=blue           % color of external links
}

\graphicspath {{./figures/}}
\newcommand{\dd}{\mathrm{d}}

\title[Order statistics of galaxy clusters]{Order statistics applied to the most massive and most distant galaxy clusters}

\author[J.-C. Waizmann, S. Ettori and M. Bartelmann]{J.-C. Waizmann$^{1,2,3}$\thanks{E-mail:
jcwaizmann@oabo.inaf.it}, S. Ettori$^{2,3}$ and M. Bartelmann$^{4}$\\
$^{1}$Dipartimento di Fisica e Astronomia, Universit\`{a} di Bologna, viale Berti Pichat 6/2, I-40127 Bologna, Italy\\
$^{2}$INAF - Osservatorio Astronomico di Bologna, via Ranzani 1, 40127 Bologna, Italy\\
$^{3}$INFN, Sezione di Bologna, viale Berti Pichat 6/2, 40127 Bologna, Italy\\
$^{4}$Zentrum f\"ur Astronomie der Universit\"at Heidelberg, Institut f\"ur 
Theoretische Astrophysik, Albert-Ueberle-Str.~2, 69120 Heidelberg, Germany}

\begin{document}

\date{Received 2011}

\pagerange{\pageref{firstpage}--\pageref{lastpage}} \pubyear{2012}

\maketitle

\label{firstpage}

\begin{abstract}
In this work we present for the first time an analytic framework for calculating the individual 
and joint distributions of the $n$-th most massive or $n$-th highest redshift galaxy cluster 
for a given survey characteristic allowing to formulate $\Lambda$CDM exclusion criteria. 
We show that the cumulative distribution functions steepen with increasing order, giving them a 
higher constraining power with respect to the extreme value statistics. Additionally, we find that 
the order statistics in mass (being dominated by clusters at lower redshifts) is sensitive to the 
matter density and the normalisation of the matter fluctuations, whereas the order statistics in 
redshift is particularly sensitive to the geometric evolution of the Universe. For a fixed cosmology, 
both order statistics are efficient probes of the functional shape of the mass function at the high 
mass end. To allow a quick assessment of both order statistics, we provide fits as a function 
of the survey area that allow percentile estimation with an accuracy better than two per cent.
Furthermore, we discuss the joint distributions in the two-dimensional case and find that for 
the combination of the largest and the second largest observation, it is most likely to find them 
to be realised with similar values with a broadly peaked distribution. When combining 
the largest observation with higher orders, it is more likely to find a larger gap between the 
observations and when combining higher orders in general, the joint pdf peaks more strongly. 

Having introduced the theory, we apply the order statistical analysis to the SPT massive cluster 
sample and MCXC catalogue and find that the ten most massive clusters in the sample are 
consistent with $\Lambda$CDM and the Tinker mass function. For the order statistics in redshift, 
we find a discrepancy between the data and the theoretical distributions, which could in principle 
indicate a deviation from the standard cosmology. However, we attribute this deviation to the 
uncertainty in the modelling of the SPT survey selection function.  In turn, by assuming the 
$\Lambda$CDM reference cosmology, order statistics can also be utilised for consistency checks of 
the completeness of the observed sample and of the modelling of the survey selection function.
\end{abstract}
%------------------------------------------
\section{Introduction}\label{sec:intro}
%------------------------------------------
Clusters of galaxies represent the top of the hierarchy of gravitationally bound structures in the 
Universe and can be considered as tracers of the rarest peaks of the initial density field. This feature 
renders their abundance across the cosmic history a valuable probe of cosmology 
\citep[for an overview of cluster cosmology see e.g.][and references therein]{Voit2005, Allen2011}. 
The recent years brought significant advances to the field from an observational point of view. 
Past and present surveys, like e.g. the \textit{ROSAT} All Sky Survey \citep[RASS;][]{Voges1999}, 
the Massive Cluster Survey \citep[MACS;][]{Ebeling2001} and the Southpole Telescope 
\citep[SPT;][]{Carlstrom2011}, provided rich data for a multitude of massive clusters 
$(>10^{15}\,M_\odot)$. In the near future, cluster data will be drastically extended in terms of 
completeness, coverage and depth by surveys like for instance \textit{PLANCK} \citep{Tauber2010}, 
\textit{eROSITA} \citep{Cappelluti2011} and \textit{EUCLID} \citep{Laureijs2011}, allowing for 
statistical analyses of the samples with increasing quality.

A particular form of statistical analysis that recently entered focus are falsification experiments 
of the concordance $\Lambda$CDM cosmology, based on the discovery of a single (or a number) 
of cluster(s) being so massive that it (they) could not have formed in the standard picture 
\citep{Hotchkiss2011, Hoyle2011, Mortonson2011, Harrison&Coles2012, Harrison&Hotchkiss2012, 
Holz2012, Waizmann2012a, Waizmann2012b}. These studies were triggered by the discovery of 
massive clusters at high redshift \citep[see e.g.][]{Mullis2005, Jee2009, Rosati2009, Foley2011, 
Menanteau2012, Stalder2012}.

However, the usage of a single observation for such falsification experiments requires statistical 
care since several subtleties have to be taken into account. From the theoretical point of view, it 
is necessary to include the Eddington bias \citep{Eddington1913} in mass, as discussed in 
\cite{Mortonson2011} and the bias that stems from the \textit{a posteriori} choice of the redshift 
interval for the analysis \citep{Hotchkiss2011}. From the observational point of view, it might, 
particularly for very high redshift systems, be difficult to define the survey area and selection 
function that are appropriate for the statistical analysis. Combining all of these effects, recent studies 
\citep{Hotchkiss2011, Harrison&Coles2012, Harrison&Hotchkiss2012, Waizmann2012a, Waizmann2012b} 
converge to the finding that, when taken alone, none of the single most massive known clusters 
can be considered in tension with the concordance $\Lambda$CDM cosmology.

Conceptually, inference based on a single observation is not desirable, because by nature the 
extreme value might not be representative for the underlying distribution from which it is 
supposedly drawn. Thus, it is advised to incorporate statistical information from the sample of 
the most massive high redshift clusters, which in turn are also particularly sensitive to the 
underlying cosmological model since they probe the exponentially suppressed tail of 
the mass function.

In this work, we introduce order statistics as a tool for analytically deriving distribution functions 
for all members of the mass and redshift hierarchy ordered by magnitude. By dividing our analysis 
in the observables mass and redshift, we avoid the bias due to an \textit{a posteriori} definition of redshift 
intervals \citep{Hotchkiss2011} and avoid as well the arbitrariness of an \textit{a priori} choice that had been necessary in our 
previous works based on the extreme value statistics. Furthermore, the formalism also allows for 
the formulation of joint probabilities of the order statistics. In the second part of this work, we 
compare our individual and joint analytic distributions to observed samples of massive galaxy 
clusters.

This paper is structured according to the following scheme. In \autoref{sec:order}, we introduce 
the statistical branch of order statistics by discussing the basic mathematical relations in 
\autoref{sec:theory} and by applying the formalism to the distribution of massive galaxy clusters in 
mass and redshift in \autoref{sec:connection_to_cosmology}. This is followed by a discussion 
of how the order statistics of haloes in mass and redshift depends on cosmological parameters 
in \autoref{sec:dependence_on_cosmology}. In order to compare our analytic results to 
observations, we prepare observed cluster samples for the analysis in 
\autoref{sec:preparing_samples}. Afterwards, we discuss the results of the comparison for the 
case of the individual order statistic in \autoref{sec:individual_analysis} and for the joint case in 
\autoref{sec:joint_analysis}. Then, we summarise our findings in \autoref{sec:summary} 
and draw our conclusions in \autoref{sec:conclusions}. In \autoref{sec:appendixA} we give a 
more detailed overview of order statistics and in \autoref{sec:appendixB} fitting formulae for 
the order statistics in mass and redshift are presented.

Throughout this work, unless stated otherwise, we adopt the \textit{Wilkinson Microwave 
Anisotropy Probe 7--year} (WMAP7) parameters $(\Omega_{{\rm m}0}, \Omega_{\Lambda 0}, 
\Omega_{{\rm b}0}, h, \sigma_8) = (0.727, 0.273, 0.0455, 0.704, 0.811)$ \citep{Komatsu2011}.
%------------------------------------------
\section{Order statistics}\label{sec:order}
%------------------------------------------
Order statistics \citep[for an introduction, see e.g.][]{Arnold1992, David&Nagaraja2003} is the 
study of the statistics of ordered (sorted by magnitude) random variates. In this 
section, the basic mathematical relations and the connection to cosmology are introduced 
as they will be needed in remainder of this work. 
%-------------------------------------
\begin{figure*}
\centering
\includegraphics[width=0.49\linewidth]{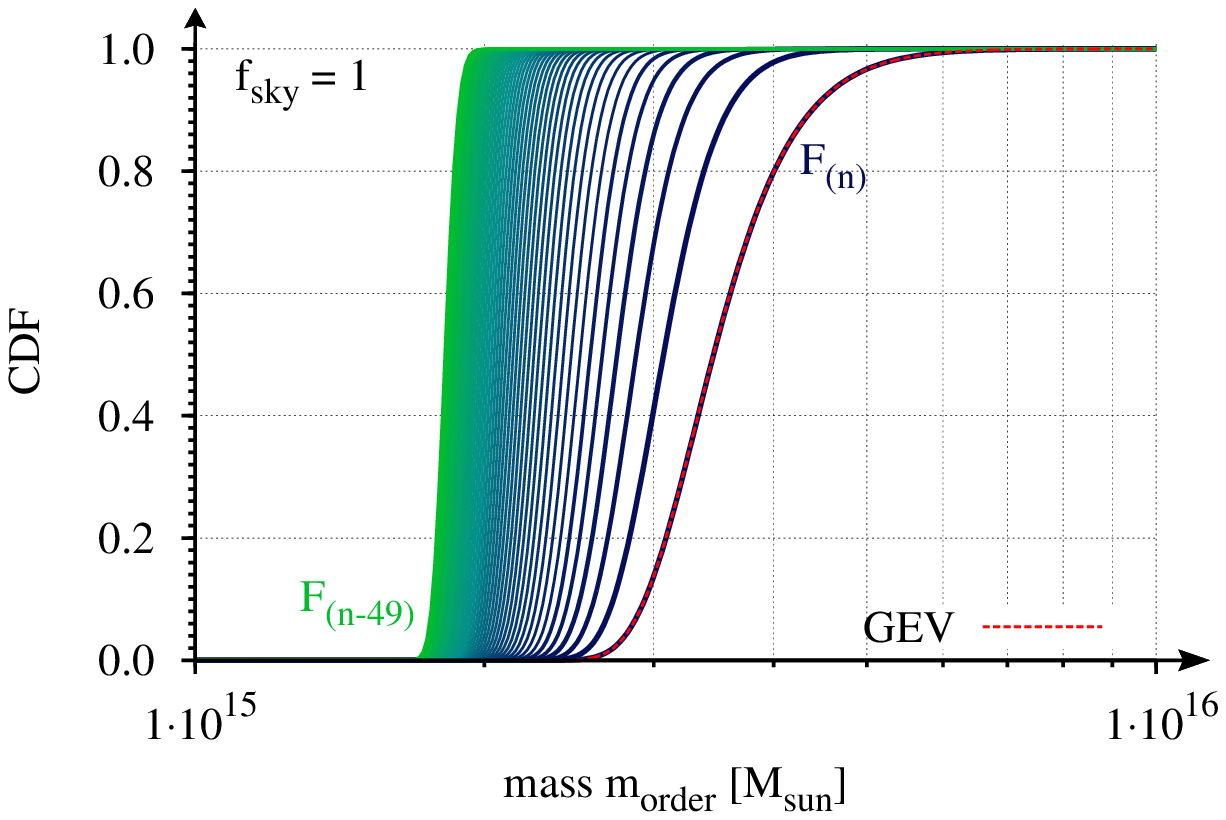}
\includegraphics[width=0.49\linewidth]{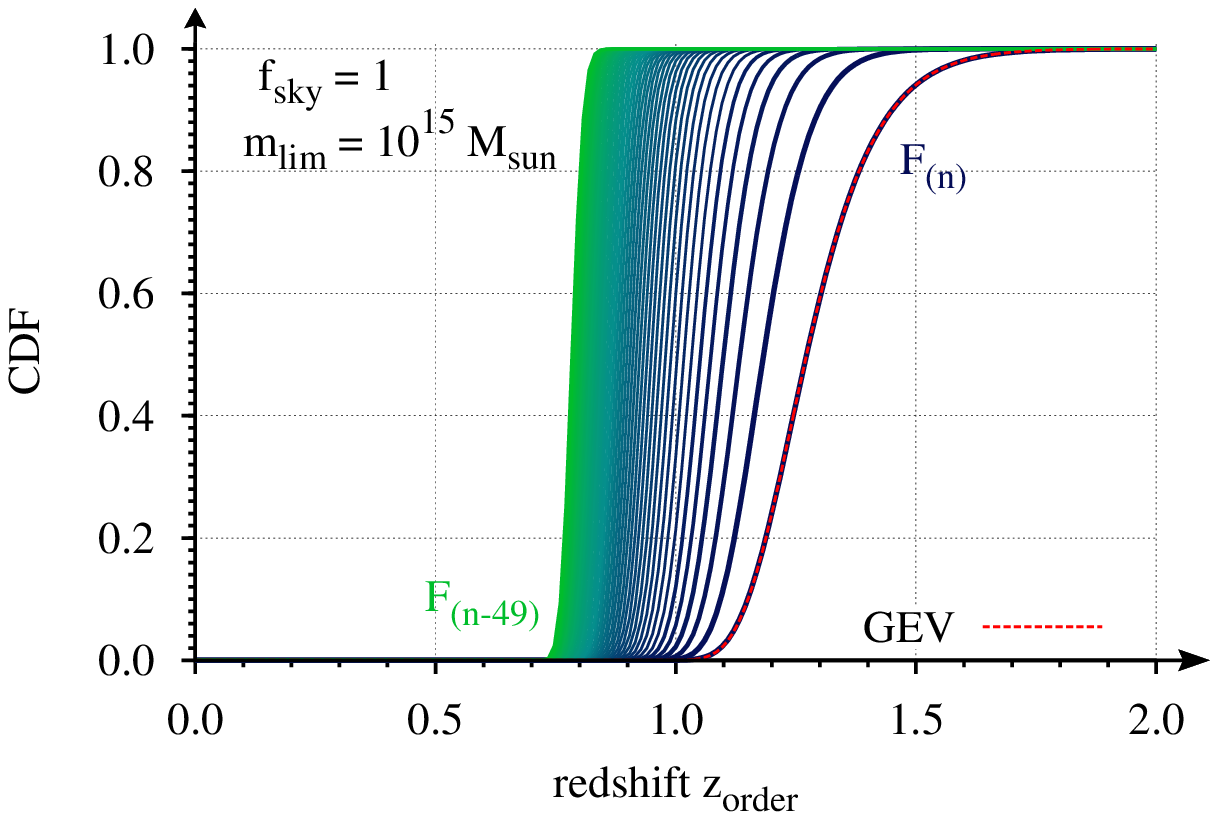}
\caption{Cumulative distribution functions of the first fifty orders from $F_{(n-49)}$ to 
$F_{(n)}$ in mass (left panel) and in redshift (right panel). For comparison, the GEV 
distribution of the maxima is shown by the red, dashed line for both cases. All distributions 
were calculated for the full sky, assuming the Tinker mass function. For the order statistics 
in redshift, a limiting mass of $m_{\rm lim}=10^{15}\,M_\odot$ has been adopted.}
\label{fig:first50}
\end{figure*}
%-------------------------------------
%------------------------------------------
\subsection{Mathematical prerequisites}\label{sec:theory}
%------------------------------------------
Let $X_1,X_2,\dots ,X_n$ be a random sample of a continuous population with the probability 
density function (pdf), $f(x)$, and the corresponding cumulative distribution function (cdf), 
$F(x)$. Further, let $X_{(1)}\le X_{(2)}\le \cdots \le X_{(n)}$ be the order statistic, the random 
variates ordered by magnitude, where $X_{(1)}$ is the smallest (minimum) and $X_{(n)}$ denotes 
the largest (maximum) variate. It can be shown (see \autoref{sec:appendixA_individual}) that the 
pdf of $X_{(i)}\,(1\le i\le n)$ is given by 
\begin{equation}\label{eq:f_order}
f_{(i)}(x)=\frac{n!}{(i-1)!(n-i)!}\left[F(x)\right]^{i-1}\left[1-F(x)\right]^{n-i}f(x)\, .
\end{equation}
The corresponding cdf of the $i$-th order reads then
\begin{equation}\label{eq:F_order}
F_{(i)}(x)=\sum_{k=i}^n\binom{n}{k}\left[F(x)\right]^k\left[1-F(x)\right]^{n-k}\, ,
\end{equation}
and the distribution function of the smallest and the largest value are found to be
\begin{equation}\label{eq:cdf_min}
F_{(1)}(x)=1-\left[1-F(x)\right]^n\, ,
\end{equation}
and 
\begin{equation}\label{eq:cdf_max}
F_{(n)}(x)=\left[F(x)\right]^n\, .
\end{equation}
In the limit of very large sample sizes both $F_{(n)}(x)$ and $F_{(1)}(x)$ can be described by a 
member of the general extreme value (GEV) distribution \citep{Fisher1928,  Gnedenko1943}
%---------------------------
\begin{equation}
 G(x) = \exp{\left\lbrace -\left[1+\gamma \left(\frac{x-\alpha}{\beta}\right)\right]^
 {-1/\gamma}\right\rbrace}\, ,
 \label{eq:cdf_gev}
\end{equation}
where $\alpha$ is the location-, $\beta$ the scale- and $\gamma$ is the shape-parameter. 
Usually these parameters are obtained directly from the data or from an underlying model 
(see for instance \cite{Coles2001}).

Apart from the distributions of the single order statistics, it is very interesting to derive joint 
distribution functions for several orders. The joint pdf of the two order statistics 
$X_{(r)},X_{(s)}\;(1\le r < s \le n)$ is for $x<y$ given by (see \autoref{sec:appendixA} for a 
more detailed discussion)
\begin{align}
f_{(r)(s)}(x,y)=&\frac{n!}{(r-1)!(s-r-1)!(n-s)!}\nonumber\\
 &\times\left[F(x)\right]^{r-1}\left[F(y)-F(x)\right]^{s-r-1}\left[1-F(y)\right]^{n-s}\nonumber\\
 &\times f(x)f(y)\, .
 \label{eq:joint_pdf_2d}
\end{align}
The joint cumulative distribution function can e.g. be obtained by integrating the pdf above or 
by a direct argument and is found to be given by
\begin{align}
F_{(r)(s)}(x,y)=& \sum_{j=s}^{n}\sum_{i=r}^{j}\frac{n!}{i!(j-i)!(n-j)!}\nonumber\\
						&\times\,\left[F(x)\right]^{i}\left[F(y)-F(x)\right]^{j-i}\left[1-F(y)\right]^{n-j}\, .
 \label{eq:joint_cdf_2d}
\end{align}
Analogously the above relations can be generalised to the joint pdf of $X_{n_1},\dots ,X_{n_k}\;
(1\le n_1 <\cdots < n_k\le n)$ for $x_1\le \cdots\le x_k$, which is given by
\begin{align}
f_{(x_1)\cdots(x_k)}(x_1,\dots, x_k)=
	& \frac{n!}{(n_1-1)!(n_2-n_1-1)!\cdots (n-n_k)!} \nonumber\\
	&\times\left[F(x_1)\right]^{n_1-1}f(x_1)\left[F(x_2)-F(x_1)\right]^{n_2-n_1-1}\nonumber \\
	&\times f(x_2)\cdots\left[1-F(x_k)\right]^{n-n_k} f(x_k)\, .
	 \label{eq:joint_pdf_Nd}
\end{align}
Further details and derivations concerning order statistics can be found in the 
\autoref{sec:appendixA}. In the remainder of this work we will repeatedly make use of 
percentiles. In statistics, a percentile is defined as the value of a variable 
below which a certain percentage, $p$, of observations fall. Percentiles can be directly 
obtained from the inverse of the cdf and will be hereafter denoted as $Qp$.
%-------------------------------------
\subsection{Connection to cosmology}\label{sec:connection_to_cosmology}
%-------------------------------------
As outlined in the previous subsection, the only quantity that is needed for calculating 
the cdfs, $F_{(i)}(x)$, of the order statistics (see \autoref{eq:F_order}) is the cdf, $F(x)$, 
of the underlying distribution from which the sample is drawn. 
Assuming the random variates, $X_i$, to be the masses of galaxy clusters, then the cdf, 
$F(m)$, can be calculated (see e.g. \cite{Harrison&Coles2012}) by means of
\begin{equation}
F(m)=\frac{f_{\rm sky}}{N_{\rm tot}}\left[ \int_{0}^{\infty}\int_{0}^{m}\dd z\,\dd M\,
\frac{\dd V}{\dd z}\frac{n(M,z)}{\dd M}\right]\; ,
\end{equation}
where the total number of clusters, $N_{\rm tot}$, is given by
\begin{equation}
N_{\rm tot}=f_{\rm sky}\left[ \int_{0}^{\infty}\int_{0}^{\infty}\dd z\,\dd M\,
\frac{\dd V}{\dd z}\frac{n(M,z)}{\dd M}\right]\; .
\end{equation}
Here, $f_{\rm sky}$ is the fraction of the full sky that is observed, $(\dd V / \dd z)$ is 
the volume element and $n(m,z)$ is the halo mass function. If needed, the corresponding 
pdf can always be obtained by $f(m)=\dd F(m) / \dd m$.

Analogously, the order statistics can be calculated as well for the redshift instead of the mass. 
In this case the cdf reads
\begin{equation}
F(z)=\frac{f_{\rm sky}}{N_{\rm tot}}\left[ \int_{0}^{z}\int_{m_{\rm lim}(z)}^{\infty}\dd z\,\dd M\,
\frac{\dd V}{\dd z}\frac{n(M,z)}{\dd M}\right]\; ,
\end{equation}
where
\begin{equation}
N_{\rm tot}=f_{\rm sky}\left[ \int_{0}^{\infty}\int_{m_{\rm lim}(z)}^{\infty}\dd z\,\dd M\,
\frac{\dd V}{\dd z}\frac{n(M,z)}{\dd M}\right]\; .
\end{equation}
For the latter, the order statistics does no longer depend only on the survey area via $f_{\rm sky}$ 
but, in addition the selection function of the survey has to be included via a limiting survey 
mass, $m_{\rm lim}(z)$. In this work, we do not attempt to model the possible redshift 
dependence of $m_{\rm lim}$ and assume it to be constant throughout the remainder of 
this work.

With the distributions $F(m)$ and $F(z)$ at hand, we can now easily derive the cdfs of the 
corresponding order statistics. Since we will focus in this work on the few largest values, 
we will refer to the distribution of the maximum, $F_{(n)}(x)$, as first order, to the second 
largest as second order and so on. 

We calculated the distributions of the first fifty orders from $F_{(n)}(x)$ to $F_{(n-49)}(x)$, 
where $n=N_{\rm tot}$, and present the results in \autoref{fig:first50} for the mass 
(left panel) and redshift (right panel). In both panels the color decodes the order of the 
distribution, ranging from the blue for $F_{(n)}(x)$ to the green for $F_{(n-49)}(x)$. For both 
cases we assumed $f_{\rm sky}=1$, a redshift range of $0\le z\le\infty$ and the 
\cite{Tinker2008} mass function. In the case of the order statistics in redshift, we assume 
a limiting survey mass of $m_{\rm lim}=10^{15}\,M_\odot$. It can be nicely seen how, with 
increasing order (from blue to green), the cdfs shift in both cases to smaller values of the 
mass or redshift. 

A first important result is that, with the increasing order, the 
cdfs steepen, which results in an enhanced constraining power, since small shifts in the 
mass or redshift may yield large differences in the derived probabilities. In this sense the higher 
orders will be more useful for falsification experiments than the extreme value distribution 
which, due to its shallow shape, requires extremely large values of the observable to 
statistically rule out the underlying assumptions. Since higher orders encode information 
from the $n$ most extreme objects, deviations from the expectation are statistically more 
significant for $n$ values instead of a single extreme one.

In addition, we compare the distribution of the maxima $F_{(n)}(m)$ and $F_{(n)}(z)$ to those 
obtained from a extreme value approach \citep[][Metcalf \& Waizmann in prep.]
{Davis2011, Waizmann2012a} based on the void probability \citep{White1979}, using 
\autoref{eq:cdf_gev}. For both cases presented in \autoref{fig:first50}, the red, dashed 
curve of the GEV distribution, $G(x)$, agrees very well with the directly calculated 
$F_{(n)}(x)$.

In order to allow a quick estimation of the distributions of the order statistics, we provide 
in the \autoref{sec:appendixB} also fitting formulae for $F(x)$ as a function of the survey area 
for the cases of mass and redshift. The fitting formulae for the distribution in mass allow an 
estimation of the quantiles in the range from the 2-percentile, $Q2$, to the 98-percentiles, 
$Q98$, with an accuracy better than one per cent for $A_{\rm s} \gtrsim 200\,{\rm deg}^2$ 
and for the ten largest masses. In the instance of the order statistics in redshift, the quality 
of the fits depends on $m_{\rm lim}$ as well. For $m_{\rm lim}=10^{15}\;M_\odot$ an 
accuracy of better than two per cent can be achieved for $A_{\rm s}\gtrsim  2000\,{\rm deg}^2$ 
and for $m_{\rm lim}=5\times 10^{14}\;M_\odot$ the same accuracy is obtained down to 
$A_{\rm s}=100\,{\rm deg}^2$. A more detailed discussion of the fitting functions and their 
performance can be found in \autoref{sec:appendixB}.

In the remaining part of this work, we will discuss how the underlying cosmological model 
affects the order statistics and confront the theoretically derived order statistics with 
observations, afterwards. 
%-------------------------------------
\begin{figure*}
\centering
\includegraphics[width=0.49\linewidth]{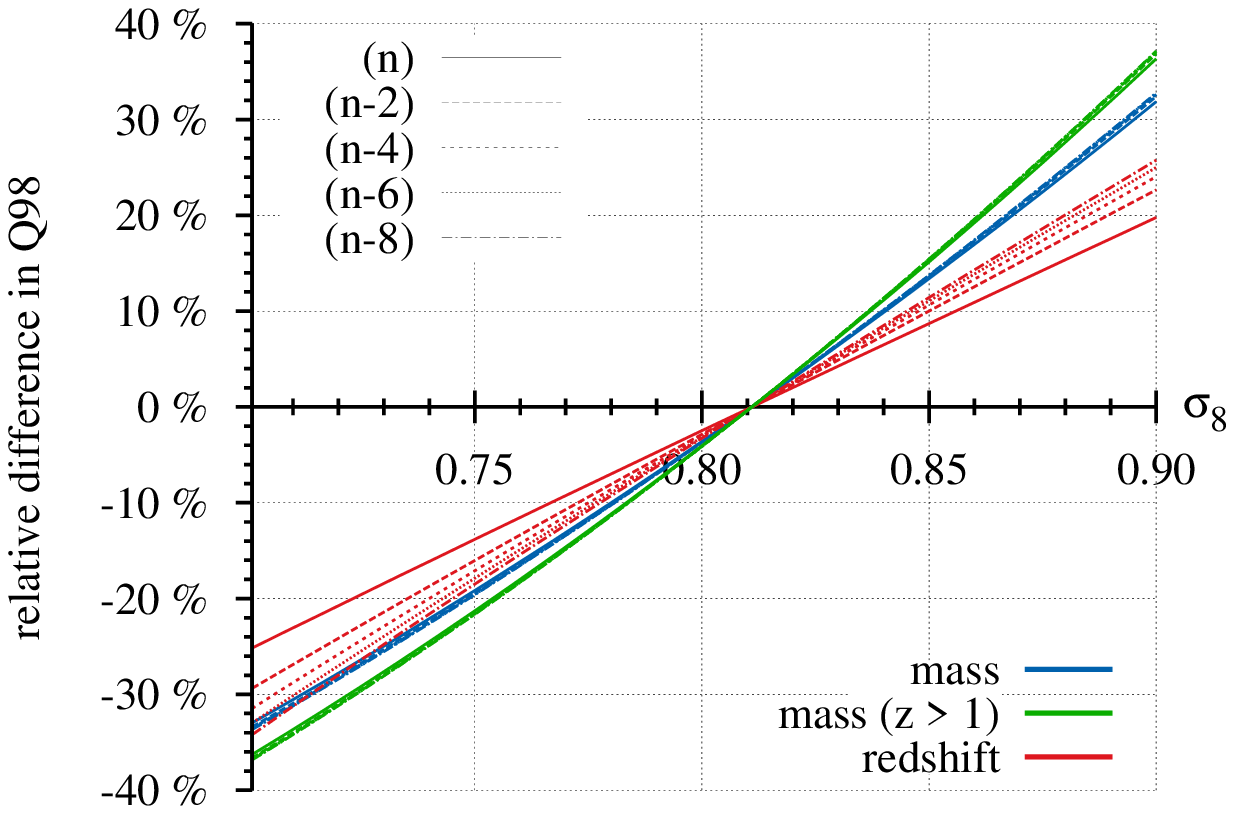}
\includegraphics[width=0.49\linewidth]{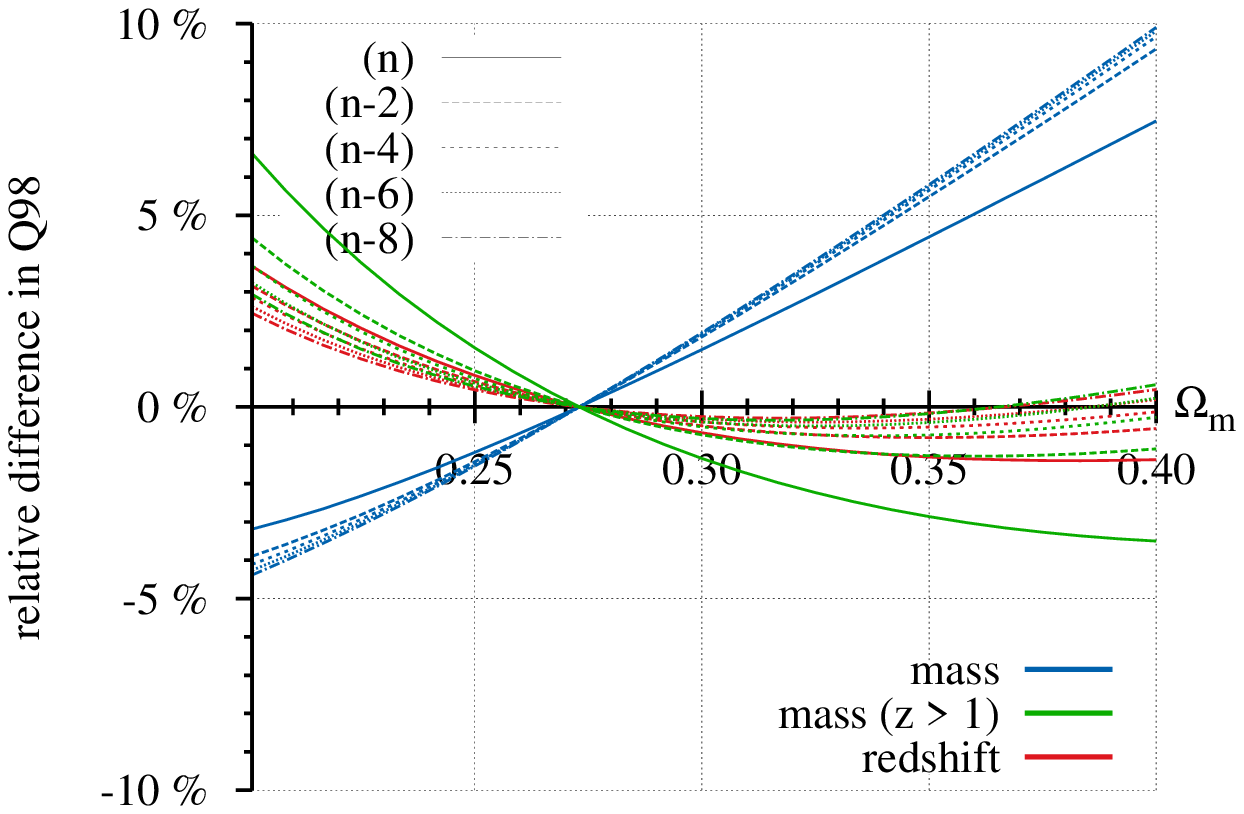}\\
\includegraphics[width=0.49\linewidth]{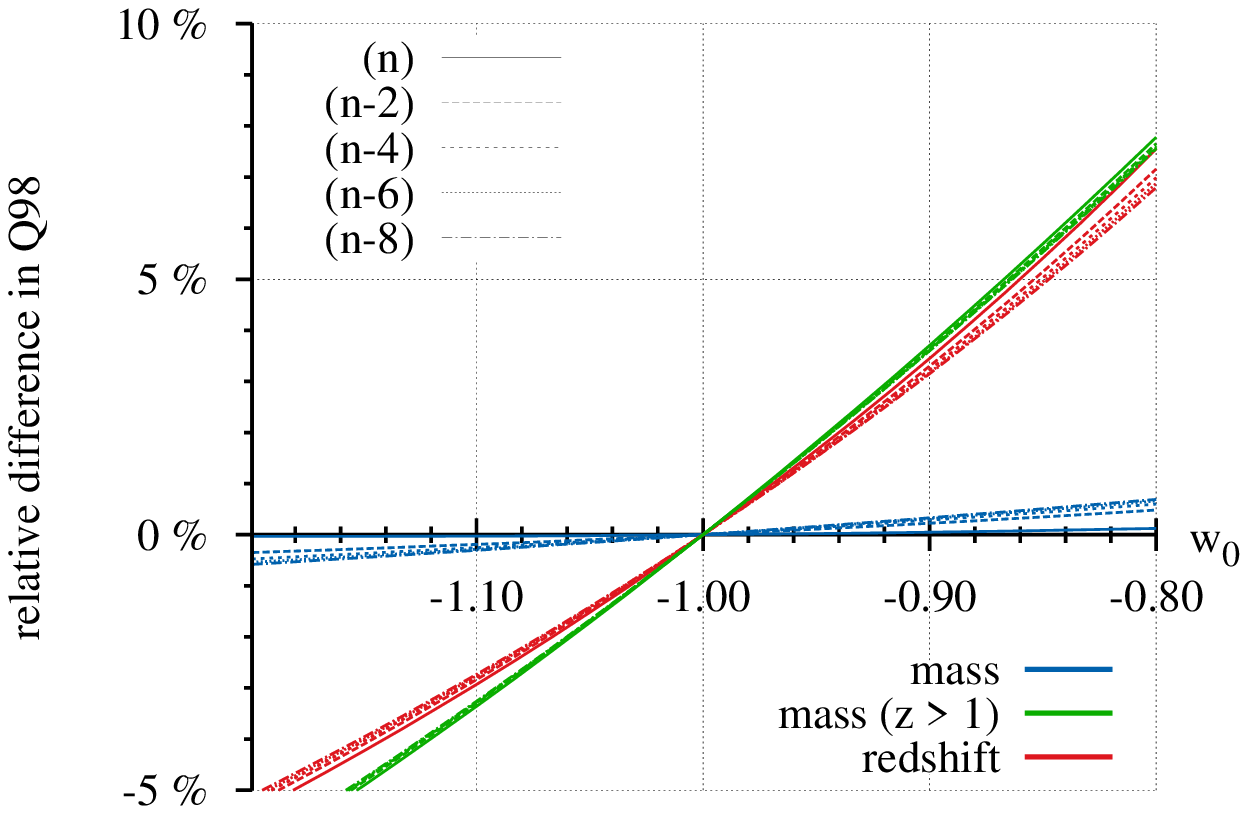}
\includegraphics[width=0.49\linewidth]{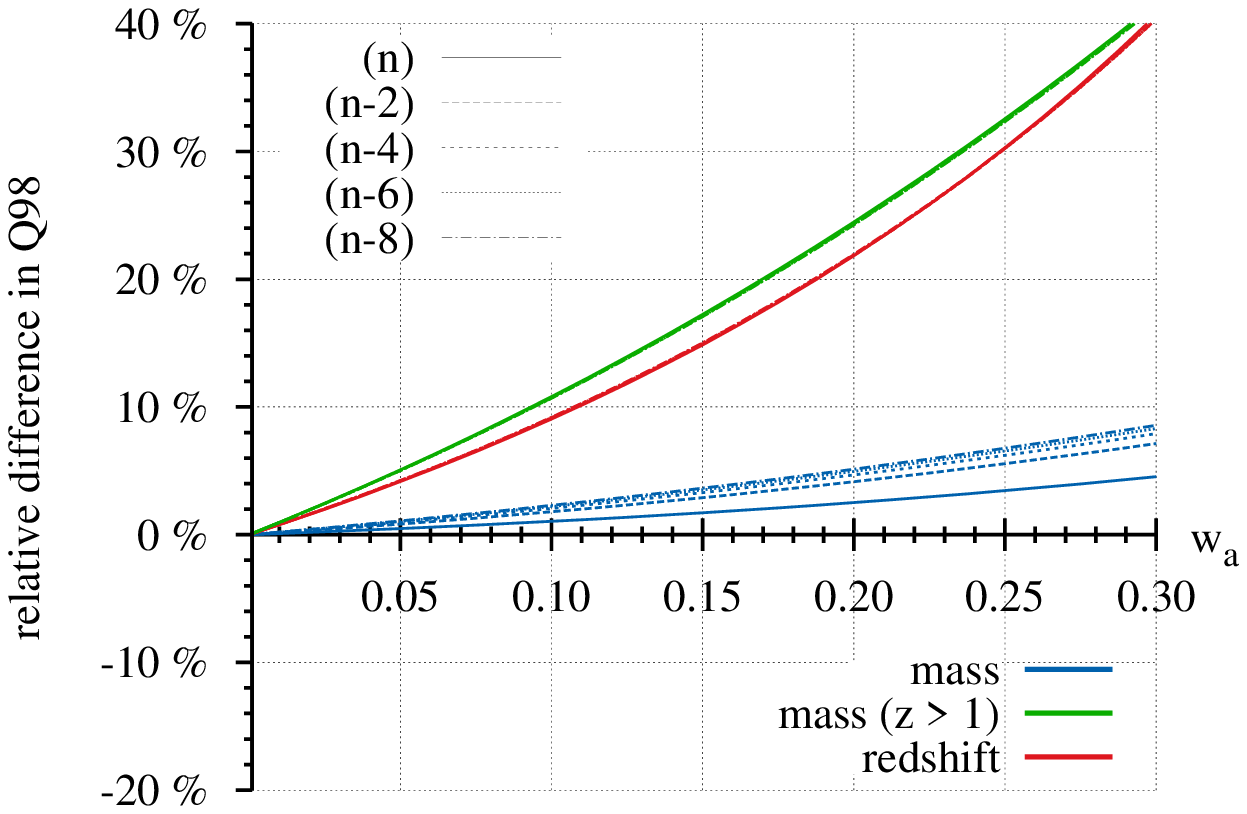}
\caption{Relative differences in the 98 percentile, $Q98$, with respect to the $\Lambda$CDM 
case for the order statistics in mass (blue lines), in mass with a lower redshift limit of $z=1$ 
(green lines) and in redshift (red lines) as a function of different cosmological parameters. The 
upper left panel shows the variation with $\sigma_8$, the upper right panel the one with 
$\Omega_{\rm m}$, the lower left one the one with the constant equation of state parameter 
$w_0$ and the lower right one shows the variation with the derivative of a linearised model for 
a time dependent equation of state $w_a$. The different line-styles denote different orders as 
indicated in the individual panels. For all calculations, the full sky and the Tinker mass function 
were assumed.}\label{fig:impact_of_cosmology}
\end{figure*}
%-------------------------------------
%-------------------------------------
\begin{figure*}
\centering
\includegraphics[width=0.45\linewidth]{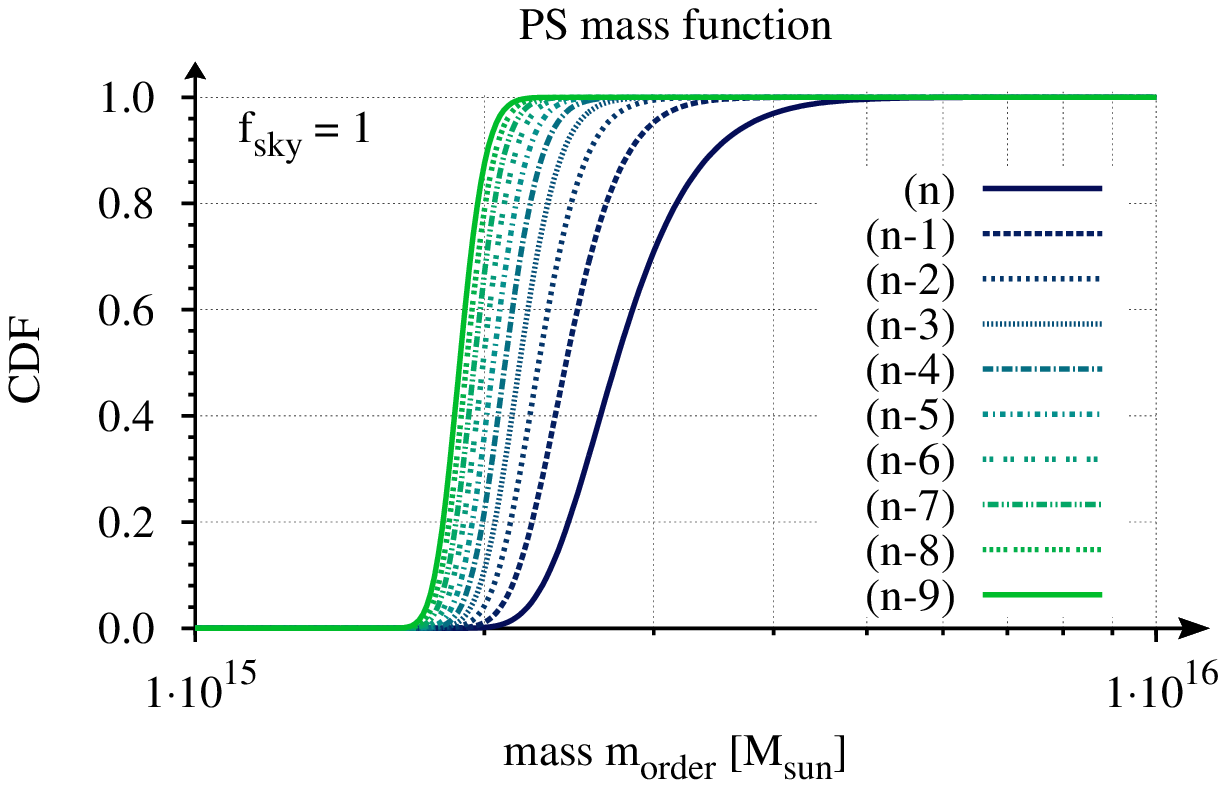}
\includegraphics[width=0.45\linewidth]{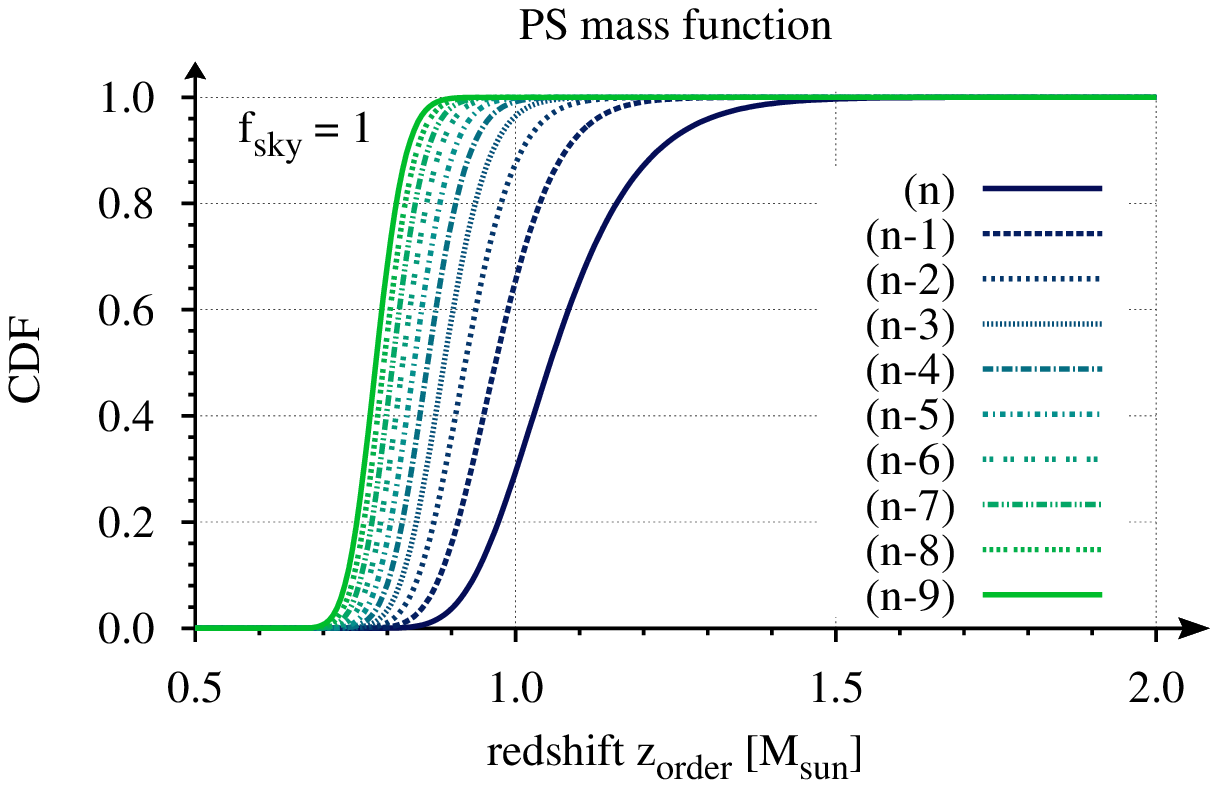}\\
\includegraphics[width=0.45\linewidth]{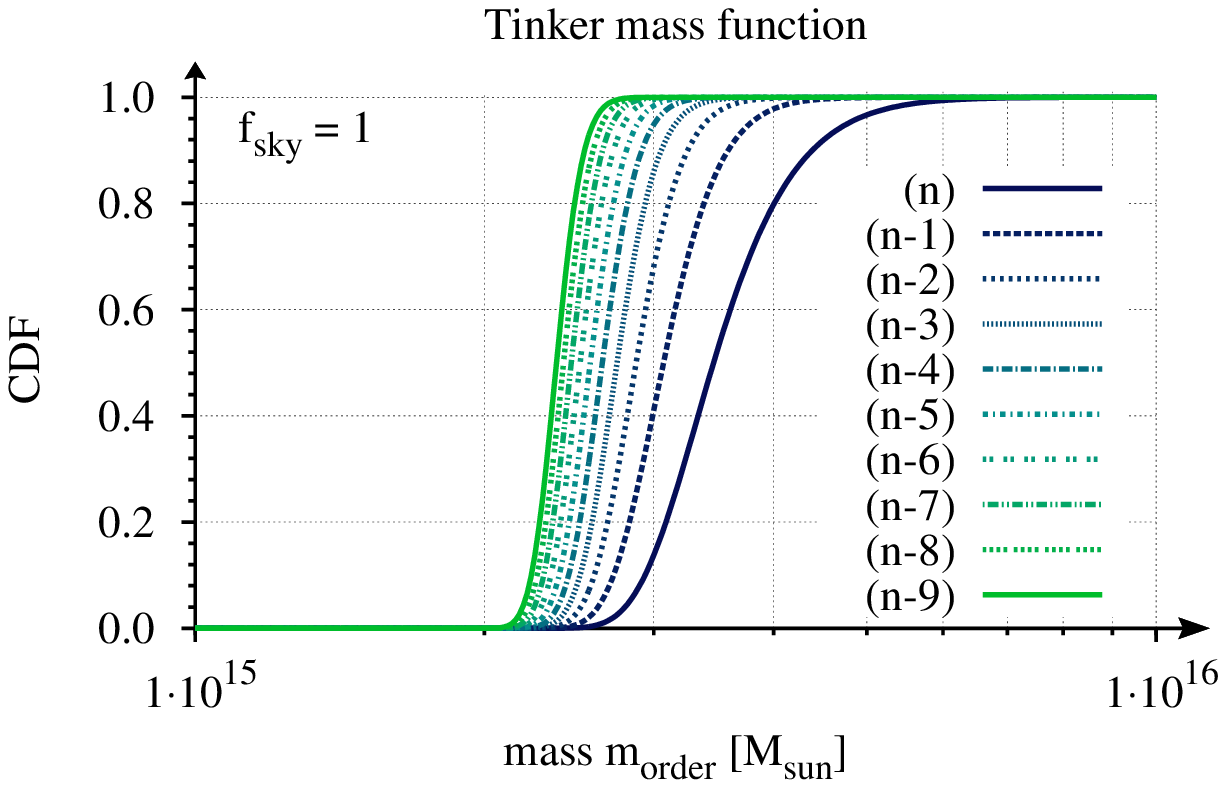}
\includegraphics[width=0.45\linewidth]{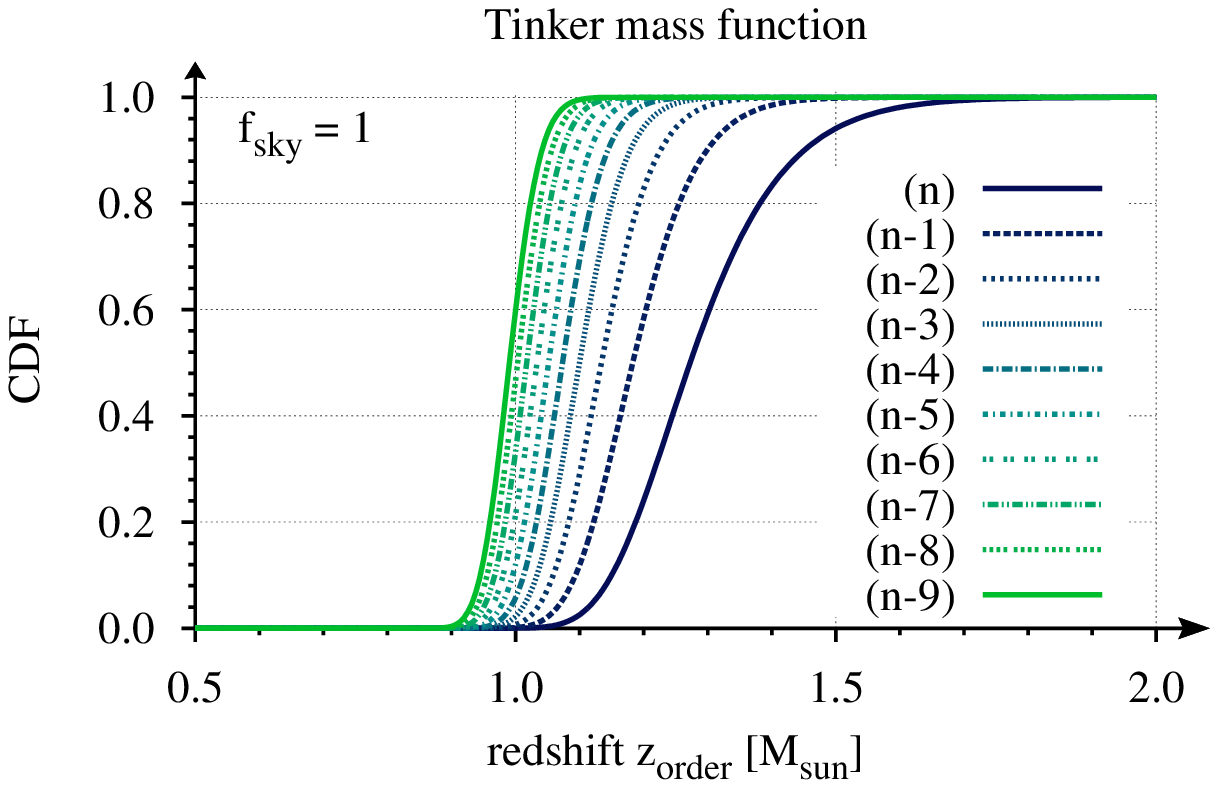}\\
\includegraphics[width=0.45\linewidth]{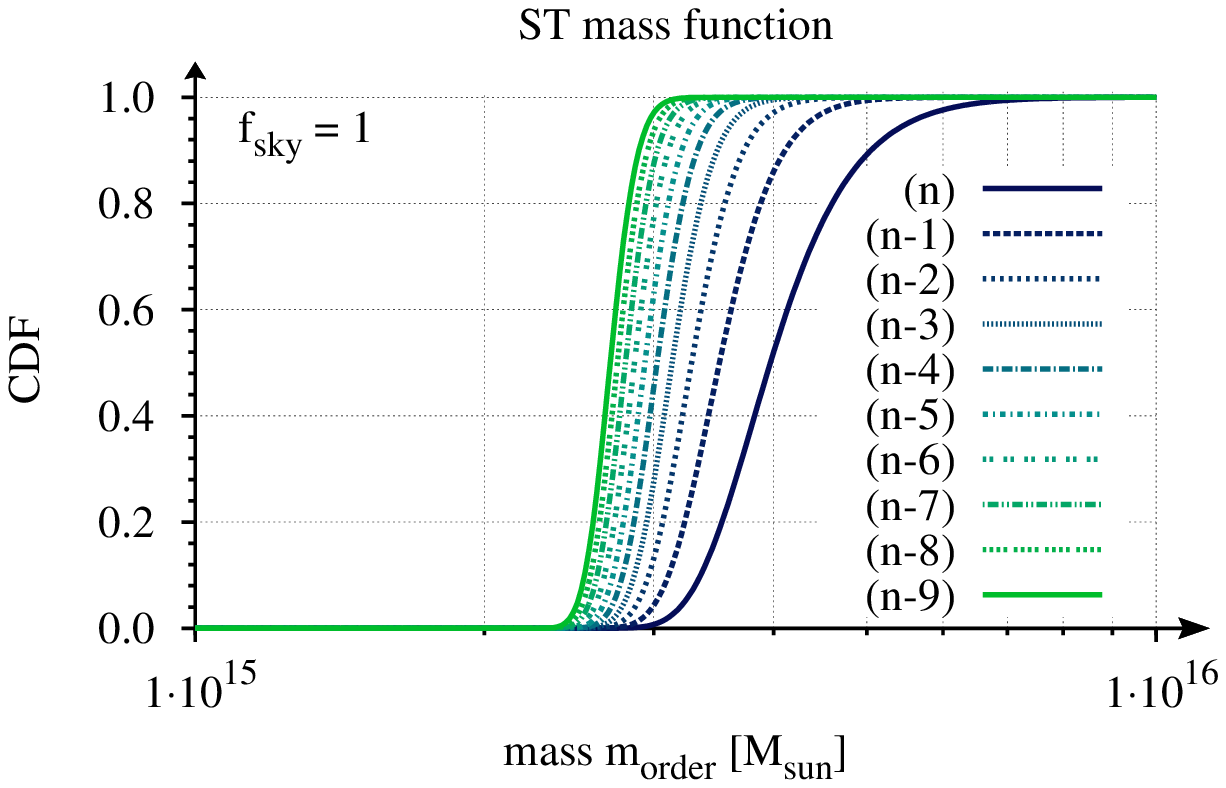}
\includegraphics[width=0.45\linewidth]{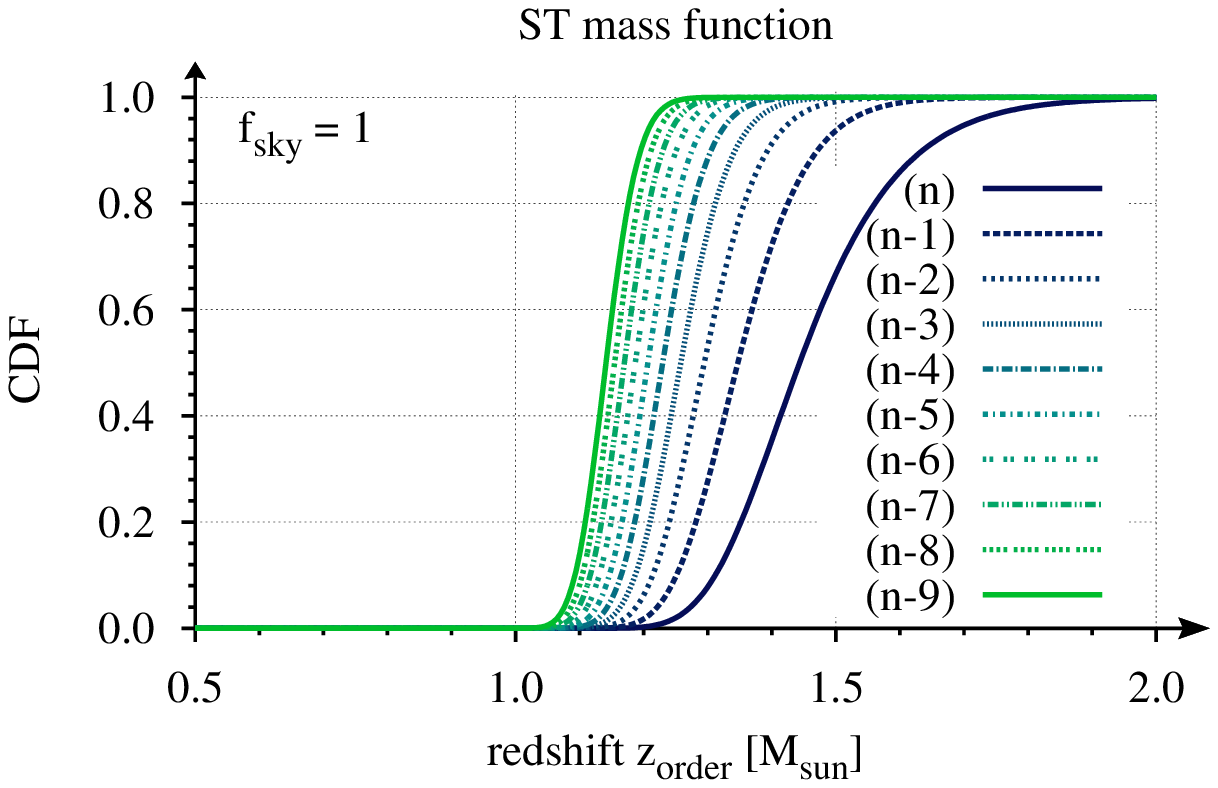}
\caption{Impact of the different mass function on the cdfs of the first ten orders $F_{(n)},
\dots ,F_{(n-9)}$ in mass (left column) and redshift (right column). All distributions were 
computed for the full sky and three different mass functions comprising the Press-Schechter 
(PS), the Tinker and the Sheth\& Tormen (ST) ones, ordered from top to bottom. For the 
distributions in redshift, a limiting survey mass of $m_{\rm lim}=10^{15}\,M_\odot$ has 
been assumed.}
\label{fig:impact_massfunction}
\end{figure*}
%-------------------------------------
%------------------------------------------
\section{Dependence of the order statistics on the underlying cosmology}\label{sec:dependence_on_cosmology}
%------------------------------------------
Eventually, the order statistics in mass and redshift is determined by the number of galaxy 
clusters in a given cosmic volume. The quantities that impact on this number can be categorised 
into two classes. The first one contains all effects that modify structure formation itself, like the 
choice of the mass function or the amplitude of the mass fluctuations, $\sigma_8$, for instance. 
These effects manifest themselves most strongly in the exponentially suppressed tail of the mass 
function, hence at high masses. The second class contains all the effects that modify the 
geometric evolution of the Universe. By changing the evolution of the cosmic volume, the number 
of clusters in a given redshift range can be substantially different, even if both cosmologies yield 
the same the number density of objects of a given mass \citep[see e.g.][]{Pace2010}.
%-------------------------------------
\subsection{Impact of cosmological parameters}
%-------------------------------------
In order to quantify the impact of different cosmological parameters on the order statistics in 
mass and redshift, we study the effect on the $98$-percentile, $Q98$, which we use to define 
possible outliers from the underlying distribution. In \autoref{fig:impact_of_cosmology}, we 
present the relative difference in $Q98$ as a function of four different cosmological parameters comprising 
$\sigma_8$, $\Omega_{\rm m}$ (assuming the flatness constraint), the equation of state parameter, 
$w_0$, and the derivative $w_{a}$ from the relation $w(a)=w_0+w_a(1-a)$, where $a$ denotes the 
scale factor. A non-vanishing value of the latter indicates a time-varying equation of state. In each 
panel of \autoref{fig:impact_of_cosmology}, we show the relative differences for $5$ different orders, 
of the order statistic in mass with $z\in[0,\infty]$ (blue lines) and $z\in[1,\infty]$ (green lines), as 
well as in redshift (red lines) assuming $m_{\rm lim}=10^{15}\,M_\odot$. For all calculations we 
assumed the full sky and the \cite{Tinker2008} mass function.

It can be seen that order statistics is very sensitive to $\sigma_8$, such that the relative 
differences in $Q98$ would amount to $\sim 7$ per cent for the range allowed by WMAP7 of 
$(\sigma_8=0.811\pm 0.023)$. All three order statistics exhibit the same functional behaviour, 
with the mass-based ones being more sensitive than the redshift-based one. This can be 
understood by the fact that the mass-based order statistics probe the most massive clusters and 
hence the exponential tail of the mass function which is highly sensitive to $\sigma_8$.

For modifications of the matter density, $\Omega_{\rm m}$, assuming the flatness constraint 
$\Omega_\Lambda=1-\Omega_{\rm m}$, the situation is substantially different from the previous 
case (see upper right panel of \autoref{fig:impact_of_cosmology}). Overall, the order statistics are 
less sensitive and they do not exhibit the same functional behaviour. The order statistics in mass 
(blue lines) performs best for larger value of $\Omega_{\rm m}$ because the most massive clusters 
will reside at rather low redshifts. At high redshifts (green and red lines), the increase in 
$\Omega_{\rm m}$ and hence, the decrease in $\Omega_{\Lambda}$, yields a smaller number of 
very massive clusters. Despite the increase in the matter density, the decrease in volume is 
dominating for the range of  $\Omega_{\rm m}$ shown in the plot and, thus, the relative difference 
decreases. In this sense the volume effects dominate at high redshifts over the increase in matter 
density, whereas at low redshifts the increase in matter density dominates.

The lower left panel of \autoref{fig:impact_of_cosmology} shows the sensitivity of the order statistics 
to changes in the constant equation of state, $w_0$. Evidently, the most massive clusters at low 
redshifts (blue line) have no sensitivity to $w_0$, whereas at high redshifts (green and red line) the 
sensitivity is better. The volume effects are, compared to modifications in $\Omega_{\rm m}$, less 
important and the observed increase in the relative difference in $Q98$ with decreasing $w_0$ is 
dominated by modifications of the exponential tail of the mass function \citep[for a more thorough 
discussion, see e.g.][]{Pace2010}.

When assuming a time-dependent equation of state, modelled by $w(a)=w_0+w_a(1-a)$, as 
presented in the lower right panel of \autoref{fig:impact_of_cosmology}, the observed functional 
behaviour can be explained by identical arguments as before. The results exhibit again the high 
sensitivity of the high redshift order statistics on modifications of $w_a$. It should be noted that 
we fixed $w_0=-1.0$ for all cases.

It can be summarised that for modifications that strongly affect the structure formation, like 
$\sigma_8$ for instance, the order statistics in mass for $z\in[0,\infty[$ is comparable in its 
sensitivity to the redshift based order statistics. Modifications that strongly alter the geometric 
evolution of the Universe affect more strongly the order statistics in redshift. However, one should 
keep in mind that in the case of the order statistics in mass, the relative differences are on the same 
level as the inaccuracies in cluster mass estimates. This problem does not occur for redshifts, which 
can be measured to a very high accuracy. Of course, in this case the observational challenge is 
transferred to compiling a sample with a precise mass limit. Apart from the cosmological 
parameters also the choice of the mass function is expected to have a strong effect on the order 
statistics as will be discussed in the following subsection. 
%-------------------------------------
\begin{table*}
% \begin{minipage}{126mm}
\caption{Compilation of the ten most massive galaxy clusters from the SPT massive cluster sample 
\citep{Williamson2011} and the MCXC catalogue \citep{Piffaretti2011}, respectively. The masses 
$M_{200\rm m}$ and $M_{200\rm m}^{\rm Edd}$ are with respect to the mean background 
density before and after the correction for the Eddington bias based on the estimated mass 
uncertainty $\sigma_{\ln M}$. The last column lists the references for the values of the observed 
mass, on which the analysis is based on.}
\begin{tabular}{lcccccc} \hline
Rank & Cluster & $z$ & $M_{200\rm m}$ in units of $M_\odot$ & $\sigma_{\ln M}$ & $M_{200\rm m}^{\rm Edd}$ in units of $M_\odot$ & Reference \\ 
\hline 
\multicolumn{7}{l}{\textbf{SPT catalogue $\mathbf{(A_{\rm s}=2500\,{\rm deg}^2)}$}}\\
$1^{\rm st}$	& SPT-CL J0658-5556 & 0.296 	& $(3.12	\pm 1.15)\times 10^{15}$ & 0.39 & $1.99_{-0.64}^{+0.95}\times 10^{15}$ & \cite{Williamson2011}\\
$2^{\rm nd}$	& SPT-CL J2248-4431 & 0.348 	& $(2.90\pm 1.03)\times 10^{15}$ & 0.37 & $1.91_{-0.59}^{+0.86}\times 10^{15}$ & " \\
$3^{\rm rd}$	& SPT-CL J0102-4915	& 0.870	& $(2.16\pm 0.32)\times 10^{15}$	& 0.15 & $1.98_{-0.28}^{+0.32}\times 10^{15}$	& \cite{Menanteau2012}\\
$4^{\rm th}$	& SPT-CL J0549-6204	& 0.320	& $(1.99\pm 0.67)\times 10^{15}$	& 0.35 & $1.48_{-0.44}^{+0.62}\times 10^{15}$	& \cite{Williamson2011}\\
$5^{\rm th}$	& SPT-CL J0638-5358	& 0.222	& $(1.91\pm 0.62)\times 10^{15}$	& 0.34 & $1.50_{-0.44}^{+0.61}\times 10^{15}$	& "\\
$6^{\rm th}$	& SPT-CL J0232-4421	& 0.284	& $(1.88\pm 0.59)\times 10^{15}$	& 0.32 & $1.48_{-0.41}^{+0.56}\times 10^{15}$	& "\\
$7^{\rm th}$	& SPT-CL J0645-5413	& 0.167	& $(1.81\pm 0.60)\times 10^{15}$	& 0.34 & $1.43_{-0.41}^{+0.58}\times 10^{15}$	& "\\
$8^{\rm th}$	& SPT-CL J0245-5302	& 0.098	& $(1.70\pm 0.46)\times 10^{15}$	& 0.25 & $1.48_{-0.33}^{+0.42}\times 10^{15}$	& "\\
$9^{\rm th}$	& SPT-CL J2201-5956	& 0.300	& $(1.70\pm 0.42)\times 10^{15}$	& 0.28 & $1.48_{-0.36}^{+0.48}\times 10^{15}$	& "\\
$10^{\rm th}$	& SPT-CL J2344-4243	& 0.450	& $(1.65\pm 0.38)\times 10^{15}$	& 0.31 & $1.28_{-0.34}^{+0.47}\times 10^{15}$	& "\\ 
\hline
\multicolumn{7}{l}{\textbf{MCXC catalogue $\mathbf{(A_{\rm s}=27490\,{\rm deg}^2)}$}} \\ 
$1^{\rm st}$	& J0417.5-1154	& 0.4430 & $(3.86\pm 0.62)\times 10^{15}$ & 0.15	 & $3.55_{-0.49}^{+0.57}\times 10^{15}$ & \cite{Piffaretti2011} \\
$2^{\rm nd}$	& J2211.7-0349	& 0.3970 & $(3.21\pm 0.51)\times 10^{15}$ & 0.15	 & $2.98_{-0.42}^{+0.48}\times 10^{15}$ & " \\
$3^{\rm rd}$	& J2243.3-0935	& 0.4470 & $(3.04\pm 0.49)\times 10^{15}$ & 0.15	 & $2.82_{-0.39}^{+0.46}\times 10^{15}$ & " \\
$4^{\rm th}$	& J0308.9+2645	& 0.3560 & $(2.95\pm 0.47)\times 10^{15}$ & 0.15	 & $2.75_{-0.38}^{+0.45}\times 10^{15}$ & " \\
$5^{\rm th}$	& J1504.1-0248	& 0.2153 & $(2.36\pm 0.35)\times 10^{15}$ & 0.14	 & $2.25_{-0.29}^{+0.34}\times 10^{15}$ & " \\
$6^{\rm th}$	& J1347.5-1144	& 0.4516 & $(2.30\pm 0.41)\times 10^{15}$ & 0.16	 & $2.14_{-0.32}^{+0.37}\times 10^{15}$ & " \\
$7^{\rm th}$	& J1731.6+2251	& 0.3890 & $(2.28\pm 0.37)\times 10^{15}$ & 0.15	 & $2.14_{-0.30}^{+0.35}\times 10^{15}$ & " \\
$8^{\rm th}$	& J0717.5+3745	& 0.5460 & $(2.21\pm 0.35)\times 10^{15}$ & 0.15	 & $2.06_{-0.29}^{+0.33}\times 10^{15}$ & " \\
$9^{\rm th}$	& J2248.7-4431	& 0.3475 & $(2.13\pm 0.35)\times 10^{15}$ & 0.15	 & $2.01_{-0.28}^{+0.33}\times 10^{15}$ & " \\
$10^{\rm th}$ & J1615.7-0608	& 0.2030 & $(2.14\pm 0.30)\times 10^{15}$ & 0.13	 & $2.03_{-0.25}^{+0.28}\times 10^{15}$ & " \\ 
\hline
\end{tabular}\label{tab:clusters_mass}
%\end{minipage} 
\end{table*}
%-------------------------------------
%-------------------------------------
\subsection{Impact of the choice of the mass function}
%-------------------------------------
When performing a falsification experiment of $\Lambda$CDM using the $n$ most massive 
or $n$ highest redshift clusters, then one has to specify the reference model against which 
the observations have to be compared with. Apart from the cosmological parameters that 
are usually fixed to the obvious choice of the WMAP7 values, a halo mass function has to be 
chosen as well. As mentioned earlier, this is particularly important for galaxy clusters since 
the exponentially suppressed tail of the mass function is naturally very sensitive to 
modifications. 

In order to quantify the impact of different mass functions on the order statistics in mass 
and redshift, we computed the cdfs, $F_{(n-9)},\dots ,F_{(n)}$, for the \cite{Press1974} (PS), 
the \cite{Tinker2008} and the \cite{Sheth&Tormen1999} (ST) mass functions for 
$f_{\rm sky}=1$ and present them from top to bottom in \autoref{fig:impact_massfunction}. 
Comparing the panels to each other reveals the tremendous sensitivity of the distributions to 
the choice of the mass function. Taking the Tinker mass function as a reference, the median, 
$Q50$, changes for both types of order statistics by $-20$ per cent for the PS case and by 
$+15$ percent for the ST case. These differences can be explained by the fact that the ST 
mass function leads to a substantial increase in the number of haloes, particularly at the high 
mass end, whereas the PS mass function results in much fewer haloes in the mass and redshift range of 
interest. For the remainder of this paper we will use the Tinker mass function as reference 
because the halo masses are defined as spherical overdensities with respect 
to the mean background density, a definition that is closer to theory and actual observations 
than friend-of-friend masses.

However, considering that due to statistical limitations, current fits for the mass function are 
still not very accurate for the highest masses $(>3\times 10^{15}\,M_\odot)$ and that 
systematic uncertainties allow even smaller masses an accuracy of a few per cent at most 
\citep{Bhattacharya2011}, one has to be very cautious with falsification experiments that are 
based on extreme objects. The uncertainty in the mass function alone will allow a rather wide 
range of distributions.
%------------------------------------------
\section{Suitable samples of galaxy clusters for an order statistical analysis}\label{sec:preparing_samples}
%------------------------------------------
Having introduced the order statistics of the most massive or the highest redshift 
clusters, we intend now to compare observed clusters with the theoretical distributions. 
To do so, it is necessary to select suitable samples of galaxy clusters, which we will discuss 
in the following.
%------------------------------------------
\subsection{General considerations}
%------------------------------------------
The selection of a suitable sample of galaxy clusters for an order statistical analysis is 
by no means a trivial task. The necessary ordering of the quantities mass and redshift 
by magnitude requires that they have been derived in an identical 
way across the sample. Otherwise, systematics and biases, like the differences between lensing 
and X-ray mass estimates for instance \citep[see e.g.][]{Mahdavi2008, Zhang2010, 
Planck_early_III_2012, Meneghetti2010, Rasia2012}, will render the ordering meaningless. 
Despite an increasing amount of data from different surveys, a lack of large homogeneous 
samples persists. Thus, we decided to base our comparison on clusters that stem from 
catalogues like the SPT massive cluster sample \citep{Williamson2011} and the MCXC 
cluster catalogue \citep{Piffaretti2011}, which will be discussed in further detail below.
%------------------------------------------
\subsection{The SPT massive cluster sample}
%------------------------------------------
The SPT survey \citep{Carlstrom2011} is ideally suited for the intended purpose of an order 
statistical analysis. Being based on the Sunyaev Zeldovich (SZ) effect \citep{Sunyaev1972, 
Sunyaev1980} the SPT survey is able to detect massive galaxy clusters up to high redshifts. 
The fact that the limiting mass of SZ surveys varies weakly with redshift \citep{Carlstrom2002} 
allows in principle to construct mass limited cluster catalogues. However, it should be 
emphasised that the assumption of an $m_{\rm lim}$ independent of redshift depends 
critically on the sensitivity and the beam width of an actual survey.

For this work, we take the catalogue of \cite{Williamson2011} which comprises the 26 most 
significant detections in the full survey area of $A_{\rm s}^{\rm SPT}=2500\,{\rm deg}^2$. 
Ensuring a constant mass limit of $M_{200\rm m}\approx10^{15}\,M_\odot$, clusters were 
selected on the basis of a signal-to-noise (S/N) threshold in the filtered SPT maps. For all 
26 catalogue members, either photometric or spectroscopic redshifts were determined as 
well. The cluster masses given in the catalogue are defined with respect to the mean cosmic 
background density and need no further conversion to match the mass definition of the 
reference \cite{Tinker2008} mass function. To each cluster of the sample we assign the error 
bars that we obtained by adding the reported statistical and systematic errors in quadrature.
%------------------------------------------
\subsection{The MCXC cluster catalogue} \label{sec:MCXC_catalogue}
%------------------------------------------
The MCXC catalogue \citep{Piffaretti2011} is based on the publicly available compilation of 
clusters' detections from ROSAT All-Sky Survey (NORAS, REFLEX, BCS, SGP, NEP, MACS, and CIZA) 
and other serendipitous surveys (160SD, 400SD, SHARC, WARPS, and EMSS), and provides the 
physical properties of 1743 galaxy clusters systematically homogenised to an overdensity of 
500 (with respect to the cosmic critical density).
This meta-catalogue is not complete in any sense, but it is constituted by X-ray flux-limited 
samples that ensure that the X-ray brightest objects in the nearby ($z \la 0.3$) Universe, and 
therefore the most massive X-ray detected clusters, are all included.

We have then simply ranked the objects accordingly to their estimated $M_{200\rm m}$, that is 
obtained from the tabulated $M_{500\rm c}$ as
\begin{equation}
M_{200\rm m} = M_{500\rm c} \frac{200 \Omega_z}{500} \left( \frac{R_{200\rm m}}{R_{500{\rm c}}} \right)^3
\end{equation}
where $\Omega_z = \Omega_{\rm m} (1+z)^3 / E_z^2$, $E_z = (\Omega_{\rm m}(1+z)^3 +
\Omega_{\Lambda})^{1/2}$, and the ratio between the radii at different overdensities has been 
obtained by assuming an NFW profile \citep{nfw1996} with $c_{200}=4$.
%------------------------------------------
\subsection{Preparations of the ordered samples}
%------------------------------------------
We order the SPT and MCXC catalogues by magnitude of the observed mass and present the 
ten most massive systems in \autoref{tab:clusters_mass}. For statistical comparisons the 
observed masses have to be corrected for the Eddington bias \citep{Eddington1913} in mass. 
As a result of the exponentially suppressed tail of the mass function and the substantial 
uncertainties in the mass determination of galaxy clusters, it is more likely that lower mass 
systems scatter up while higher mass systems scatter down, resulting in a systematic shift. 
Thus, before an observed mass can be compared to a theoretical distribution, this shift has 
to be corrected for. To do so, we follow \cite{Mortonson2011} and shift the observed masses, 
$M_{\rm obs}$, to the corrected masses, $M_{\rm corr}$, according to
\begin{equation}
\ln M_{\rm corr}=\ln M_{\rm obs}+\frac{1}{2}\epsilon\sigma_{\ln M}^2,
\end{equation} 
where $\epsilon$ is the local slope of the mass function ($\dd n/\dd \ln M\propto M^\epsilon$) 
and $\sigma_{\ln M}$ is the uncertainty in the mass measurement. We corrected the observed 
masses in both, the SPT and the MCXC catalogues, using the values of $\sigma_{\ln M}$ listed in 
the fifth column of \autoref{tab:clusters_mass} which we deduced from the reported uncertainties 
in the nominal masses. The larger the observational errors are, the larger is the correction towards 
lower masses.

As an exemplary exception from the SPT catalogue, we used for the mass of 
SPT-CL J0102-4915 the value reported by \cite{Menanteau2012}, which is based on a 
combined SZ+X-rays+optical+infrared analysis. The multi-wavelength study shifts 
$M_{\rm obs}=(1.89\pm 0.45)\times 10^{15}\,M_\odot$ \citep{Williamson2011} to a larger value of 
$M_{\rm obs}=(2.16\pm 0.32)\times 10^{15}\,M_\odot$, changing the rank from the fifth to the 
third most massive. This shows that with the expected increase in the quality of cluster mass estimates, the 
ordering of the most massive cluster will undergo significant changes. We expect that the reshuffling will 
affect more strongly the most massive clusters due to the fact that the large error bars will cause lower 
ranked clusters to scatter up. We will discuss the impact of the reshuffling in more detail in 
\autoref{sec:individual_analysis_mass}.

In addition, we sorted the SPT catalogue by redshift and list the ten highest redshift clusters above 
$m_{\rm lim}\approx 10^{15}\,M_\odot$ in \autoref{tab:clusters_redshift}.
%-------------------------------------
\begin{table}
\caption{Compilation of the ten highest redshift clusters from the SPT massive cluster sample 
\citep{Williamson2011}. Here, (s) and (p) denote the spectroscopic and photometric redshifts, 
respectively.}
\begin{tabular}{lccc} \hline
Rank & Cluster & $z$ & $M_{200\rm m}$ in units of $M_\odot$  \\ 
\hline 
$1^{\rm st}$ & SPT-CL J2106-5844 &	1.132 (s)& $(1.27	\pm 0.21)\times 10^{15}$ \\
$2^{\rm nd}$ & SPT-CL J0615-5746 & 0.972 (s) & $(1.32\pm 0.40)\times 10^{15}$ \\
$3^{\rm rd}$	& SPT-CL J0102-4915 & 0.870 (s)	& $(2.16\pm 0.32)\times 10^{15}$\\
$4^{\rm th}	$& SPT-CL J2337-5942 & 0.775 (s)	& $(1.99\pm 0.20)\times 10^{15}$\\
$5^{\rm th}	$& SPT-CL J2344-4243 & 0.620 (p)& $(1.91\pm 0.50)\times 10^{15}$\\
$6^{\rm th}	$& SPT-CL J0417-4748 & 0.620 (p)& $(1.88\pm 0.20)\times 10^{15}$\\
$7^{\rm th}	$& SPT-CL J0243-4833 & 0.530 (p)	& $(1.81\pm 0.23)\times 10^{15}$\\
$8^{\rm th}	$& SPT-CL J0304-4401 & 0.520 (p)	& $(1.70\pm 0.33)\times 10^{15}$\\
$9^{\rm th}	$& SPT-CLJ0438-5419 & 0.450 (p)	& $(1.70\pm 0.38)\times 10^{15}$\\
$10^{\rm th}$	& SPT-CLJ0254-5856 & 0.438 (s)& $(1.65\pm 0.25)\times 10^{15}$\\ 
\hline
\end{tabular}\label{tab:clusters_redshift}
\end{table}
%-------------------------------------
%-------------------------------------
\begin{figure*}
\centering
\includegraphics[width=0.33\linewidth]{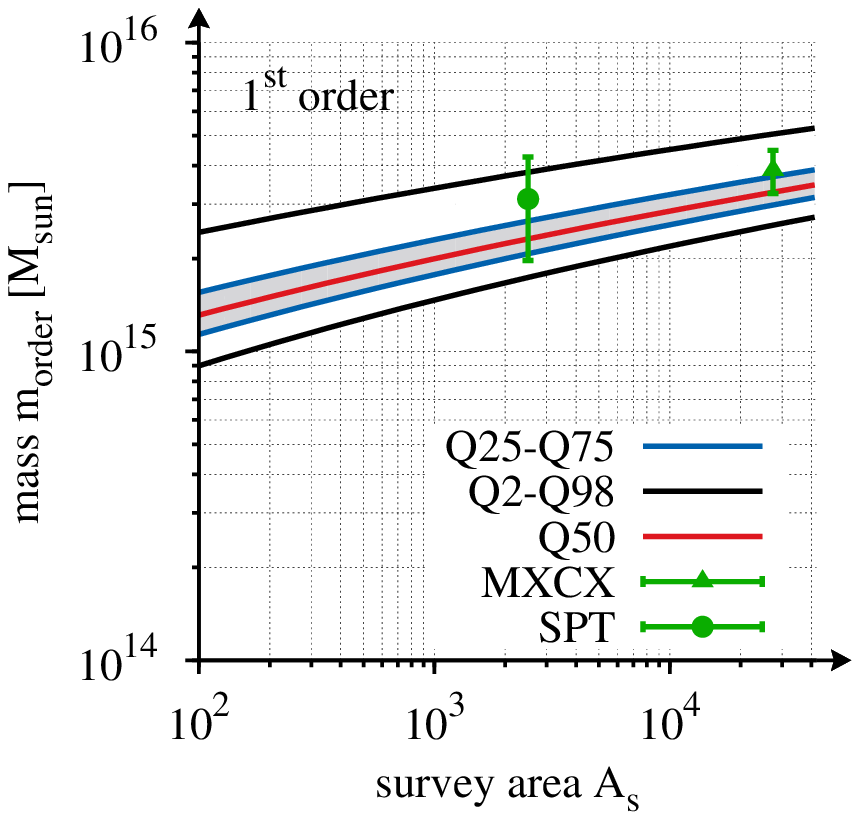}
\includegraphics[width=0.33\linewidth]{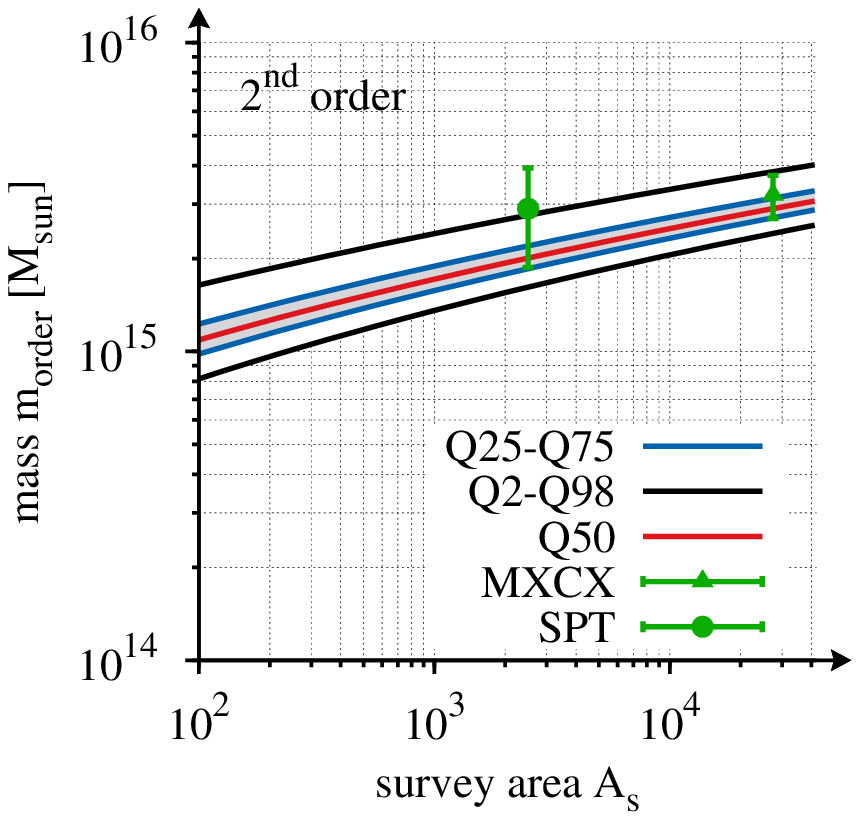}
\includegraphics[width=0.33\linewidth]{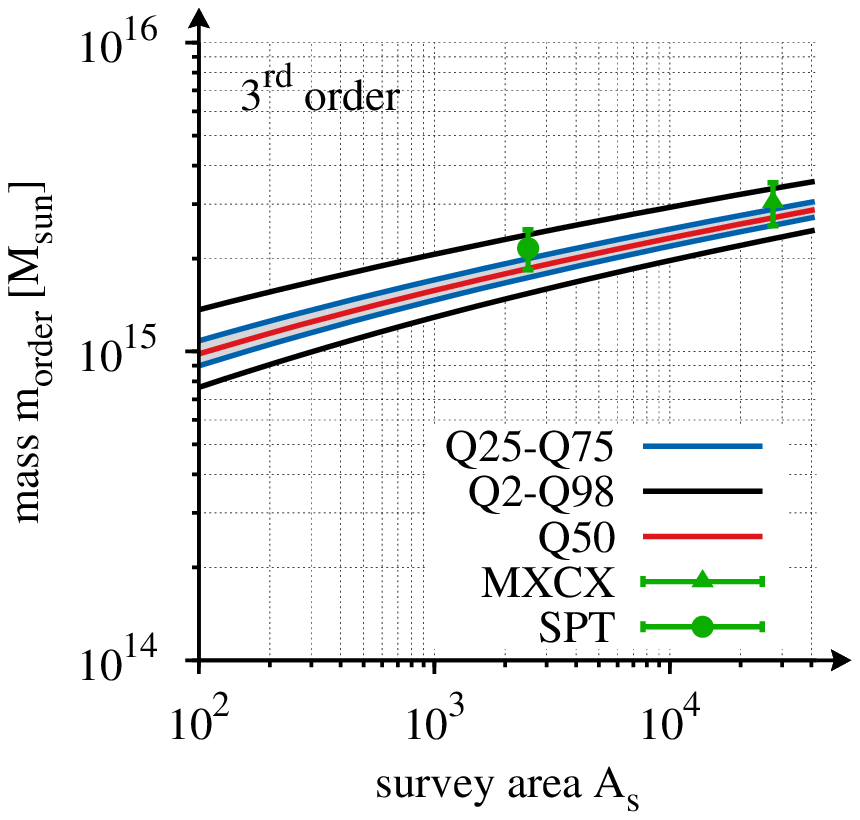}\\
\includegraphics[width=0.33\linewidth]{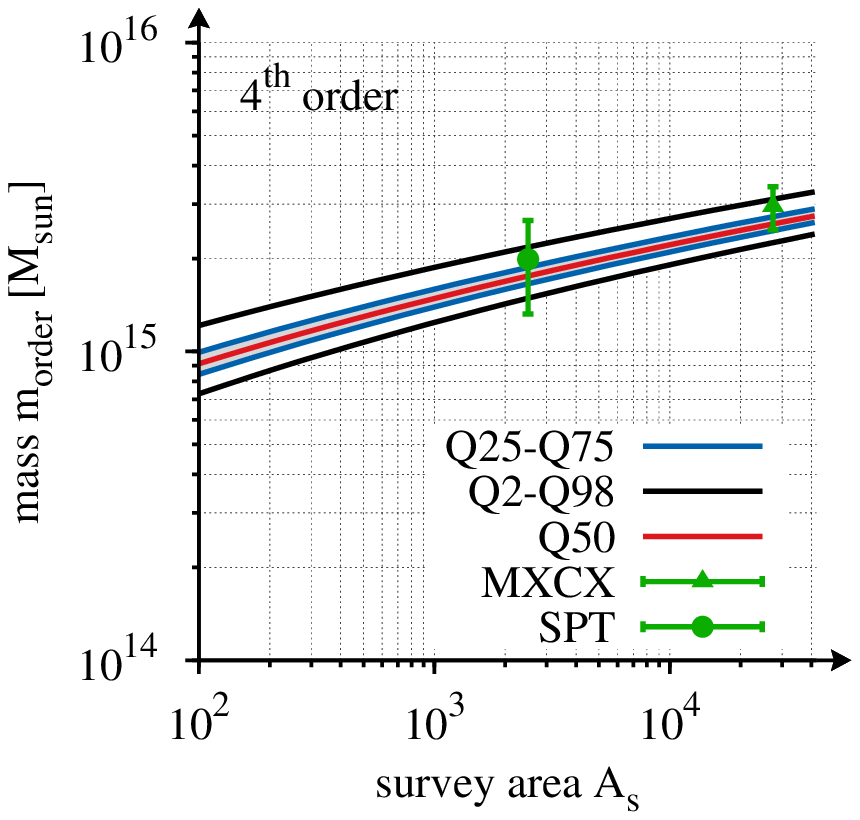}
\includegraphics[width=0.33\linewidth]{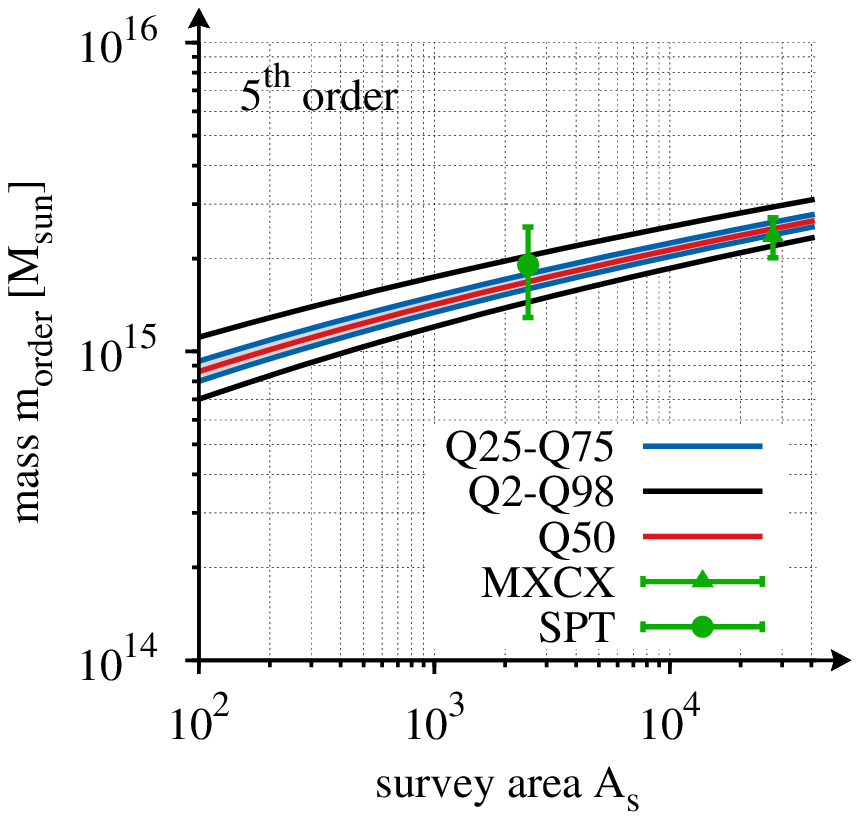}
\includegraphics[width=0.33\linewidth]{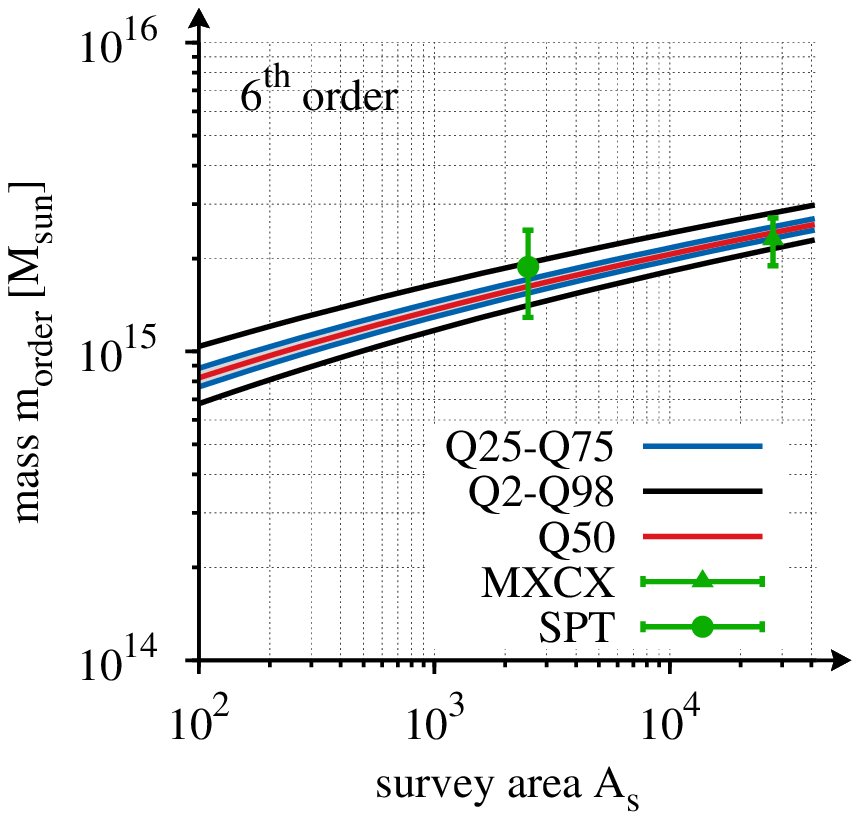}\\
\includegraphics[width=0.33\linewidth]{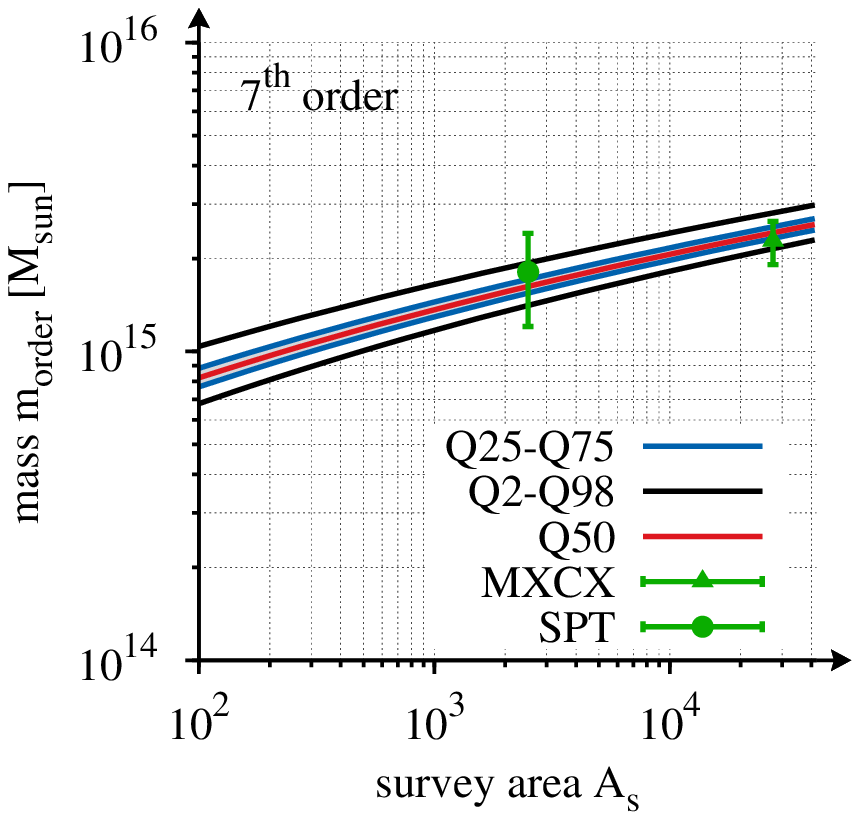}
\includegraphics[width=0.33\linewidth]{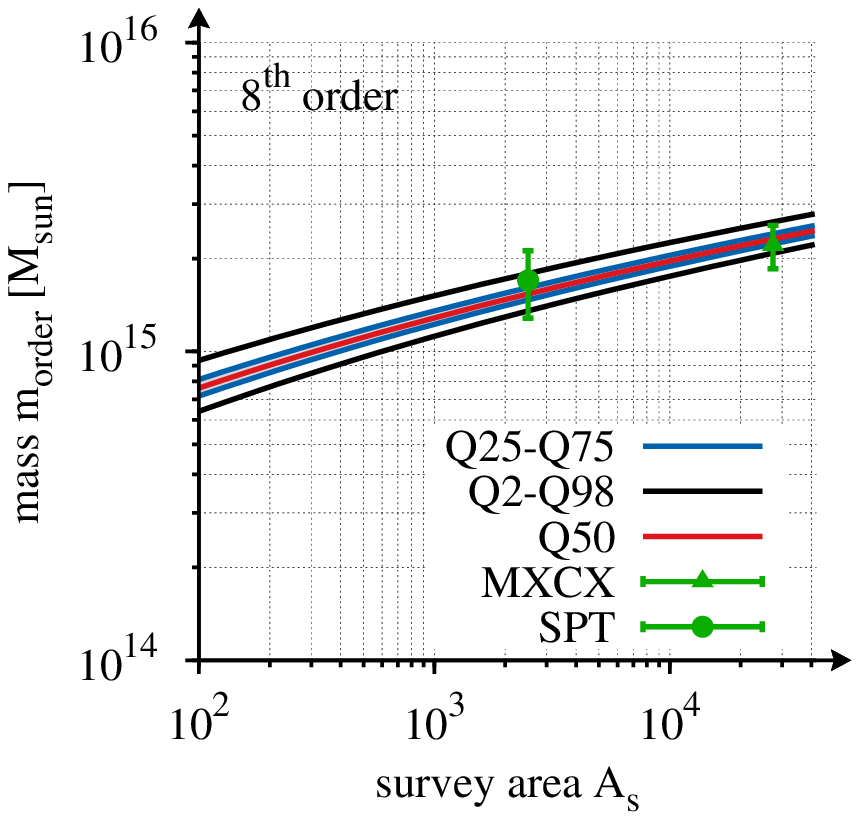}
\includegraphics[width=0.33\linewidth]{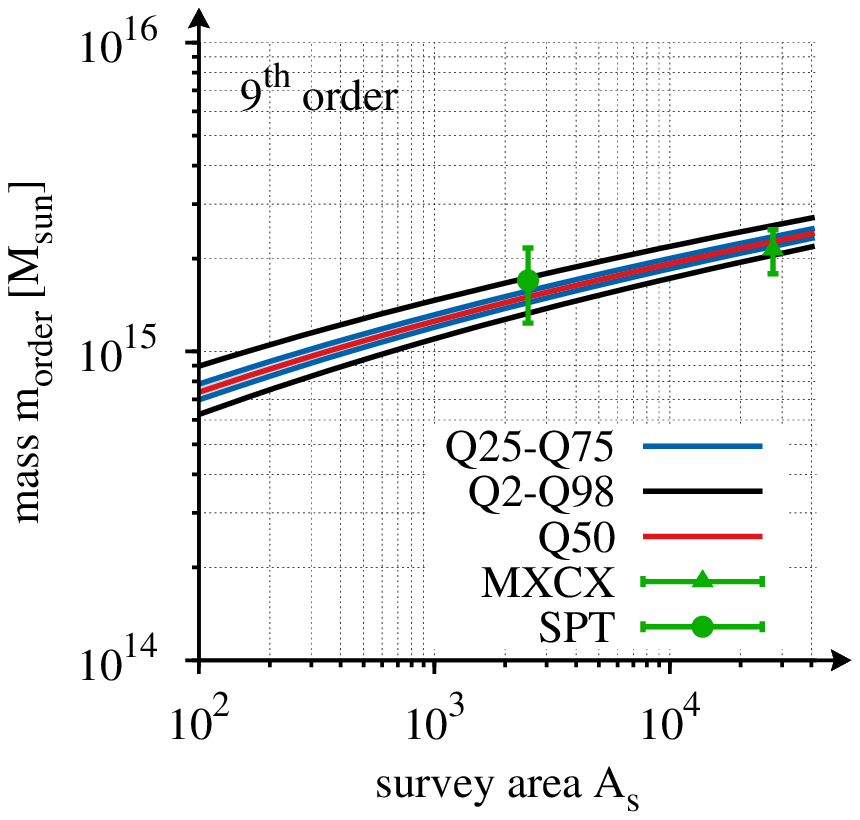}
\caption{Functional box plots for the first nine orders in the observable mass as indicated 
in the individual panels. Here, the red line denotes the median ($Q$50), the blue-bordered, 
grey region the interquartile range (IQR) and the black lines denote the 2 and the 98-percentile 
($Q$2, $Q$98). The green error bars show the corresponding observed masses, 
$M_{200\rm m}$, from the SPT (green circles) and the MCXC (green triangles) 
catalogues (see \autoref{tab:clusters_mass}) for their respective survey areas of 
$A^{\rm SPT}_{\rm s}=2500\,{\rm deg^2}$ and $A^{\rm MCXC}_{\rm s}=27490\,{\rm deg^2}$.}
\label{fig:quantile_mass_area}
\end{figure*}
%-------------------------------------
%-------------------------------------
\begin{figure*}
\centering
\includegraphics[width=0.45\linewidth]{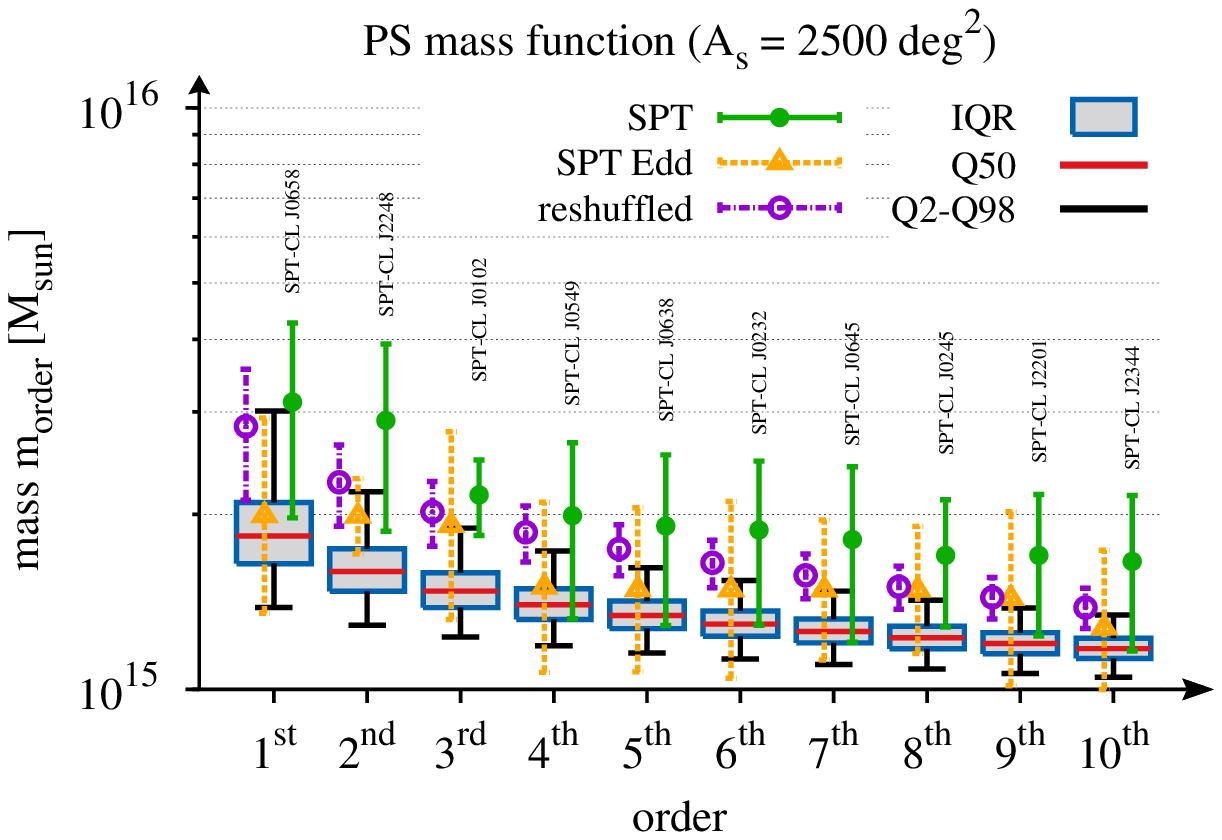}
\includegraphics[width=0.45\linewidth]{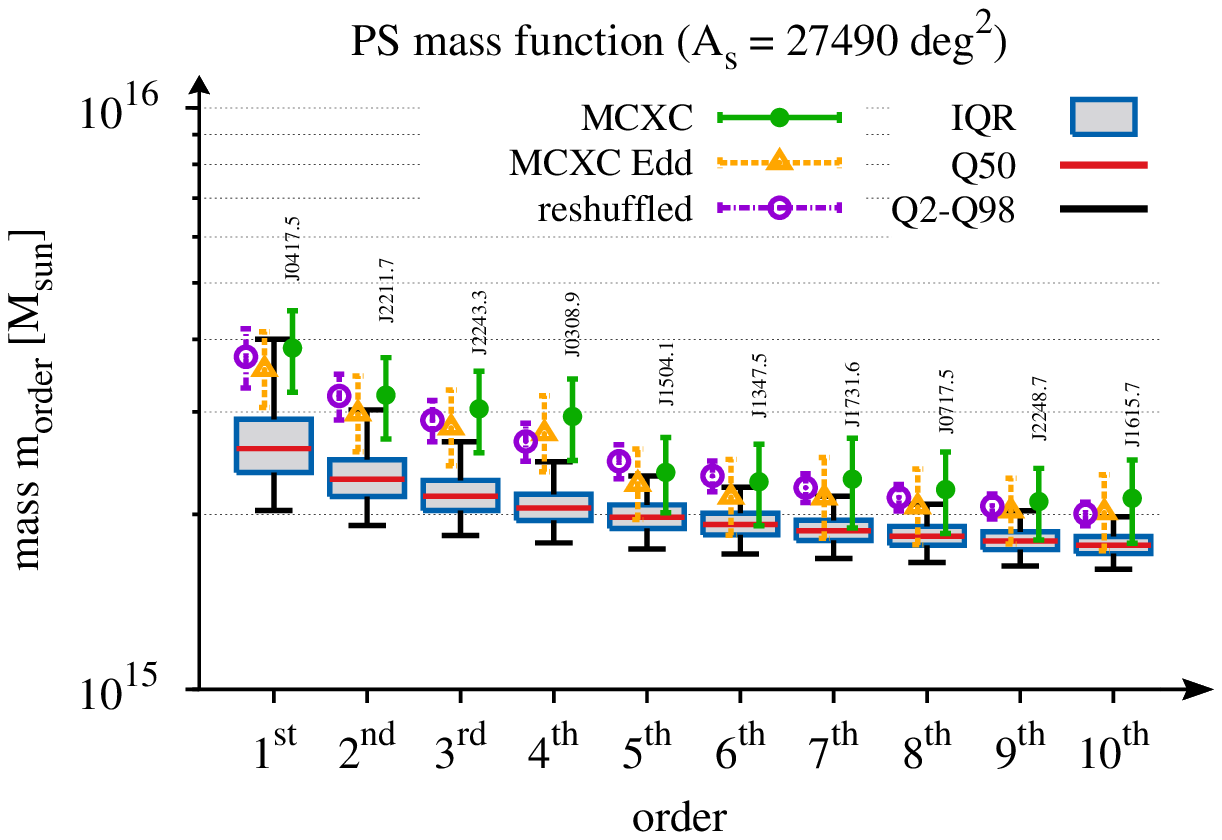}\\
\includegraphics[width=0.45\linewidth]{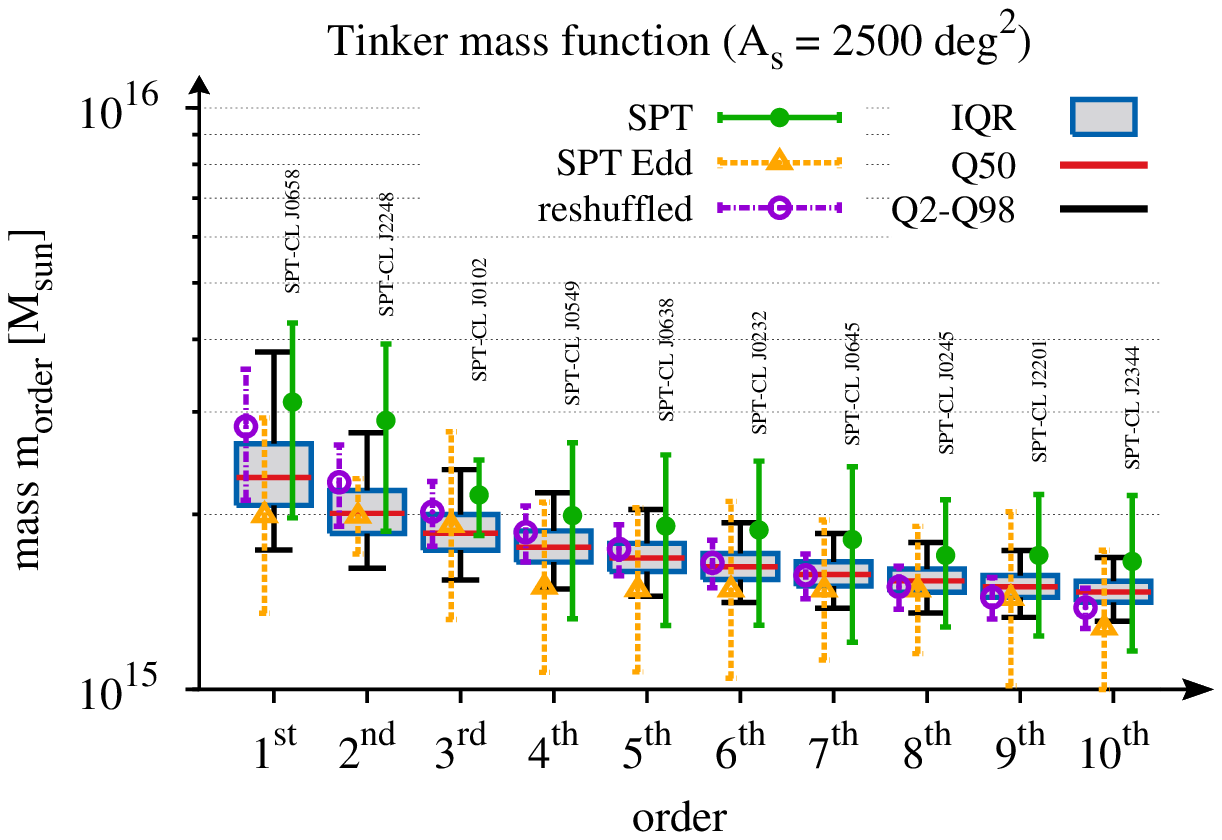}
\includegraphics[width=0.45\linewidth]{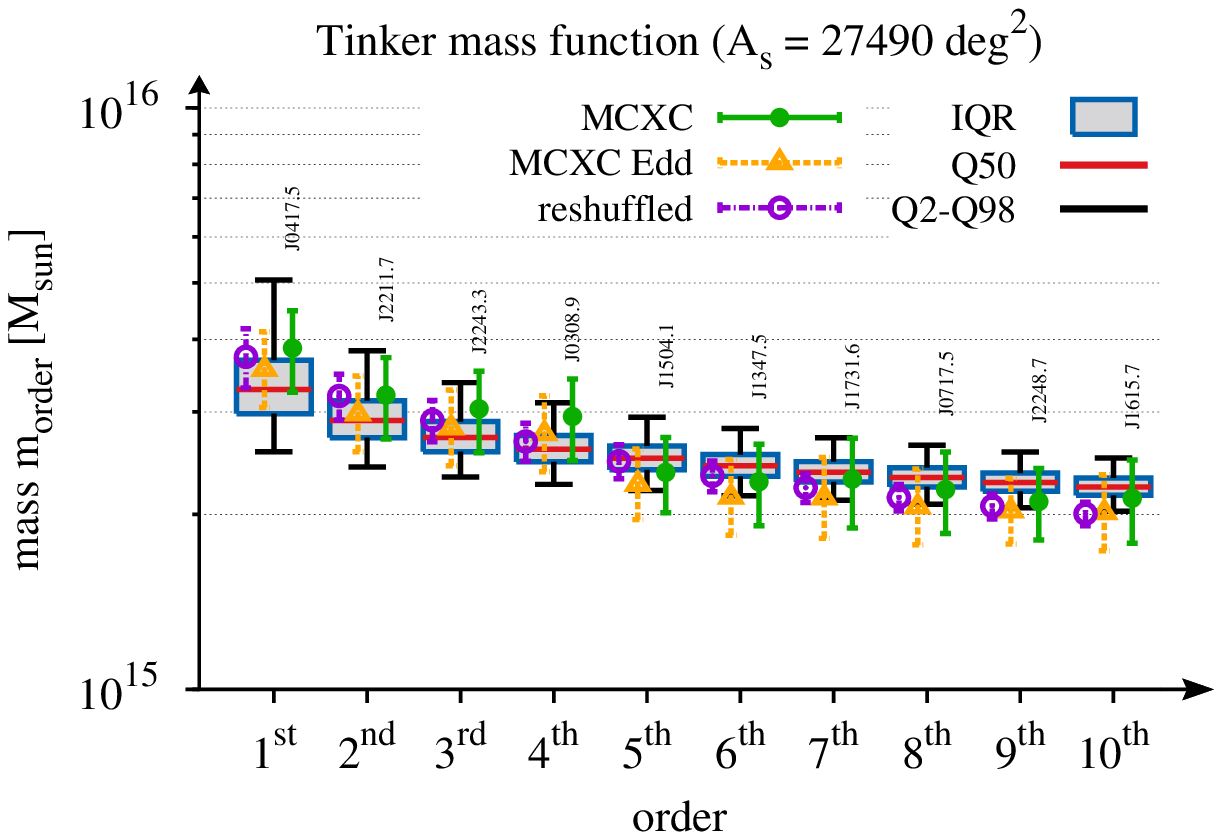}\\
\includegraphics[width=0.45\linewidth]{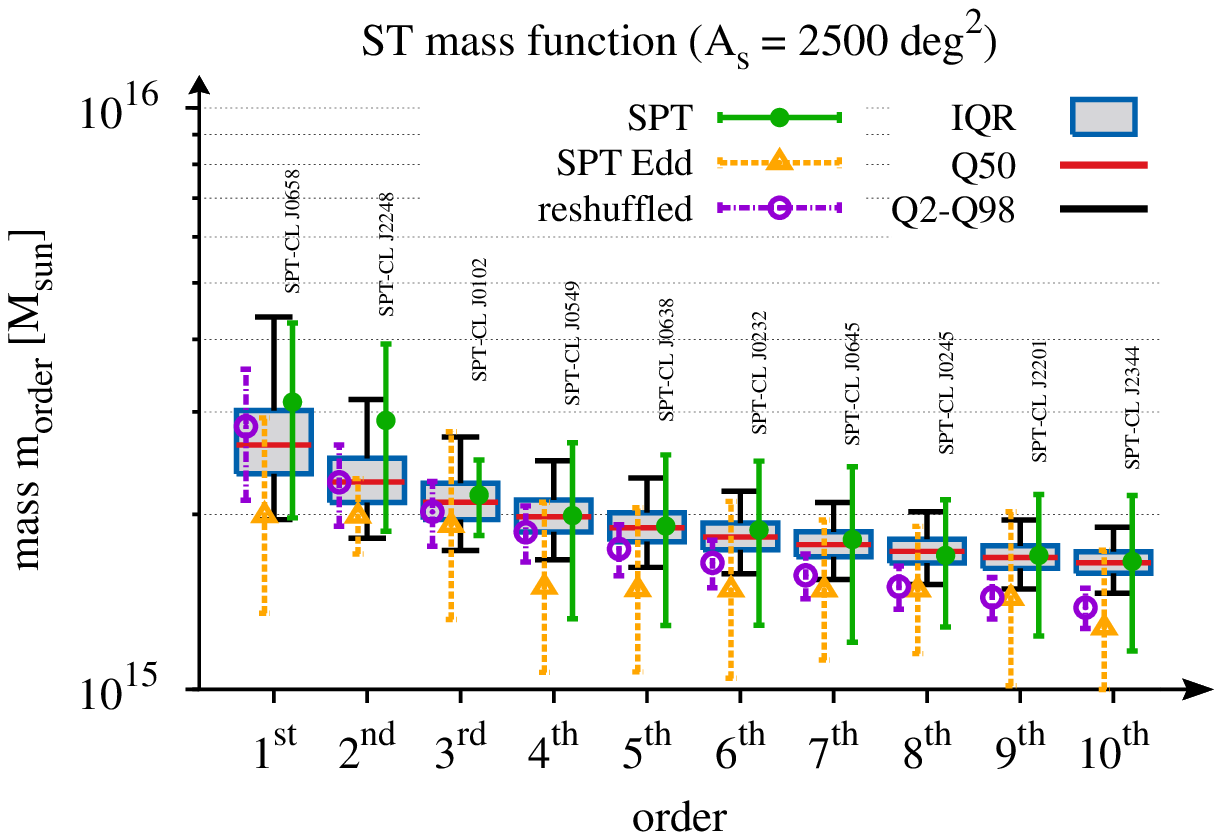}
\includegraphics[width=0.45\linewidth]{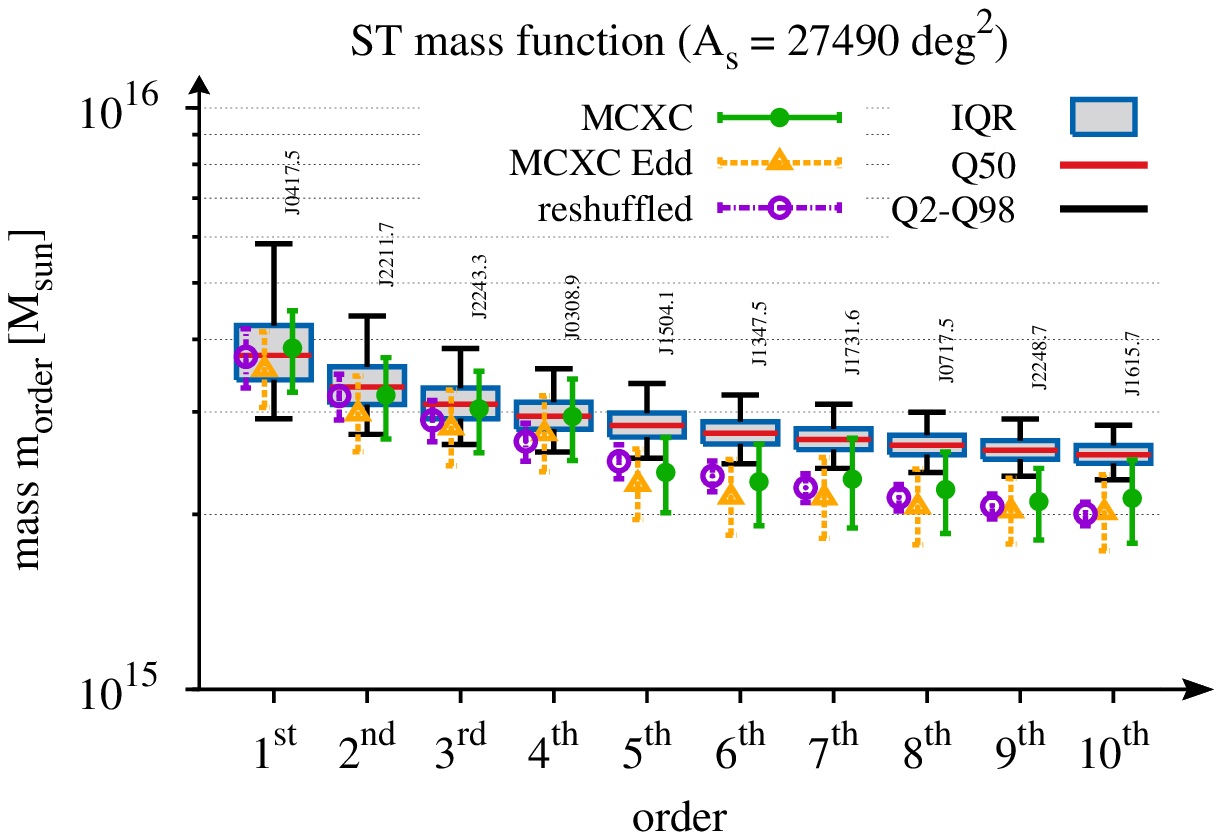}
\caption{Box-and-whisker diagram of the ten most massive clusters from the SPT survey 
(left column) and the MCXC catalogue (right column) for three different choices of the 
mass function as denoted in the title of each panel. For each order, the red lines denote 
the median ($Q$50), the blue-bordered, grey boxes give the IQR and the black whiskers mark 
the range between the $2$ and $98-$percentile ($Q$2, $Q$98) of the theoretical distribution. 
The green, filled circles denote the nominal observed cluster masses, $M_{200\rm m}$, the orange, 
empty triangles the ones that are corrected for the Eddington bias in mass and the violet, empty 
circles are the results of the Monte Carlo reshuffling of the ranks. All error bars denote the 
$1\sigma$ range.}
\label{fig:box_and_whisker_mass}
\end{figure*}
%-------------------------------------
%-------------------------------------
\begin{figure}
\centering
\includegraphics[width=0.99\linewidth]{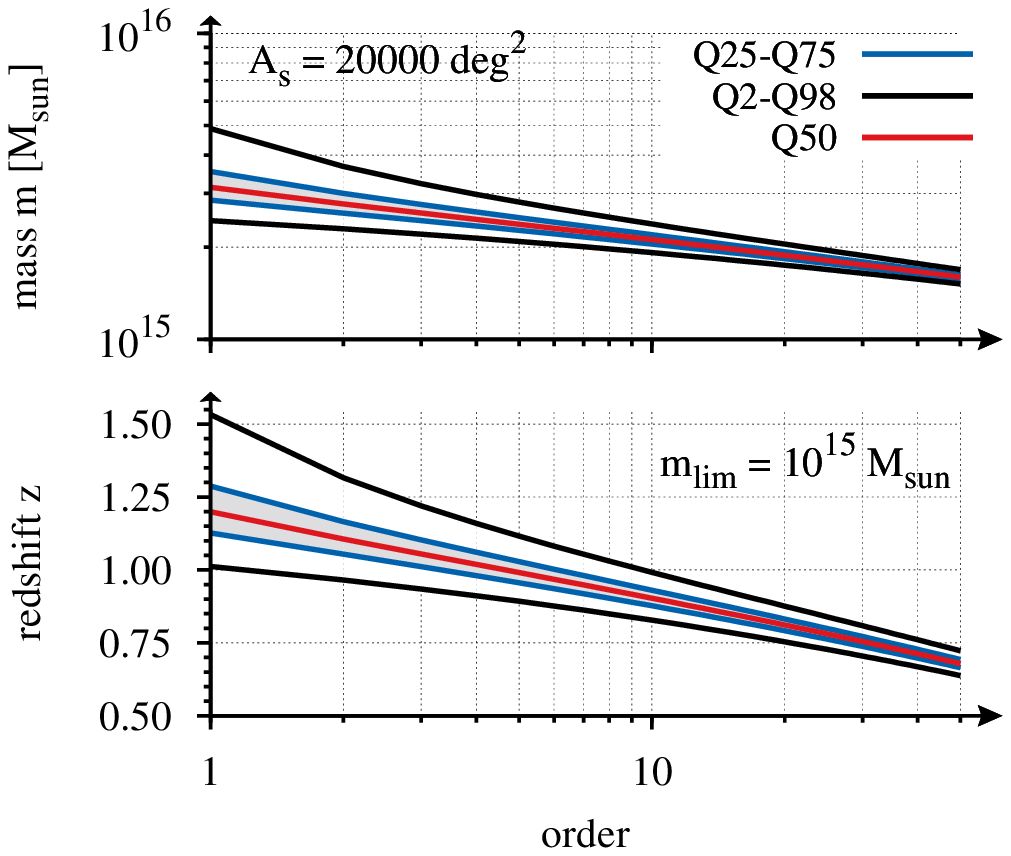}
\caption{Dependence of different percentiles on the order for the order statistics in mass (upper 
panel) and redshift (lower panel). For both cases a survey area of $A_{\rm s}=20\,000\,{\rm deg}^2$ 
is assumed. Further, a limiting mass of $m_{\rm lim}=10^{15}\,M_\odot$ has been adopted for the 
order statistics in redshift. All percentiles are denoted by the same line styles as used in the previous 
figures and are given in the key.}
\label{fig:exclusion_order}
\end{figure}
%-------------------------------------
%------------------------------------------
\section{Comparison of the individual order statistics with observations}\label{sec:individual_analysis}
%------------------------------------------
In this section we will compare the individual ranked systems listed in 
\autoref{tab:clusters_mass} for the mass and in \autoref{tab:clusters_redshift} for the redshift with the individual
distributions for each rank, as e.g. shown in \autoref{fig:impact_massfunction}.
%------------------------------------------
\subsection{Order statistics in cluster mass}\label{sec:individual_analysis_mass}
%------------------------------------------
In order to demonstrate the impact of the survey area on the distributions of the order 
statistics in mass, we show in \autoref{fig:quantile_mass_area} the dependence of different 
quantiles ($Q$2, $Q$25, $Q$50, $Q$75 and $Q$98) on the survey area for the nine most massive clusters. 
In addition, the green error bars show the clusters from the SPT and MCXC catalogues listed in 
\autoref{tab:clusters_mass} for the respective survey areas of $A^{\rm SPT}_{\rm s}=2500\,
{\rm deg^2}$ and $A^{\rm MCXC}_{\rm s}=27490\,{\rm deg^2}$.  

From the individual panels in \autoref{fig:quantile_mass_area} it can be inferred that, as expected, 
a larger survey area yields a larger expected mass for the individual rank. Furthermore, with 
increasing rank towards higher orders, the interquantile range, like (Q2-Q98), narrows. A 
behaviour that can also be seen in \autoref{fig:first50} as steepening of the cdf with increasing 
rank. Therefore, the largest mass (first order) is expected to be realised in a much wider mass 
range than the higher orders.

We will now compare the observations in more detail with the theoretical expectations in the form 
of box-and-whisker diagrams as shown in \autoref{fig:box_and_whisker_mass}. Here, the 
blue-bordered, grey filled box denotes the interquartile range (IQR) which is bounded by the 25 and 
75-percentiles ($Q$25, $Q$75) and the median ($Q$50) is depicted as a red line. The black whiskers 
denote the 2 and 98-percentiles ($Q$2,$Q$98) and we follow the convention that observations 
that fall outside are considered as outliers. As before, the nominal observed cluster masses are denoted 
as green error bars where for the left column the SPT catalogue and for the right column the MCXC 
catalogue was used. In addition we plot the Eddington bias corrected masses, $M_{200\rm m}^{\rm 
Edd}$, from the sixth column of \autoref{tab:clusters_mass} as orange triangles with dashed error bars. 
We performed the analysis for three different mass functions, comprising from the top to the bottom 
panel, the PS, the Tinker and the ST mass functions. In addition to the Eddington bias in mass, we expect 
a shift to larger masses caused by the reshuffling of orders due to the uncertainties in mass. In 
order to quantify this effect, we Monte Carlo (MC) simulated $10\,000$ realisations of the 26 SPT and 123 
MCXC (with $M>10^{15}\,M_\odot$) cluster masses after their correction for the Eddington bias and order 
them by mass. The masses were randomly drawn from the individual error interval, assuming Gaussian 
distributions. We present the results as violet, empty circles with dash-dotted $1\sigma$ error bars in 
\autoref{fig:box_and_whisker_mass}. It can be seen that the highest ranks are more strongly affected by 
the reshuffling than the lower ones and that they are on average shifted to larger values. Of course, the 
amount of this effect will depend on the size of the error bars. Further, the reshuffling yields mass values 
that fall between the nominal (green error bars) and the Eddington bias corrected ones (orange error bars). 

For the SPT catalogue, it can be seen from the top left panel of \autoref{fig:box_and_whisker_mass} 
that the outdated PS mass function seems to be disfavoured by the reshuffled and the nominal masses of the 
ten largest objects. However, the error bars are large and do not allow an exclusion of the PS mass function. 
For the Tinker and the ST mass function, the boxes indicating the theoretical 
distributions move to larger mass values and therefore they match the observed masses better than 
the PS mass function. In particular, the third ranked (second ranked after Eddington bias correction) 
system SPT-CL J0102 with its smaller errors and, hence, giving the tightest constraints, is consistent with 
$\Lambda$CDM for both mass functions. All other ranks are consistent as well due to their large error 
bars. The reshuffled sample matches perfectly the Tinker mass function consolidating the conclusion 
that the most massive clusters of the SPT sample are in agreement with the statistical expectations. 
The conclusions for the MCXC catalogue are identical, however the jump between the fourth and the fifth 
largest order yield to an inconsistency of the observed higher orders with the expectations based on the 
ST mass function. This jump is clearly caused by the incompleteness of the MCXC catalogue and, thus, 
the inclusion of the missing clusters would most certainly move the observed sample to higher masses in 
the direction of the results we obtained from the SPT sample. In this sense we do not see any indication 
of a substantial difference between the small and wide field survey.

The analysis of the SPT sample illustrates the potential of utilising the $n$ most massive galaxy clusters 
to test underlying assumptions, like e.g. the mass function. For instance, a multi-wavelength study of 
the 26 SPT clusters would reduce the error bars to the level of SPT-CL J0102 (the nominal third ranked cluster 
in the left column of \autoref{fig:box_and_whisker_mass} ), which would significantly tighten the 
constraints on the underlying assumptions like e.g. the halo mass function. In turn, by assuming the 
$\Lambda$CDM reference cosmology, the comparison of the observed masses with the individual 
order distributions allows to check the completeness of the observed sample. 

In the upper panel of \autoref{fig:exclusion_order}, we present the dependence of different 
percentiles ($Q2$,$Q25$,$Q50$,$Q75$ and $Q98$) on the order for a survey 
area of $A_{\rm s}=20\,000\,{\rm deg}^2$. Choosing the $Q98$ percentile as exclusion criterion, 
one would need roughly to find ten clusters with $m\gtrsim 2.5\times 10^{15}\,M_\odot$, three clusters 
with $m\gtrsim 3.2\times 10^{15}\,M_\odot$ or one cluster with $m\gtrsim 5\times 10^{15}\,M_\odot$ 
in order to report a significant deviation from the $\Lambda$CDM expectations. Of course, the observed 
masses might have to be corrected for the Eddington bias in mass and a possible reshuffling as previously 
demonstrated. In general, exclusion criteria based on order statistics extend previous works 
\citep{Mortonson2011, Waizmann2012a} from statements about single objects to statements about 
object samples which considerably improves the reliability of the entire study.
%-------------------------------------
\begin{figure*}
\centering
\includegraphics[width=0.33\linewidth]{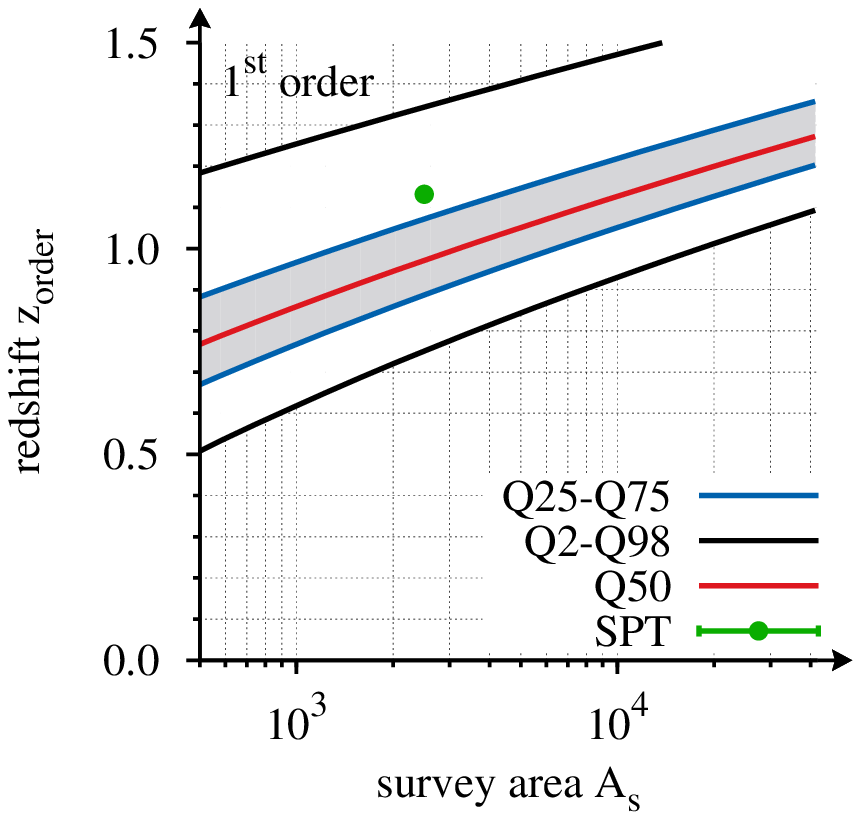}
\includegraphics[width=0.33\linewidth]{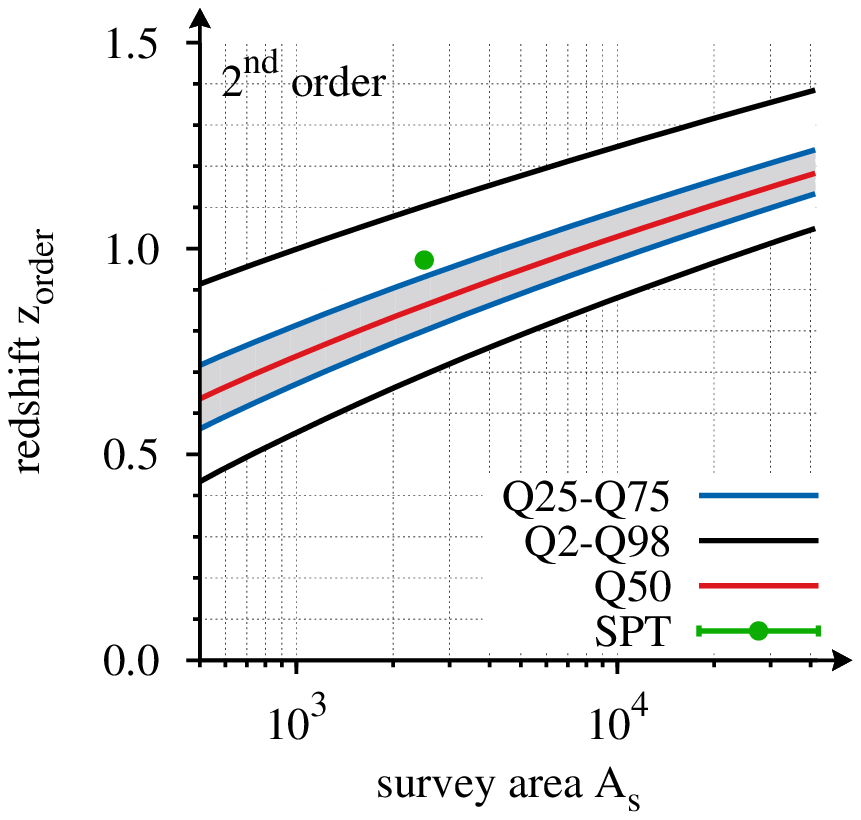}
\includegraphics[width=0.33\linewidth]{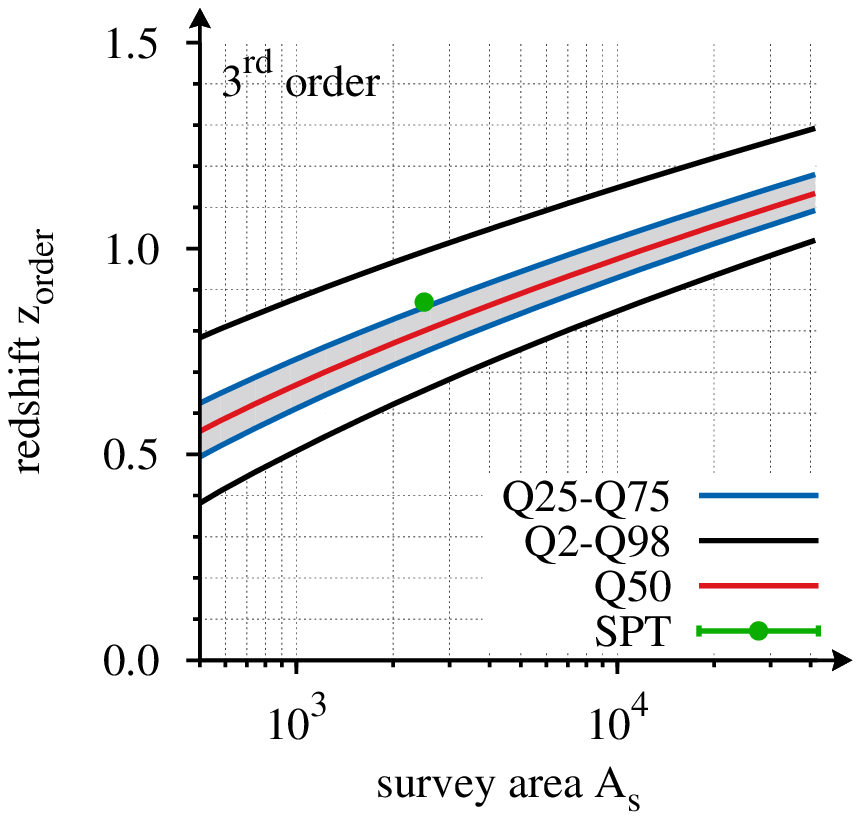}\\
\includegraphics[width=0.33\linewidth]{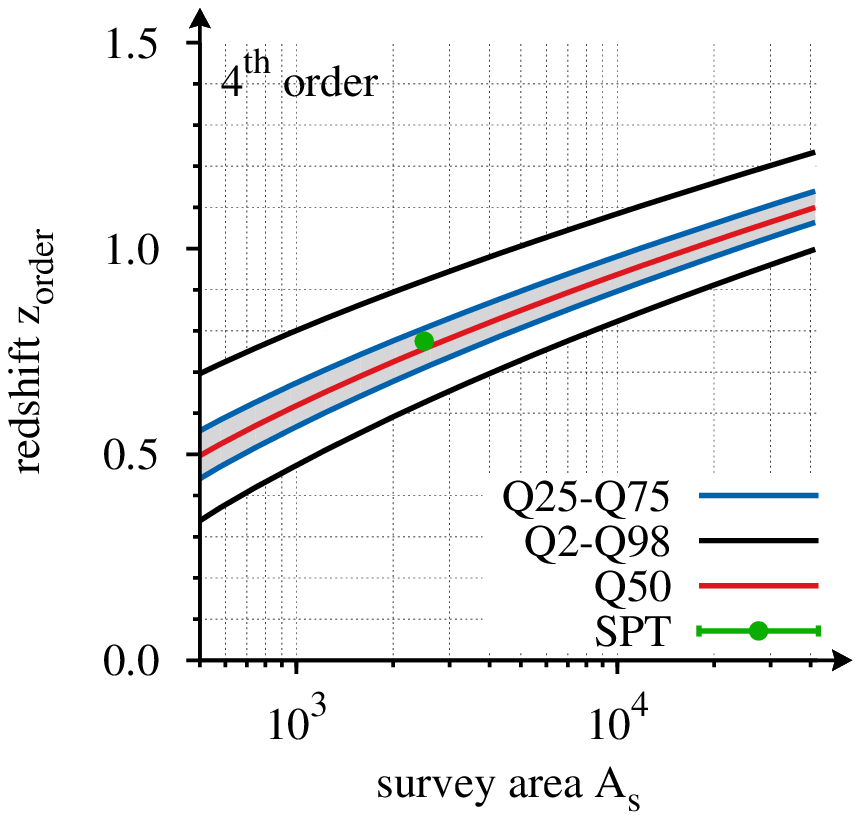}
\includegraphics[width=0.33\linewidth]{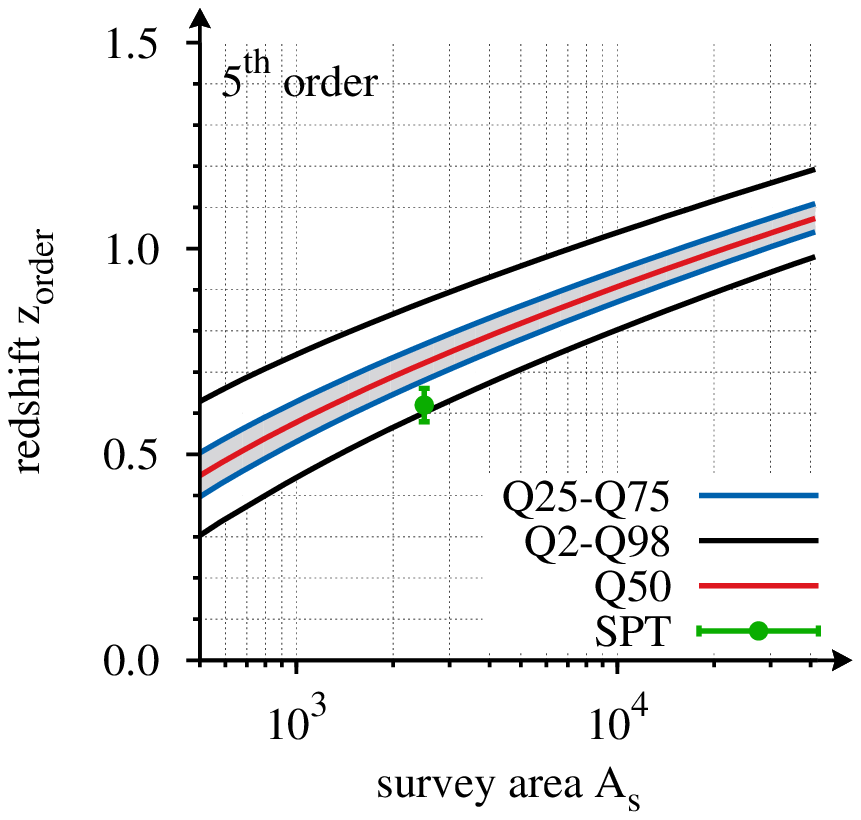}
\includegraphics[width=0.33\linewidth]{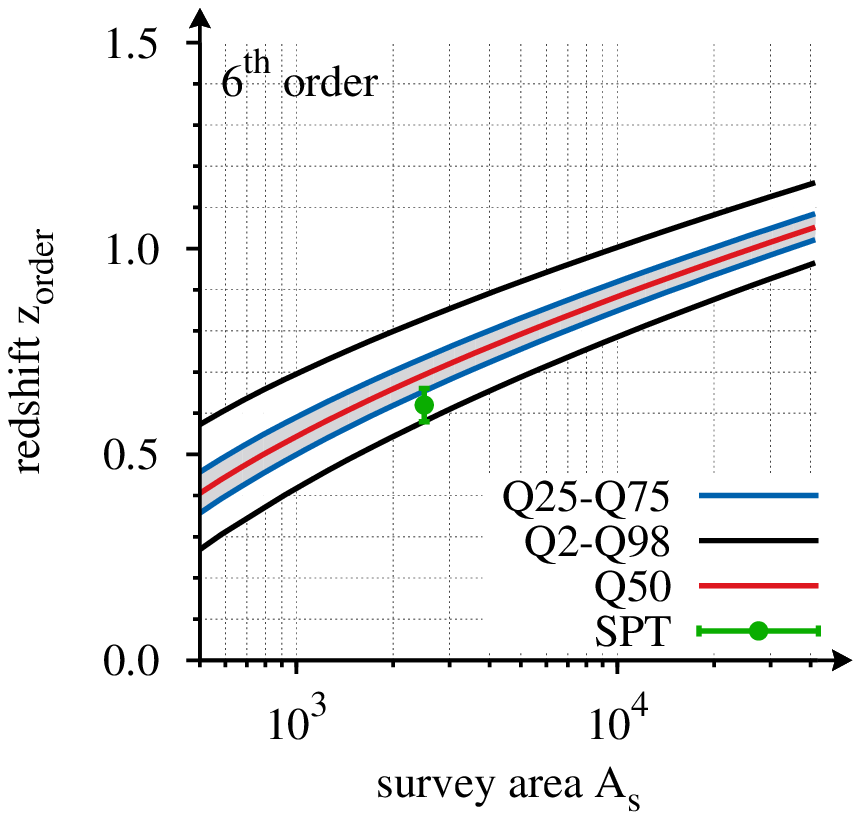}\\
\includegraphics[width=0.33\linewidth]{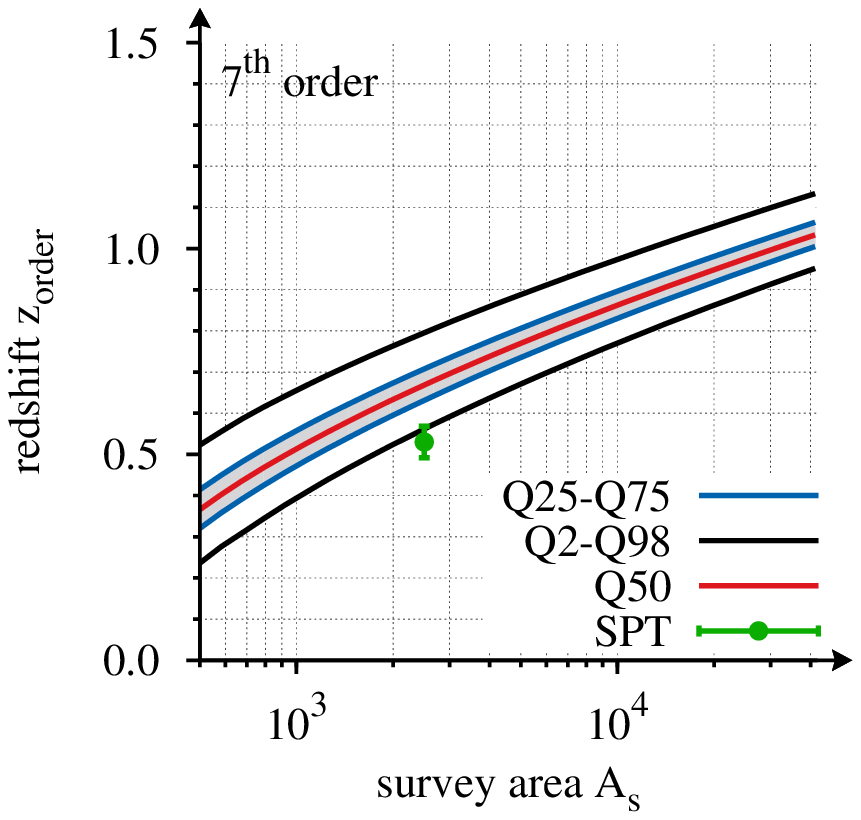}
\includegraphics[width=0.33\linewidth]{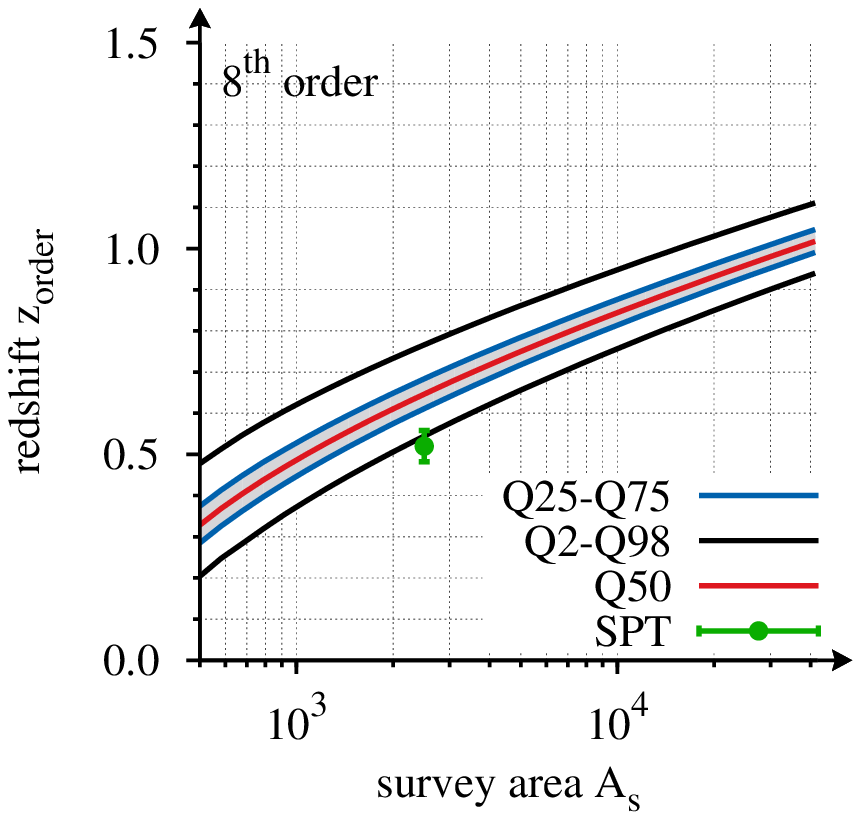}
\includegraphics[width=0.33\linewidth]{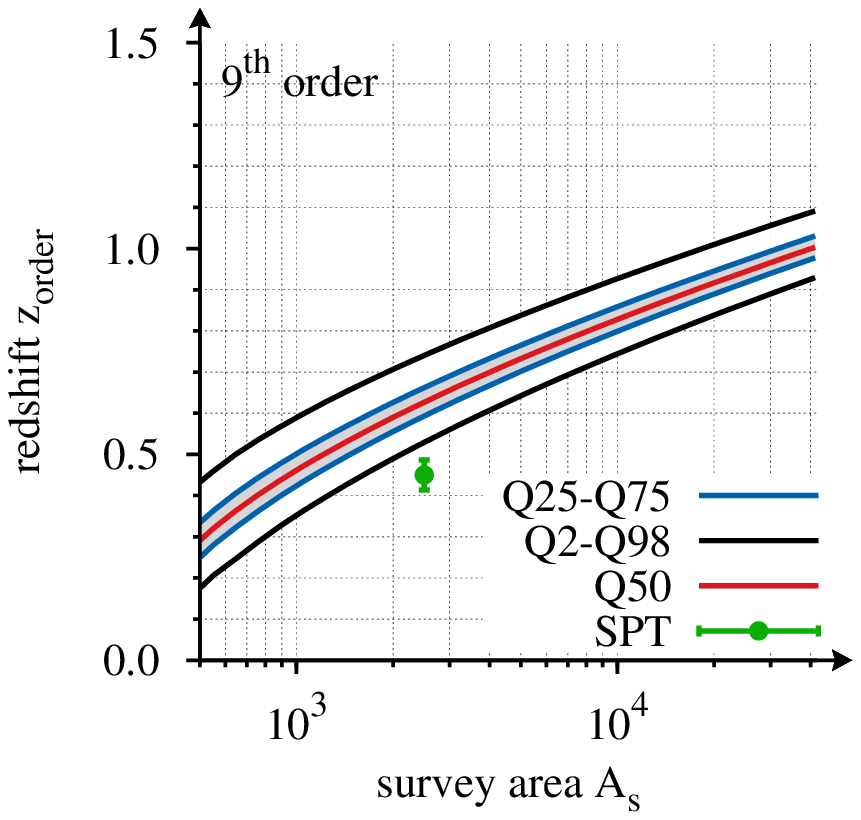}
\caption{Functional box plot for the first nine orders in the observable redshift as indicated 
in the individual panels. Here, the red line denotes the median ($Q$50), the blue-bordered, grey 
region the interquartile range (IQR) and the black lines denote the 2 and the 98-percentile ($Q$2, $Q$98). 
The green circles (with error bars in the case of photometric redshifts) denote the redshifts from 
the SPT catalogue as listed in \autoref{tab:clusters_redshift}.}
\label{fig:quantile_redshift_area}
\end{figure*}
%-------------------------------------
%------------------------------------------
\subsection{Order statistics in cluster redshift}
%------------------------------------------
We performed an identical analysis for the individual order statistics for the SPT massive 
cluster catalogue ranked by redshift listed in \autoref{tab:clusters_redshift}. For the 
theoretical distributions we assume a limiting mass of $m_{\rm lim}=10^{15}\,M_\odot$ 
and a survey area of $A^{\rm SPT}_{\rm s}=2500\,{\rm deg^2}$. As before, we present 
in \autoref{fig:quantile_redshift_area} the dependence of the order statistical distributions 
on the survey area for the first nine orders. Again, an increase in the survey area yields a 
shift of the theoretical distributions to higher redshifts and, as shown in the right panel 
of \autoref{fig:first50}, the cdfs steepen for the higher ranks, resulting in a shrinking 
interquantile range.

\begin{figure}
\centering
\includegraphics[width=0.95\linewidth]{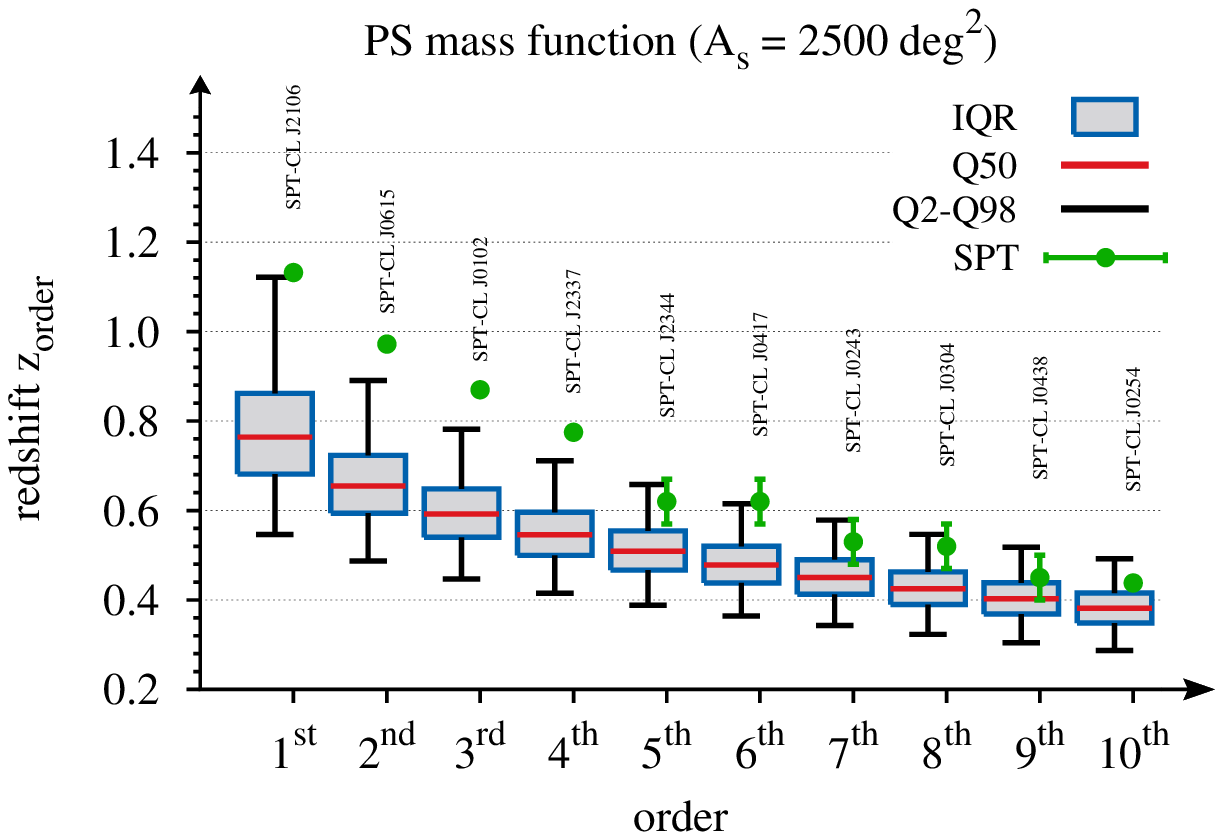}\\
\includegraphics[width=0.95\linewidth]{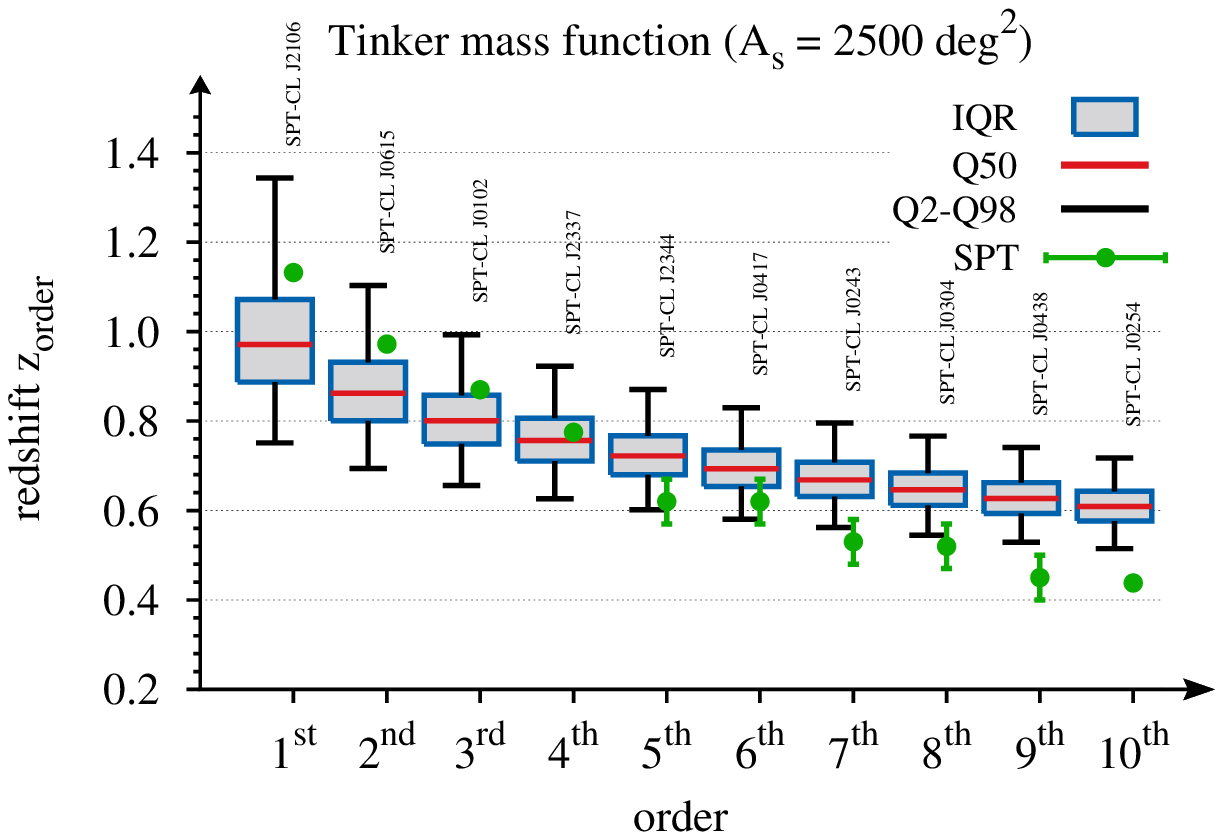}\\
\includegraphics[width=0.95\linewidth]{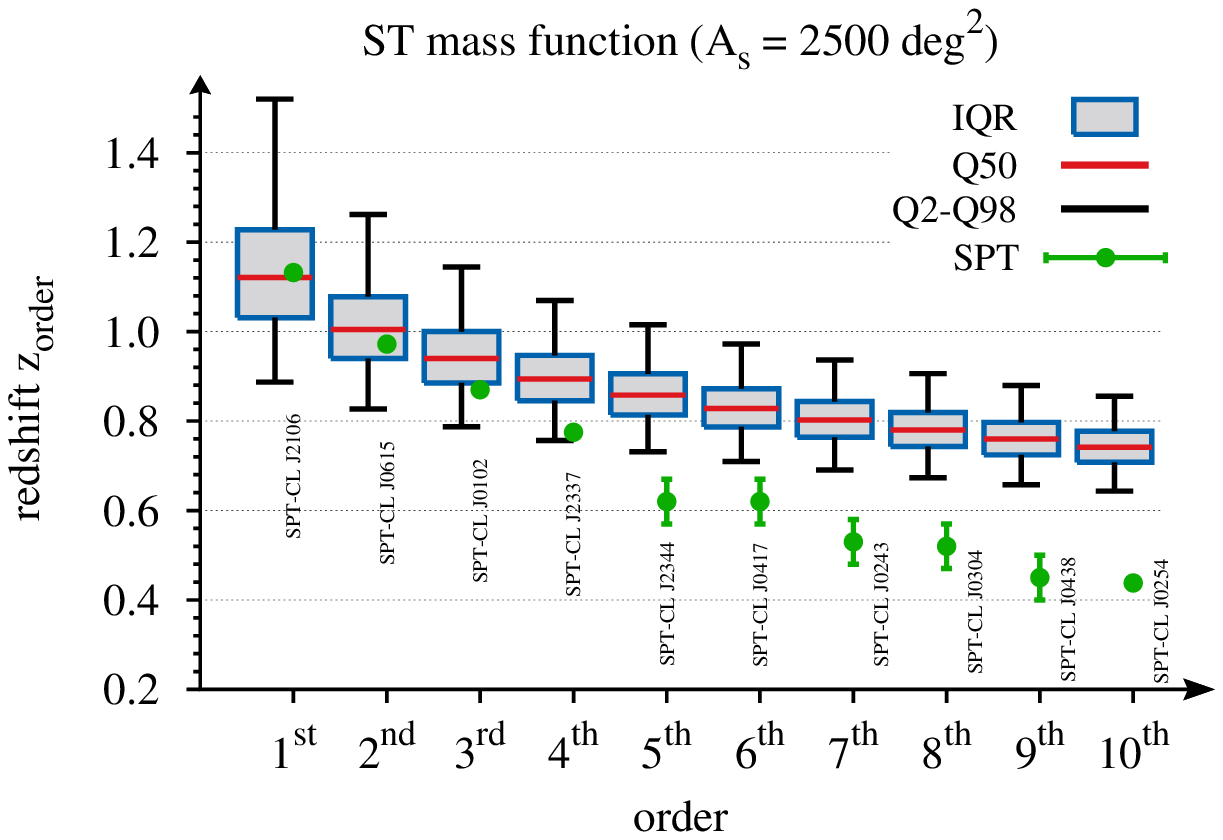}
\caption{Box-and-whisker diagram of the ten highest redshift clusters from the SPT catalogue 
as listed in \autoref{tab:clusters_redshift} for three different choices of the mass function as 
denoted in the title of each panel. For each order, the red lines denote the median ($Q$50), the 
blue-bordered, grey boxes give the IQR and the black whiskers mark the range between the $2$ 
and $98-$percentile ($Q$2, $Q$98). The green circles (with error bars in the case of photometric 
redshifts) represent the observed cluster redshifts.}
\label{fig:box_and_whisker_redshift}
\end{figure}

In \autoref{fig:box_and_whisker_redshift}, we present the box-and-whisker diagram in 
redshift, again for the PS, the Tinker and the ST mass functions (from top to bottom). The 
definition of boxes and whiskers remains unchanged with respect to 
\autoref{fig:box_and_whisker_mass}. Again, the data from \autoref{tab:clusters_redshift} 
is denoted by green error bars, which are negligibly small in the case of spectroscopic redshifts. 
Thus, we abstained from the MC simulation of the reshuffling in the case of redshift. 
While for the order statistics in mass the results only depended on the choice of the survey area,  
the situation is different for the order statistics in redshift. Here, a constant survey limiting 
mass is assumed, which will be subject to uncertainties for a real survey and, furthermore, will 
also exhibit some redshift dependence. Thus, the theoretical distributions are intrinsically less 
accurate than the ones with respect to cluster mass. Indeed, the comparison with the data 
in \autoref{fig:box_and_whisker_redshift} exhibits a different behaviour with respect to the one 
in \autoref{fig:box_and_whisker_mass}. Here, first four orders seem to be fit better by the 
Tinker mass function while the higher orders seem to favour the PS mass function. Taking the 
Tinker mass function as reference it seems that a few systems with $M>10^{15}\,M_\odot$ are 
missing at redshifts $z\gtrsim 0.7$. The difference with respect to the findings for the order 
statistics in mass for the same sample could, along the lines of 
\autoref{sec:dependence_on_cosmology}, be interpreted as a signature of a deviation from the 
reference $\Lambda$CDM model. However, considering the previously mentioned simplifying 
assumptions in the modelling of the theoretical distributions, we do not infer any cosmological 
conclusions and leave a better, more realistic, modelling of $m_{\rm lim}(z)$ of the SPT survey 
to a future work.

In the lower panel of \autoref{fig:exclusion_order}, we present the dependence of different 
percentiles ($Q2$,$Q25$,$Q50$,$Q75$ and $Q98$) on the order for a survey 
area of $A_{\rm s}=20\,000\,{\rm deg}^2$ and a constant limiting mass of 
$m_{\rm lim}=10^{15}\,M_\odot$. Taking the $Q98$ percentile as exclusion criterion, one 
would need to find ten clusters with $z\gtrsim1$, three clusters with $z\gtrsim1.2$ or one 
cluster with $z\gtrsim1.55$ in order to report a significant deviation from the $\Lambda$CDM 
expectations. Currently, SPT-CL J2106 is the only known cluster of such a high mass having a 
redshift $z>1$. With an assigned survey area of $A_s=2800\,{\rm deg}^2$ (ACT+SPT), it might 
from a statistical point of view still be possible to find ten objects that massive at $z>1$ in the 
larger survey area. The method presented in this work allows to construct similar exclusion 
criteria for any kind of survey design.
%-------------------------------------
%-------------------------------------
%------------------------------------------
\section{Comparison of the joint order statistics with observations}\label{sec:joint_analysis}
%------------------------------------------
Having studied the individual order statistics in mass and redshift in the previous section, 
we turn now to the study of the joint distributions of the order statistics as introduced in 
\autoref{sec:theory}.

\begin{figure*}
\centering
\includegraphics[width=0.497\linewidth]{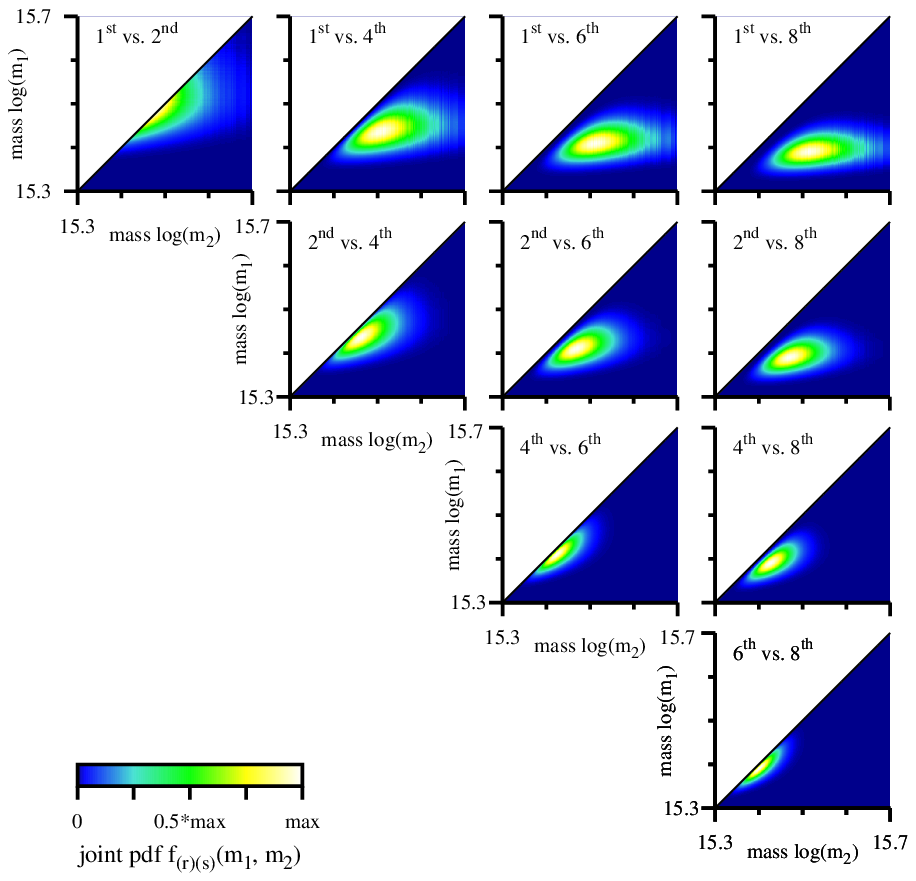}
\includegraphics[width=0.497\linewidth]{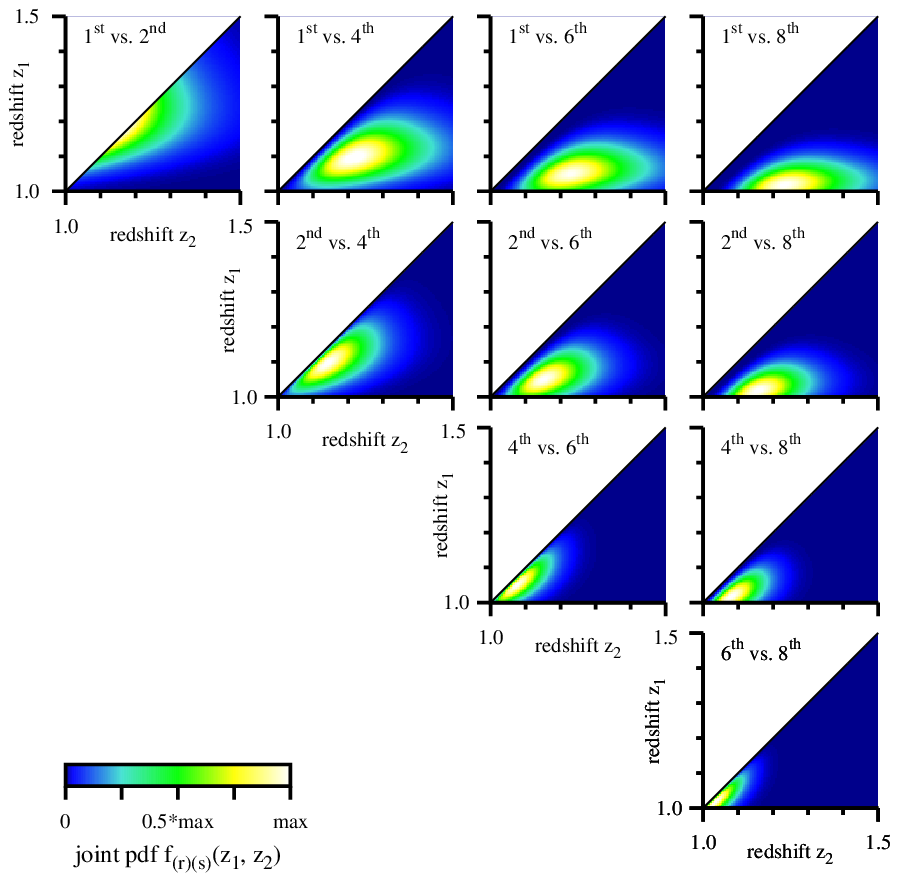}
\caption{Joint pdf $f_{(r)(s)}(x,y)$ (see \autoref{eq:joint_pdf_2d}) for the observable mass 
(left panel) and redshift (right panel) for different combinations of rank as indicated in the upper 
left of each small panel. The distributions are calculated for the fullsky and a constant limiting 
survey mass of $m_{\rm lim}=10^{15}\,M_\odot$ has been assumed for the joint distribution 
in redshift. The color bar is set to range from 0 to the maximum of the joint pdf for each rank 
combination.}
\label{fig:multi_comb}
\end{figure*}

The simplest case of a joint order distribution is two-dimensional. In this case the pdf and cdf 
are given by \autoref{eq:joint_pdf_2d} and \autoref{eq:joint_cdf_2d}, respectively. Starting with 
the joint pdf, we present in \autoref{fig:multi_comb} the joint distributions in mass (left panel) 
and redshift (right panel) for several order combinations as denoted in the individual panels. All 
calculations assume the full sky and the Tinker mass function. In the case of the joint distributions 
in redshift, we assume a constant limiting survey mass of $m_{\rm lim}=10^{15}\,M_\odot$. Due 
to the condition that $x<y$, all distributions are limited to a triangular domain.

An inspection of the different pdfs in \autoref{fig:multi_comb} reveals that, for the combination 
of the first and the second largest order (upper leftmost panel), the most likely combination of 
the observables is very close to the diagonal. This means that it is more likely to find the two 
largest values close to each other, at absolute values that are smaller than the extreme value 
statistics would imply for the maximum alone. Then, when moving to combinations of the first 
with higher orders (first row), it can be seen that the peaks of the pdfs move away from the 
diagonal and that they extend to larger values for the larger observable. This indicates that it is 
more likely to find the two systems with a larger separation in the observable when the 
difference between the considered orders is larger. Accordingly, for higher order combinations 
(lower rows), the peaks of the joint pdf move to smaller values of the observables. It 
should also be noted that the peaks steepen for higher order combinations, confining the pdfs to 
smaller and smaller regions in the observable plane. As an example, the first and second largest 
observations (upper leftmost panel) can be realised in much larger area than the sixth and eighth 
largest one (lower rightmost panel).

\begin{figure*}
\centering
\includegraphics[width=0.497\linewidth]{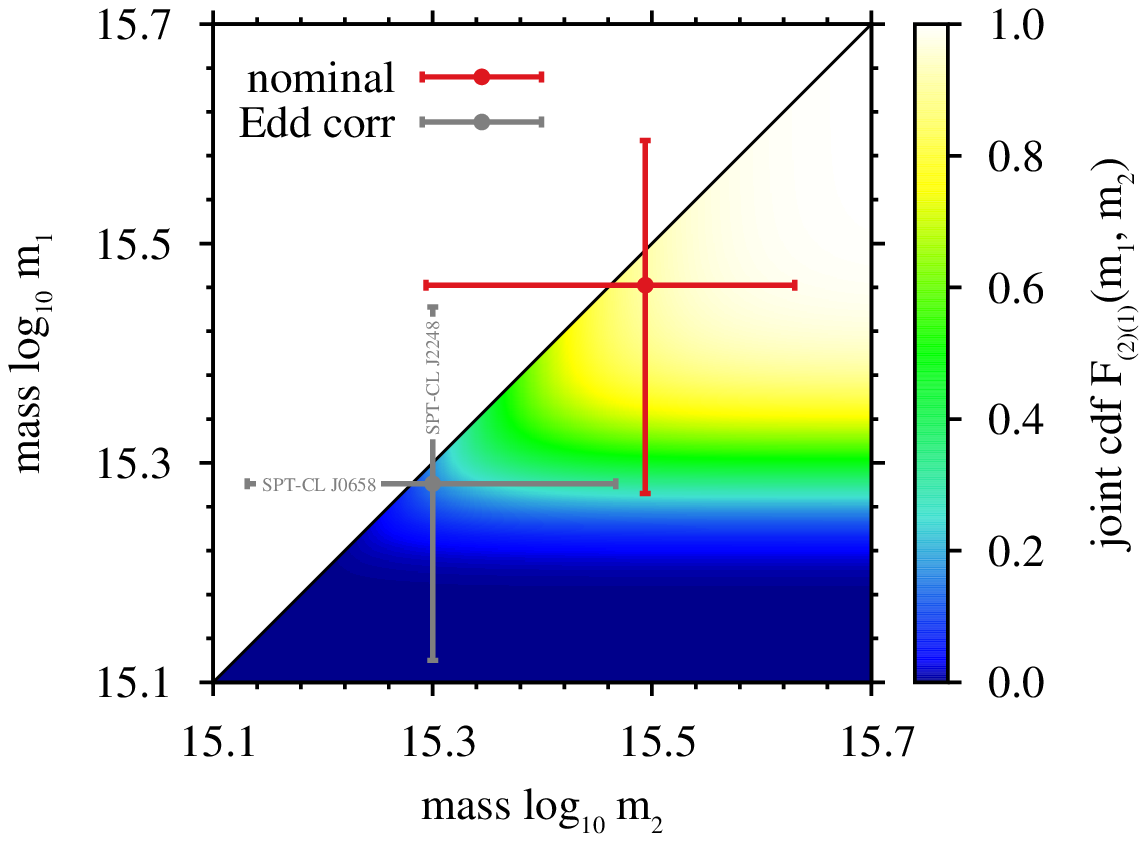}
\includegraphics[width=0.497\linewidth]{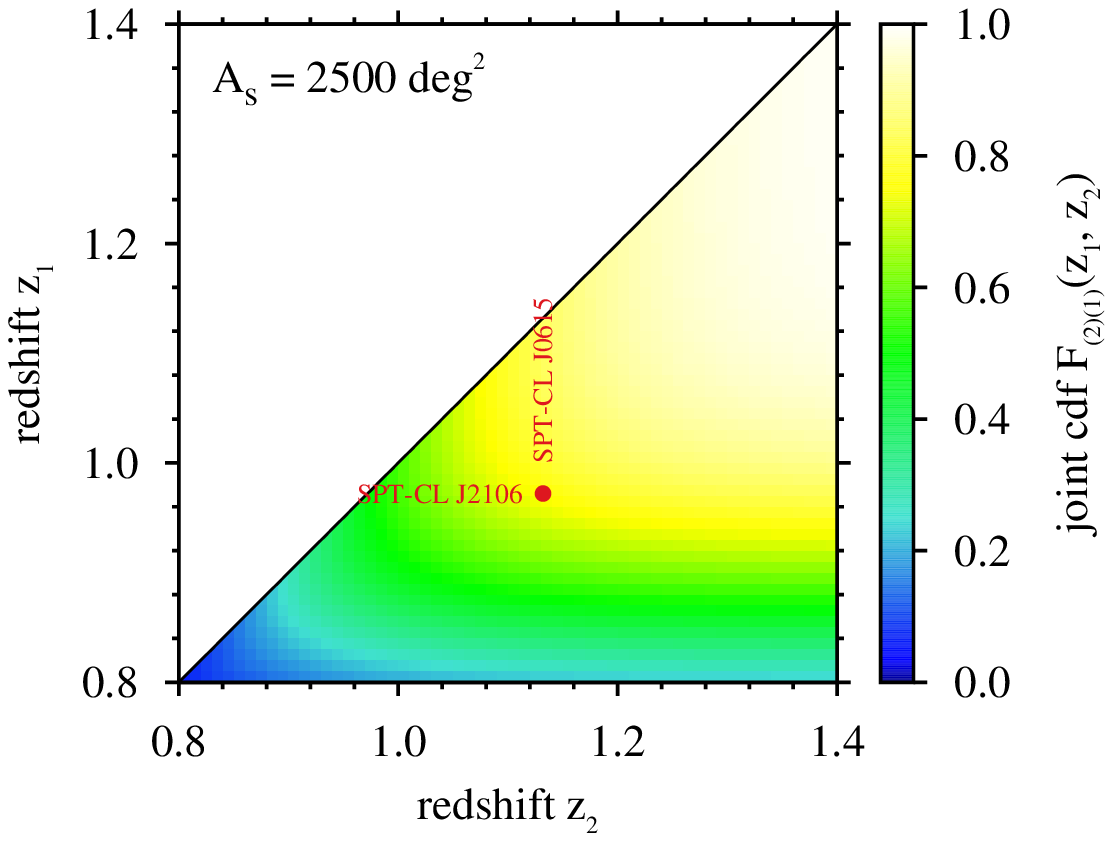}
\caption{Joint cdf $F_{(r)(s)}(x,y)$ (see \autoref{eq:joint_cdf_2d}) for the observable mass 
(left panel) and redshift (right panel) for the combination of the largest and second largest 
observation, assuming a survey area of $A_{\rm s}^{\rm SPT}=2500\,{\deg}^2$ and a constant limiting 
survey mass of $m_{\rm lim}=10^{15}\,M_\odot$ for the joint distribution in redshift. The red 
and grey error bars denote the nominal and the Eddington bias corrected values, $M_{200\rm m}$ 
and $M_{200\rm m}^{\rm Edd}$, for the SPT catalogue as listed in \autoref{tab:clusters_mass} 
for the mass and in \autoref{tab:clusters_redshift} for the redshift.}
\label{fig:cdf_2D_comparison}
\end{figure*}

Apart from the joint pdfs, it is also instructive to study the joint cdfs as presented in 
\autoref{fig:cdf_2D_comparison} for the observed mass (left panel) and redshift (right panel). 
In order to add observational data from the SPT catalogue, we assume a survey area of 
$A_{\rm s}^{\rm SPT}=2500\,{\deg}^2$ and a $m_{\rm lim}=10^{15}\,M_\odot$ for the joint 
distribution in redshift. Additionally, we added the two largest nominal observed (red error bars) 
and the Eddington bias corrected masses (grey error bars) from \autoref{tab:clusters_mass} to 
the left panel and the two highest redshifts of the SPT massive cluster sample from 
\autoref{tab:clusters_redshift} to the left panel. In the case of the mass, we find 
$F_{(n-1)(n)}\approx 0.92$ for the nominal and $F_{(n-1)(n)}\approx 0.1$ for the Eddington bias 
corrected masses. Hence, using the central values, in $\sim (8-90)$ percent of the cases a mass 
larger than the one of SPT-CL J0658 and a mass larger than the one of SPT-CL J2248 are observed. 
Thus, also the joint cdf confirms that the two largest masses do not exhibit any tension with the 
concordance cosmology. The same conclusion applies in the case of the joint distribution in redshift. 

By means of \autoref{eq:joint_pdf_Nd} these results can be extended to the $n$-dimensional case, 
allowing the formulation of a likelihood function of the ordered sample of the $n$ most massive 
or highest redshift clusters.
%-------------------------------------
%-------------------------------------
%-------------------------------------
%-------------------------------------
%------------------------------------------
\section{Summary}\label{sec:summary}
%------------------------------------------
In this work, we studied the application of order statistics to the mass and redshifts of galaxy clusters and 
compared the theoretically derived distributions with observed samples of galaxy clusters. Our work extends 
previous studies that hitherto considered only the extreme value distributions in mass or redshift. 

On the theoretical side, our results can be summarised as follows.
\begin{enumerate}
\item We introduce all relations necessary to calculate pdfs and cdfs of the individual and joint order 
statistics in mass and redshift. In particular, we find a steepening of the cdfs for higher order distributions 
with respect to the extreme value distribution of both mass and redshift. This steepening corresponds to 
a higher constraining power from distributions of the $n$-largest observations. The presented method 
extends previous works to include exclusion criteria based on the $n$ most massive or $n$ highest 
redshift clusters for a given survey set-up.
\item Conceptually, we avoid the bias due to an \textit{a posteriori} choice of the redshift interval in the 
case of the order statistics in mass by selecting the interval $0\le z\le\infty$. Hence, we study the statistics 
of the hierarchy of the most massive haloes in the Universe, which mostly stem from redshifts $z\lesssim 
0.5$. On the contrary, when choosing the order statistics in redshift, focus is laid on haloes that stem from 
the highest possible redshifts. However, the calculations will require a model of the survey characteristics 
in the form of a limiting survey mass as a function of redshift.
\item By putting the emphasis on either the most massive or on the highest redshift clusters above a 
given mass limit, the order statistics is e.g. particularly sensitive to the choice of the mass function. While 
the order statistics in mass is very sensitive to $\sigma_8$ and $\Omega_{\rm m}$ due to the domination 
of low redshift objects, the order statistics in redshift proves to be very sensitive to $w_0$ and $w_a$. For 
a fixed cosmology, both order statistics are efficient probes of the functional shape of the mass function 
at the high mass end.
\item In addition to the individual order statistics, we study as example case also the joint two order 
statistics. We find that for the combination of the largest and the second largest observation, it is most likely 
to find them to be realised with very similar values with a relatively broadly peaked distribution. 
%When combining the largest observation with higher orders, it is more likely to find a larger gap between the 
%observations and when combining higher orders in general, the joint pdf peaks more strongly. 
\item In order to allow a quick estimation of the distributions of the order statistics, we provide 
in the \autoref{sec:appendixB} fitting formulae for $F(x)$ as a function of the survey area 
for the cases of mass and redshift. The fitting formulae for the distribution in mass allow for a percentile 
estimation in the range from $Q2$ to $Q98$ with an accuracy better than one per cent for $A_{\rm s} 
\gtrsim 200\,{\rm deg}^2$ and for the ten largest masses. In the case of the order statistics in redshift, 
the quality of the fits depends on the chosen $m_{\rm lim}$. However, for survey areas of $A_{\rm s}\gtrsim 
2000\,{\rm deg}^2$ accuracies better than two per cent can be achieved for large values of $m_{\rm lim}=
10^{15}\;M_\odot$ and a lowering of $m_{\rm lim}$ further improves the accuracy.
\end{enumerate}
After introducing the theoretical framework, we compared the theoretical distributions with actually 
observed samples of galaxy clusters that we ranked by the magnitude of the observables mass and redshift. 
We decided to compile two catalogues, the main one is based on the SPT massive cluster sample 
\citep{Williamson2011} and additionally we analysed the meta-catalogue of X-ray detected clusters of galaxies 
MCXC \citep{Piffaretti2011} based on publicly available flux-limited all-sky survey and serendipitous cluster 
catalogues. This meta-catalogue can be considered as complete for $z\lesssim 0.3$ and, hence, by no means 
as complete as the SPT one. The results of the comparison can be summarised as follows.
\begin{enumerate}
\item In the case of the order statistics in mass, we compared the theoretical expectations for the ten largest 
masses for the PS, the Tinker and the ST mass functions. Assuming 
\textit{WMAP7} parameters, we find that the nominal and the Eddington bias corrected values for the observed 
masses favour the Tinker and the ST mass functions. When considering the possible bias due to a reshuffling 
of the ranks caused by the large error bars (statistical + systematic errors), we find that the SPT sample matches 
the Tinker mass function very well. The constraints are expected to tighten considerably once the error bars of 
all objects are scaled down by combining several cluster observables in multi-wavelength studies.   
\item In contrast to the ranking in mass, the order statistics of the SPT clusters in redshift is less well fit by the 
theoretical distributions based on the Tinker mass function. It appears that a few systems with $M>10^{15}\,
M_\odot$ are missing at redshifts $z\gtrsim 0.7$. One explanation could be found in a non-standard 
cosmological evolution to which the order statistics in redshift is more sensitive. However, it is more likely 
that a more precise modelling (including the redshift dependence) of the true limiting survey mass of SPT will 
account for the observed deviations.
\item Instead of utilising order statistics to perform exclusion experiments, it can also be used for consistency 
checks of the completeness of the observed sample and of the modelling of the survey selection function as 
indicated by the analysis of the MCXC (mass) and the SPT (redshift) samples.
\end{enumerate}
%------------------------------------------
\section{Conclusions}\label{sec:conclusions}
%------------------------------------------
We introduced a powerful theoretical framework which allows to calculate the expected 
individual and joint distribution functions of the $n$-largest masses or the $n$-highest redshifts of 
galaxy clusters in a given survey area. This approach is more powerful than the extreme value statistics 
that focusses on the statistics of the single largest observation alone. 

As a proof of concept, we compared the theoretical distributions with observed samples of galaxy clusters. 
However, data of sufficient quantity, uniformity and completeness is still sparse such that constraints 
are not particularly tight. This situation will most certainly improve in the near and intermediate future. 
Since the emphasis of this work lies on the introduction of the theoretical framework of order statistics 
and its application to galaxy clusters, we contended ourselves with a study of cluster masses and redshifts. 
Unfortunately, the mass of a galaxy cluster is not a direct observable and subject to large scatter and 
observational biases. In a follow-up work, we intend to extend the formalism to direct 
observables, like for instance X-ray luminosities, and to include the scatter in the scaling relations into 
the theoretical distributions.
%------------------------------------------------------
\section*{Acknowledgments}
%------------------------------------------------------
We acknowledge financial contributions from contracts ASI-INAF I/023/05/0, ASI-INAF I/088/06/0, 
ASI I/016/07/0 COFIS, ASI Euclid-DUNE I/064/08/0, ASI-Uni Bologna-Astronomy Dept. Euclid-NIS 
I/039/10/0, and PRIN~MIUR~2008 \textit{Dark energy and cosmology with large galaxy surveys}. M.B. is 
supported in part by the Transregio-Sonderforschungsbereich TR33 \textit{The Dark Universe} of the 
German Science Foundation.  J.C.W. would like to thank Lauro Moscardini and Ben Metcalf for the very 
helpful discussions.
%------------------------------------------------------
%Bibliography
%------------------------------------------------------
\bibliographystyle{mn2e}
\bibliography{gev5}

\appendix
%------------------------------------------------------
\section{Order statistics}\label{sec:appendixA}
%------------------------------------------------------
In this appendix, we outline the derivation of the most important relations of the order statistics and some 
subtleties considering their implementation. For more details we refer to the excellent textbooks on the 
topic by \cite{Arnold1992} and by \cite{David&Nagaraja2003} which we closely follow for the remainder 
of this appendix.
%------------------------------------------------------
\subsection{Individual distributions} \label{sec:appendixA_individual}
\begin{figure}
\centering
\includegraphics[width=0.99\linewidth]{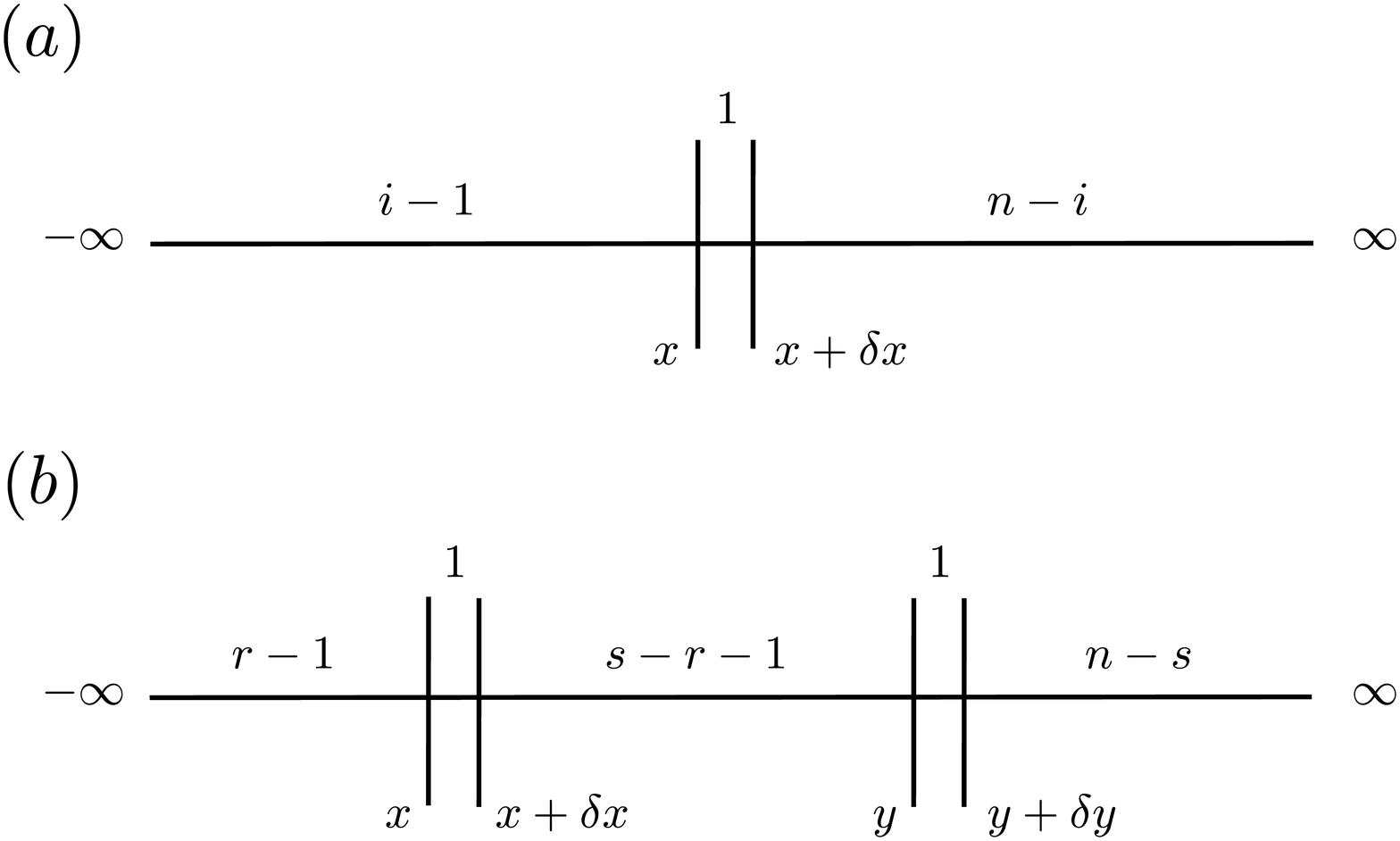}
\caption{Schematic for the derivation of $f_{(i)}(x)$ and $f_{(r)(s)}(x,y)$.}
\label{fig:orderstat_derivation}
\end{figure}

%------------------------------------------------------
Let $X_1,X_2,\dots ,X_n$ be a random sample of a continuous population with the cumulative distribution 
function, $F(x)$. Further, let $X_{(1)}\le X_{(2)}\le \cdots \le X_{(n)}$ be the order statistic, the random 
variates ordered by magnitude, where $X_{(1)}$ is the smallest (minimum) and $X_{(n)}$ denotes the largest 
(maximum) variate. The event $x<X_{(i)}\le x+\delta x$ is the same as the one depicted in panel $(a)$ of 
\autoref{fig:orderstat_derivation} and, thus, we have $X_k\le x$ for $i-1$ of the $X_k$, exactly one $X_k$ in 
$x<X_k\le x+\delta x$ and the remaining $n-i$ of the $X_k$ in $X_k>x+\delta x$. Now, the number of 
ways how $n$ observations can be arranged in the three regimes is given by
\begin{equation}
A(n,i)=\frac{n!}{(i-1)!1!(n-i)!}\, ,
\label{eq:App_pre_factor _pdf}
\end{equation}
where each of them has a probability of
\begin{equation}
\left[F(x)\right]^{i-1}\left[F(x+\delta x)-F(x)\right]\left[1-F(x)\right]^{n-i}.
\end{equation}
Therefore, under the assumption that $\delta x$ is small, we find for the probability
\begin{equation}
Pr\lbrace x<X_{(i)}\le x+\delta x\rbrace=A(n,i)\left[F(x)\right]^{i-1}\left[1-F(x)\right]^{n-i}f(x)\delta x,
\end{equation}
neglecting terms of $\mathcal{O}(\delta x)^2$. Dividing by $\delta x$ and performing $\delta x
\rightarrow 0$ 
yields the pdf as given in \autoref{eq:f_order}
\begin{eqnarray}
f_{(i)}(x)&=&\lim_{\delta x\rightarrow 0}\left\lbrace \frac{Pr\lbrace x<X_{(i)}\le x+\delta x\rbrace}{\delta x} 
\right\rbrace \nonumber \\
&=&A(n,i)\left[F(x)\right]^{i-1}\left[1-F(x)\right]^{n-i}f(x).
\label{eq:app_pre_pdf}
\end{eqnarray}
The corresponding cdf of the $i$-th order, as given by \autoref{eq:F_order} in \autoref{sec:order}, can 
now either be obtained by integrating the above equation or by the following argument
\begin{eqnarray}
F_{(i)}(x)&=&Pr\lbrace X_{(i)}\le x\rbrace \nonumber \\
			&=&Pr\lbrace\text{at least }i\text{ of } X_{(1)},\dots, X_{(n)}\text{ are at most } x\rbrace \nonumber \\
			&=&\sum_{k=i}^n Pr\lbrace\text{exactly }k\text{ of } X_{(n)},\dots, X_{(i)}\text{ are at most } x\rbrace \nonumber \\
			&=&\sum_{k=i}^n\binom{n}{k}\left[F(x)\right]^k\left[1-F(x)\right]^{n-k},
\end{eqnarray} 
for $-\infty < x < \infty$. Hence, the cdf of $X_{(i)}$ is equivalent to the tail probability (starting from $i$) of a 
binomial distribution with $n$ trials and a success probability of $F(x)$. By setting $i=n$ or $i=1$ one 
obtains the cdfs for the smallest and the largest order statistics as given by \autoref{eq:cdf_min} and 
\autoref{eq:cdf_max}.
%-------------------------------------
%-------------------------------------
%------------------------------------------------------
\subsection{Joint distributions}\label{sec:appendixA_joint}
%------------------------------------------------------
The joint pdf of the two order statistics $X_{(r)},X_{(s)}\;(1\le r < s \le n)$ for $x<y$ can be derived by 
similar arguments as for the single order statistics. The derivation scheme is now extended according to 
panel $(b)$ of \autoref{fig:orderstat_derivation}. Analogously to \autoref{eq:app_pre_pdf} we obtain 
\begin{eqnarray}
f_{(r)(s)}(x,y)&=&\lim_{\substack{\delta x\rightarrow 0 \\ \delta y\rightarrow 0}} 
\left\lbrace\frac{Pr\lbrace x<X_{(r)}\le x+\delta x,\,y<X_{(s)}\le y+\delta y\rbrace}{\delta x\delta y}
\right\rbrace\nonumber\\
&=&A(n,r,s)\nonumber\\
& &\times\,\left[F(x)\right]^{r-1}\left[F(y)-F(x)\right]^{s-r-1}\left[1-F(y)\right]^{n-s}\nonumber\\
& &\times\, f(x)f(y)\, ,
\label{eq:App_joint_pdf_2d}
\end{eqnarray}
where
\begin{equation}
A(n,r,s)= \frac{n!}{(r-1)!(s-r-1)!(n-s)!}\, .
\end{equation}

The joint cumulative distribution function can be obtained by integrating the pdf from above or 
again by the following direct argument
\begin{eqnarray}
F_{(r)(s)}(x,y)&=&Pr\lbrace X_{(r)}\le x,\,X_{(s)}\le y\rbrace \nonumber \\
			&=&Pr\lbrace\text{at least }r\, X_{(i)}\le x\,\wedge\text{ at least } s\,X_{(i)}\le y\rbrace 
			\nonumber \\
			&=&\sum_{j=s}^{n}\sum_{i=r}^{j} Pr\lbrace\text{exactly }i\, X_{(i)}\le x\,\wedge
			\text{ exactly } j\,X_{(i)}\le y \rbrace \nonumber \\
			&=&\sum_{j=s}^{n}\sum_{i=r}^{j}\frac{n!}{i!(j-i)!(n-j)!}\nonumber \\
			& &\left[F(x)\right]^{i}\left[F(y)-F(x)\right]^{j-i}\left[1-F(y)\right]^{n-j}.
\end{eqnarray} 
This is exactly identical to the tail probability of a bivariate binominal distribution.

Following the same line of reasoning as for the joint two order statistics, the above relations can 
be generalised to the joint pdf of $X_{n_1},\dots ,X_{n_k}\;(1\le n_1 <\cdots < n_k\le n)$ for 
$x_1\le \cdots\le x_k$, which is given by
\begin{align}
f_{(x_1)\cdots(x_k)}(x_1,\dots, x_k)=& \frac{n!}{(n_1-1)!(n_2-n_1-1)!\cdots (n-n_k)!} \nonumber\\
					&\times\left[F(x_1)\right]^{n_1-1}f(x_1)\left[F(x_2)-F(x_1)\right]^{n_2-n_1-1}\nonumber \\
					&\times f(x_2)\cdots\left[1-F(x_k)\right]^{n-n_k} f(x_k)\, .
\end{align}
The right hand side of this relation can be written in a more compact form \citep{David&Nagaraja2003} as
\begin{equation}
n!\left[\prod_{j=1}^{k}f(x_j)\right]\prod_{j=0}^{k}\left\lbrace\frac{\left[F(x_{j+1})-F(x_j)\right]
^{n_{j+1}-n_j-1}}{\left(n_{j+1}-n_j-1\right)!}\right\rbrace\, ,
\end{equation}
where we defined $n_0=0$, $n_{k+1}=n+1$, $x_0=-\infty$ and $x_{k+1}=+\infty$. 
%------------------------------------------------------
\subsection{Regarding the implementation}
%------------------------------------------------------
The implementation of the order statistics for the intended application of this work, as discussed 
in \autoref{sec:theory} and \autoref{sec:connection_to_cosmology}, is rather straightforward. 
However, one important subtlety arises from the combinatoric prefactors that contain factorials of 
$n$, which due to the large number of haloes cannot be calculated directly.
However, for all prefactors the factorials of $n$ can be avoided by writing them as products and by 
dividing out common terms. As a simple example, we take the prefactor from 
\autoref{eq:App_pre_factor _pdf}. In this case the index $i$ will, depending on the order, be given 
by a term like $i=(n-j)$ with $j=0$ for the distribution of the maximum, $j=1$ for the second largest 
and so on. Thus, we obtain
\begin{align}
A(n,i=n-j)	&=\frac{n!}{(i-1)!1!(n-i)!}=\frac{n!}{(n-j-1)!(n-n+j)!}\, ,\nonumber \\ 
				&=\frac{1}{j!}\prod_{k=0}^{j}(n-k)\, ,
\end{align}
which can be calculated for rather large values of $n$. In a similar manner, all combinatoric prefactors 
can be simplified and implemented.
%-------------------------------------
\begin{figure}
\centering
\includegraphics[width=0.99\linewidth]{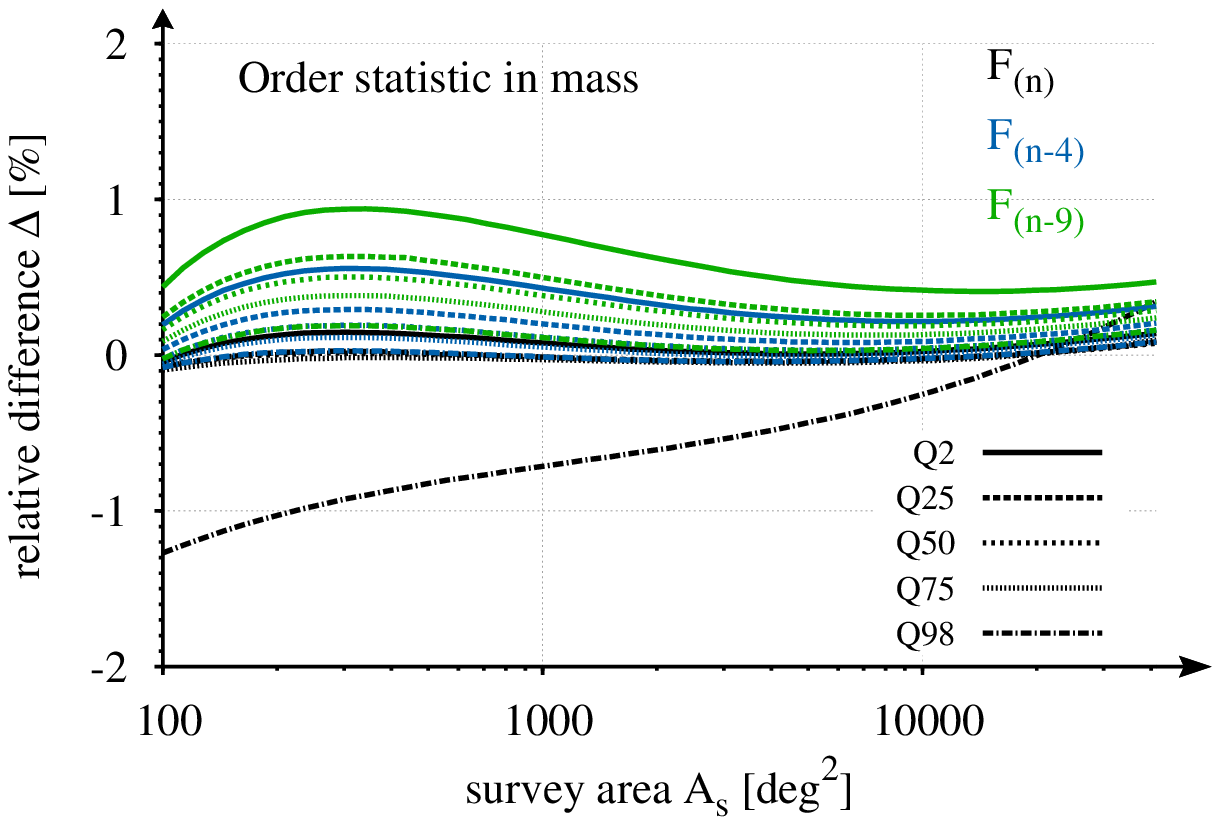}
\caption{Relative differences, $\Delta=(Q^{\rm fit}-Q^{\rm dir})/Q^{\rm dir}$, between the fitted 
and directly calculated percentiles (different line styles) as a function of the survey area for the 
order statistics in mass. The differences are shown for three different ranks, the largest (black lines), 
the fifth largest (blue lines) and the tenth largest (green lines) one.}
\label{fig:rel_diff_mass}
\end{figure}
%-------------------------------------
%------------------------------------------------------
\section{A fitting function for the order statistics}\label{sec:appendixB}
%------------------------------------------------------
In this additional section, fitting functions for the order statistics in mass and in redshift 
are defined. As functional form for the numerical fits, we will use \autoref{eq:cdf_gev} in 
combination with the relation \autoref{eq:cdf_max}, which yields
\begin{equation}
 F(x) = \left(\exp{\left\lbrace -\left[1+\gamma(y) \left(\frac{x-\alpha(y)}{\beta(y)}\right)
 \right]^{-1/\gamma(y)}\right\rbrace}\right)^{1/n(y)}\, .
 \label{eq:App_cdf_gev}
\end{equation}
Here, $x$ is the observable, either mass or redshift, and the GEV parameters $\alpha$, 
$\beta$ and $\gamma$ as well as the number of haloes\footnote{For the numerical 
calculation of $n$, we limit the mass range without loss of generality to the interval relevant 
for galaxy clusters of $10^{13}\, M_{\odot}\le m\le 10^{16}\, M_{\odot}$}, $n$, are functions 
of the survey area via the variable $y=\log_{10}(A_{\rm s})$. Once the cdf, $F(x)$, is known, 
all order statistics can be calculated by means of the relations discussed in 
the previous \autoref{sec:appendixA}. Inverting the cdfs of order statistics allows to obtain 
the percentiles which can then be utilised as $\Lambda$CDM exclusion criteria (see e.g. 
\autoref{fig:exclusion_order}).
%------------------------------------------------------
\subsection{Order statistics in mass}
%------------------------------------------------------
In order to determine the fitting function for the order statistics in mass, we calculate the GEV 
parameters according to \cite{Davis2011} and \cite{Waizmann2011} and the number of haloes, 
$n$, as a function of the survey area and fit them by the following functions
\begin{align}
\alpha(y)		&=5.99888\ln(y^{0.568634}+10.5689)\, ,\\
\beta(y)		&=0.362939 \exp(-1.11069y^{0.324255})\, ,\\
\gamma(y)	&=-0.239274\ln(y^{-0.448009}+0.747006)\, ,\\
n(y)				&=10^{y+2.94112}\, ,
\end{align}
where $y=\log_{10}(A_{\rm s})$. The observable $x$ in \autoref{eq:App_cdf_gev} is defined to be 
$x=\log_{10}(M_{200m}h)$. We present the results in \autoref{fig:rel_diff_mass} in the form 
of relative differences between the fitted and directly calculated values of five selected percentiles 
($Q$2, $Q$25, $Q$50, $Q$75 and $Q$98) as a function of the survey area. The different colors 
denote the largest order statistics, $F_{(n)}(x)$ (black lines), the fifth largest order statistics, 
$F_{(n-4)}(x)$ (blue lines) and the tenth largest order statistics, $F_{(n-9)}(x)$. The relative errors 
in the five different percentiles are for almost the complete range of survey areas on the sub-per 
cent level (only $Q$98 for $F_{(n)}(x)$ exhibits a slightly larger error for very small survey areas). 
%------------------------------------------------------
\subsection{Order statistics in redshift}
\begin{figure}
\centering
\includegraphics[width=0.99\linewidth]{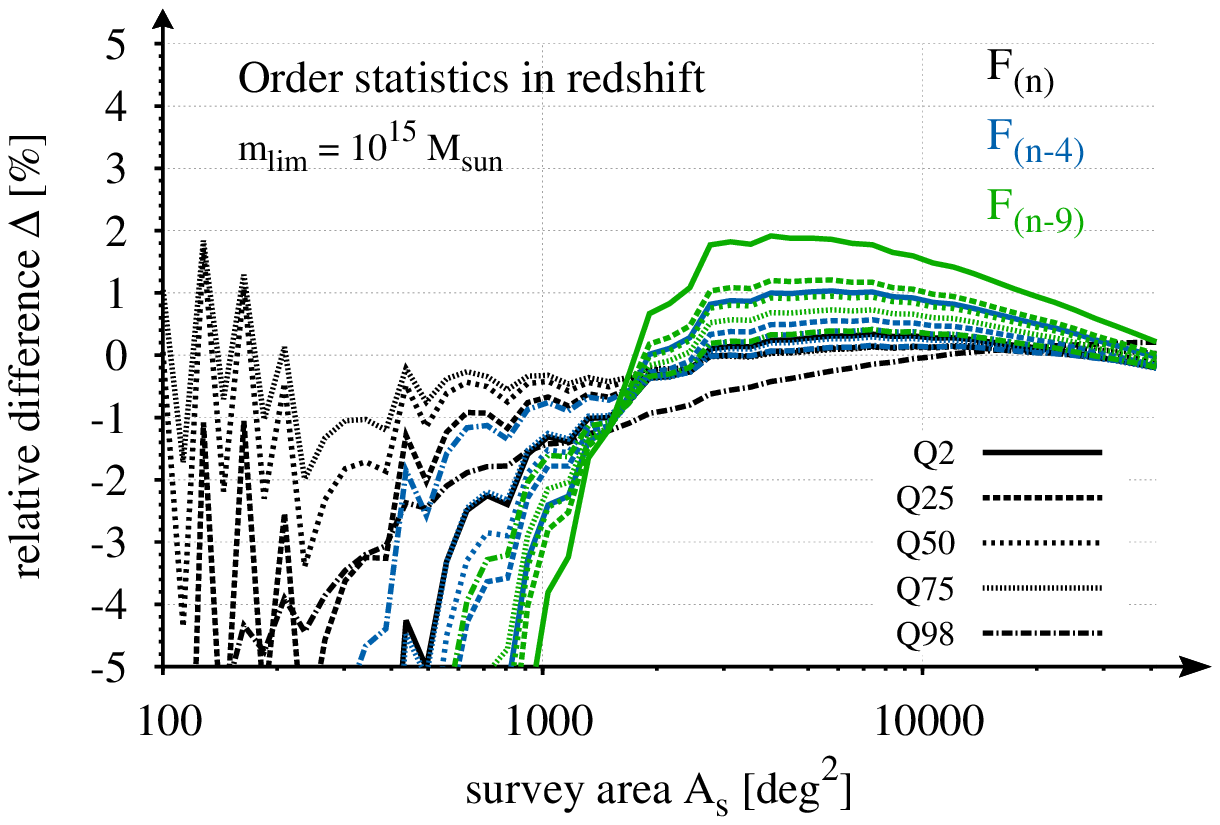}\\
\includegraphics[width=0.99\linewidth]{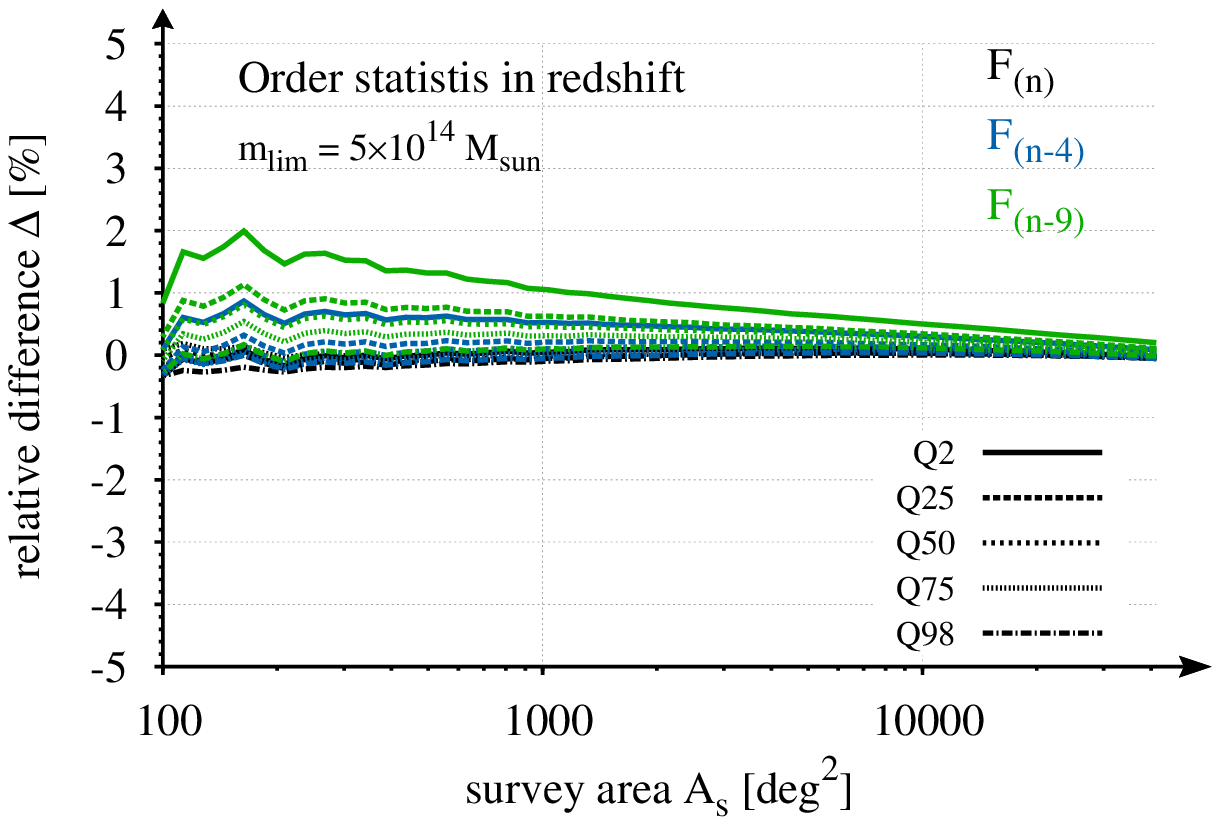}
\caption{Relative differences, $\Delta=(Q^{\rm fit}-Q^{\rm dir})/Q^{\rm dir}$, between the fitted 
and directly calculated percentiles (different line styles) as a function of the survey area for the 
order statistics in redshift assuming $m_{\rm lim}=10^{15}\,M_\odot$ (upper panel) and 
$m_{\rm lim}=5\times 10^{15}\,M_\odot$ (lower panel). The differences are shown for three 
different ranks, the largest (black lines), the fifth largest (blue lines) and the tenth largest 
(green lines) one.}
\label{fig:rel_diff_redshift}
\end{figure}

%------------------------------------------------------
For fitting the order statistic in redshift, we proceed in a similar way as for the 
mass, setting $x=z$ in \autoref{eq:App_cdf_gev}. For calculating the GEV parameters as a function 
of the survey area, we follow the approach presented in (Metcalf \& Waizmann in preparation). However, 
since in contrast to the order statistics in mass, the distributions depend on the choice of the limiting 
survey mass, we fitted the distributions for two choices of $m_{\rm lim}$. First, we set $m_{\rm lim}=
10^{15}\,M_\odot$, identical to the setup we discussed in this paper for the SPT massive cluster sample. 
Secondly, we lower the threshold to $m_{\rm lim}=5\times 10^{14}\,M_\odot$. In the first case, we obtain 
\begin{align}
\alpha(y)		&=1.13729\ln(0.567735y+0.332933)\, ,\\
\beta(y)		&=\exp[-\exp(-1.76728y^{-1.84932}+0.929307)]\, ,\\
\gamma(y)	&= -2.23597\ln(y^{-2.96376}+1.01017)\, ,\\
n(y)				&=10^{0.981095y-1.52015}\, .
\end{align}
and for the second choice we find
\begin{align}
\alpha(y)		&=2.1084\ln(0.284062y+1.09002)\, ,\\
\beta(y)		&=\exp[-\exp(-0.905364y^{-0.375228}+1.30066)]\, ,\\
\gamma(y)	&= -0.260275\ln(y^{-1.55487}+1.0592)\, ,\\
n(y)				&=10^{0.998552y-0.451364}\, ,
\end{align}
where $y=\log_{10}(A_{\rm s})$ for both cases. We present the results in \autoref{fig:rel_diff_redshift} again 
as relative differences. It can be seen that in the case of high limiting mass (upper panel), the fit performs 
poorly for survey areas smaller than $\sim 1000\,{\rm deg}^2$ due to the insufficient number of haloes that 
are expected to be found. However, above  $\sim 2000\,{\rm deg}^2$ the percentiles of the first ten orders 
can be fitted with an accuracy better than two per cent. 

If the limiting mass is lowered, the quality of the fit improves drastically as shown in the lower panel of 
\autoref{fig:rel_diff_redshift} for $m_{\rm lim}=5\times 10^{14}\,M_\odot$. In this case, sub-percent-level 
accuracy is reached for $A_{\rm s} \geq 1000\,{\rm deg}^2$ and an accuracy better than two per cent down to 
$100\,{\rm deg}^2$. 
%-------------------------------------
%-------------------------------------

\label{lastpage}
   
\end{document}